\documentclass[traditabstract]{aa} 

\usepackage{graphicx}
\usepackage[varg]{txfonts}
\usepackage{multirow} 
\usepackage[FIGTOPCAP, center, nooneline]{subfigure}
\usepackage{upgreek} 
\usepackage[]{natbib}
\newcommand{\HII}{H\,{\sc ii}}
\newcommand{\HI}{H\,{\sc i}}

\newcommand{\OI}{O\,{\sc i}}

\bibpunct{(}{)}{;}{a}{}{,}

\begin{document}
   \title{Complex organic molecules in strongly UV-irradiated gas\thanks{Based on observations obtained with the IRAM 30m telescope. IRAM is supported by INSU/CNRS (France), MPG (Germany), and IGN (Spain).}}

     \author{S. Cuadrado\inst{\ref{inst1}}\and J. R. Goicoechea\inst{\ref{inst1}}\and J. Cernicharo\inst{\ref{inst1}}\and A. Fuente\inst{\ref{inst2}}\and J. Pety\inst{\ref{inst3},\ref{inst4}}\and B. Tercero\inst{\ref{inst1}}}

   \institute{
   Grupo de Astrof\'{\i}sica Molecular. Instituto de Ciencia de Materiales de Madrid (CSIC), Sor Juana In\'es de la Cruz 3, 28049 Cantoblanco, Madrid, Spain.  \email{s.cuadrado@icmm.csic.es}\label{inst1}    
   \and Observatorio Astron\'omico Nacional, Apdo. 112, 28803 Alcal\'a de Henares, Madrid, Spain \label{inst2}
   \and Institut de Radioastronomie Millim\'etrique (IRAM), 300 rue de la Piscine, F-38406 Saint Martin
d'H\`eres, France \label{inst3}
  \and LERMA, Observatoire de Paris, CNRS UMR 8112, \'Ecole Normale Sup\'erieure, PSL research
university, 24 rue Lhomond, 75231, Paris Cedex 05, France\label{inst4}}

   \date{Received 18 January 2017 / Accepted 12 May 2017}

 
  \abstract{We investigate the presence of complex organic molecules (COMs)
	  	in strongly UV-irradiated interstellar molecular gas.  
	  	We have carried out a complete millimetre (mm) line survey using the \mbox{IRAM 30 m} telescope
	  	towards the edge of the Orion Bar photodissociation region (PDR), close to the H$_2$ dissociation front, a position irradiated by a very intense far-UV (FUV) radiation field.
		These observations have been complemented with \mbox{8.5$''$ resolution
	  	maps} of the H$_2$CO \mbox{$J_{K_{\rm a},K_{\rm c}}$ = 5$_{1,5} \rightarrow$ 4$_{1,4}$} and C$^{18}$O \mbox{$J$ = 3 $\rightarrow$ 2} 
	  	emission at 0.9\,mm. Despite being a harsh
	  	environment, we detect more than 250 lines from COMs and related
	  	precursors: H$_{2}$CO, CH$_{3}$OH, HCO, H$_{2}$CCO, CH$_{3}$CHO, 
	  	H$_{2}$CS, HCOOH, CH$_{3}$CN, CH$_{2}$NH, HNCO, H$_{2}^{13}$CO, and HC$_{3}$N 
	  	(in decreasing order of abundance). For each species, the large number of detected lines allowed
	  	us to accurately constrain their rotational temperatures ($T_{\rm rot}$) and 
	  	column densities ($N$). Owing to subthermal excitation and intricate spectroscopy 
	  	of some COMs (symmetric- and asymmetric-top molecules such as CH$_{3}$CN and H$_{2}$CO, respectively), 
	  	a correct determination of $N$ and 
	  	$T_{\rm rot}$ requires building rotational population diagrams of their
	  	rotational ladders separately. The inferred column densities are in the 
	  	\mbox{10$^{11} - $10$^{13}$\,cm$^{-2}$} range.  
	  	 We also provide accurate upper limit abundances for chemically related  molecules 
	  	 that might have been expected, but
	  	are not conclusively detected at the edge of the PDR
	  	(HDCO, CH$_{3}$O, CH$_{3}$NC, CH$_{3}$CCH, CH$_{3}$OCH$_{3}$, 
	  	HCOOCH$_{3}$, CH$_{3}$CH$_{2}$OH, CH$_{3}$CH$_{2}$CN, and CH$_{2}$CHCN).
	  	 A non-thermodynamic equilibrium excitation analysis for molecules with known collisional rate coefficients suggests that some COMs arise from different 
	  	PDR layers but we cannot resolve them spatially. 
	  	In particular, H$_2$CO and CH$_3$CN survive
	  	in the extended gas directly exposed to the strong FUV flux 
	  	(\mbox{$T_{\rm k}$ = 150 $-$ 250 K} and \mbox{$T_{\rm d}$ $\gtrsim$ 60 K}), 
	  	whereas CH$_3$OH only arises from denser and 
	  	cooler gas clumps in the more shielded PDR interior (\mbox{$T_{\rm k}$ = 40 $-$ 50 K}).
	  	The non-detection of HDCO towards the 
	   PDR edge is consistent with the minor role of pure gas-phase deuteration at very
	    high temperatures. We find a
	  	HCO/H$_2$CO/CH$_3$OH\,$\simeq$\,1/5/3 abundance ratio. 
	  	These ratios are different from those inferred in hot cores and shocks.
		Taking into account the 
	  	elevated gas and dust temperatures at the edge of the Bar (mostly mantle-free 
	  	grains), we suggest the following scenarios for the formation of COMs:
	  	 (i) hot gas-phase reactions
	  	not included in current models;
	  	(ii) warm grain-surface chemistry; or (iii) the
	  	PDR dynamics is such that COMs or precursors formed in cold icy grains deeper inside the
	  	molecular cloud
	  	desorb and advect into the PDR.}

   

   \keywords{ astrochemistry - surveys - ISM: photon-dominated region (PDR) - ISM: molecules - ISM: abundances.}

   \maketitle
%
\section{Introduction}\label{Intro}

Almost 200 molecules have been detected in the interstellar medium (ISM) and circumstellar shells. 
A very large fraction of them are complex organic molecules (COMs). COMs are traditionally defined as carbon-based molecular species with more than six atoms in their structure \citep{Herbst_2009,Caselli_2012}. To date, the largest organic molecules detected in the ISM (excluding PAHs and fullerenes) are propyl cyanide (C$_{3}$H$_{7}$CN; \citealt{Belloche_2009,Belloche_2014}) and benzene (c-C$_{6}$H$_{6}$; \citealt{Cernicharo_2001}).

COMs have been detected in the ISM since the 1970s (e.g. CH$_{3}$CHO, \citealt{Gottlieb_1973}; or HCOOCH$_{3}$, \citealt{Brown_1975}). Most of the detections have been reported towards hot cores, that is, 
dense gas surrounding high-mass protostars in massive star-forming regions such as Sgr~B2 and Orion~KL \citep[e.g.][]{Bisschop_2007,Ziurys_1993,Tercero_2013,Tercero_2015,Ikeda_2001}, and towards hot corinos, that is, the low-mass analogs of hot cores such as IRAS~16293-2422 \citep{Cazaux_2003} and NGC~1333 IRAS~4A \citep{Bottinelli_2004}. The high degree of chemical complexity (e.g. C$_{2}$H$_{5}$CN, CH$_{3}$CCH, HCOOCH$_{3}$, CH$_{3}$OCH$_{3}$...) found in these protostellar sources is thought to result from thermal desorption of the ice mantles coating dust grains.
 \citet{Herbst_2009} have reviewed the subject. They classify the observed COMs towards protostars in three different generations depending on the time they are produced. The zeroth-generation species, such as H$_{2}$CO and CH$_{3}$OH, are formed through grain-surface reactions in a previous cold interstellar phase in which icy mantles built up around granular cores of carbon and silicate grains. The first-generation species, such as HCOOCH$_{3}$, are formed during the passive warm-up of the inner envelope of the protostar. During this phase, zeroth-generation species are photodissociated producing radicals such as HCO and CH$_{3}$O, which can associate through surface reactions to form larger molecules. Finally, the second-generation species are formed once the core has become a hot core or hot corino. The dust temperature at this stage is high enough \mbox{($T_{\rm d}$  $\gtrsim$ 100 K)} to sublimate the mantles completely, and new molecules are formed through warm gas-phase reactions (ion molecule and endothermic neutral-neutral reactions).
 
  When protostellar outflows impact the ambient envelope material, grains can be eroded, and because of the high temperatures reached in shocks, icy mantles can sublimate. In sufficiently high-velocity shocks, ices  can also be sputtered and directly injected into the gas phase \mbox{(e.g. $v_{\rm s}$ $>$ 20 $-$ 25\,km\,s$^{-1}$} for water ice, \citealt{Draine_1983}).
Several observations of the L1157 outflow have revealed the existence of H$_{2}$CO, H$_{2}$CS, CH$_{3}$OH, HC$_{3}$N, HNCO, NH$_{2}$CHO, and CH$_{3}$CHO in shocked gas \citep{Bachiller_1997, Mendoza_2014, Codella_2015}. 
In addition, \citet{Requena-Torres_2006} have shown that the galactic centre contains  dense clouds that are rich in CH$_{3}$OH, HCOOCH$_{3}$, and CH$_{3}$OCH$_{3}$.
 Widespread shocks have been invoked to explain their abundance and extended spatial distribution. The relative importance of the gas-phase reactions immediately after ice mantle sublimation compared to a formation of COMs on the grain surfaces is however far from \mbox{being understood}.
 
Recent observations towards UV-shielded cold cores (e.g.~TMC\,1, L1689B, or Barnard 1-b) have revealed molecules once considered to be present only in hot molecular cores \citep[e.g.][]{Remijan_2006,Bacmann_2012,Cernicharo_2012b}. In these cold environments, COMs are thought to form on the surface of grains and to be released through non-thermal desorption processes, chemical desorption, direct desorption by cosmic ray impacts, or secondary photon induced processes \citep{Cernicharo_2012b}.
Finally, a  number of organic species have been identified in circumstellar envelopes around evolved stars \citep[e.g.~IRC+10216,][]{Cernicharo_2000}, towards extragalactic sources \citep[e.g.][]{Meier_2005, Meier_2012, Aladro_2011}, meteorites \citep[e.g.][]{Cronin_1993}, and comets \citep[e.g.][]{Bockelee_2004}.

Studies of environments permeated by stellar far-UV photons  (FUV, \mbox{6.0 eV $<$ $h\nu$ $<$ 13.6 eV}) are more scarce. 
\citet{Guzman_2014} and \citet{Gratier_2013} presented the unexpected detection of HCOOH, CH$_{2}$CO, CH$_{3}$CN, CH$_{3}$OH, CH$_{3}$CHO, and CH$_{3}$CCH in the Horsehead photodissociation region (PDR; a relatively low-FUV-flux dense PDR, \mbox{$\chi \approx$ 60} times the mean interstellar field in Draine units), finding enhanced abundances compared to a nearby cold and dense core shielded from external FUV radiation. 
\citet{Guzman_2014} proposed that owing to the cold grain temperatures, ice-mantle photodesorption processes dominate the formation of COMs 
 in the Horsehead. In lower-density translucent clouds \mbox{($\chi$ $\approx$ 1)},
only H$_2$CO has been unambiguously detected \citep{Liszt_2006}. These observations might
suggest that, in FUV-irradiated environments, the presence of COMs diminishes  as the $\chi$/$n_{\rm H}$ ratio increases. Thus, COMs might not be present in strongly FUV-irradiated gas.

In this work we test the above hypothesis and investigate the presence and abundances of COMs at the high FUV-illuminated edge of the Orion Bar, with an impinging radiation field of a few 10$^{4}$ times 
the mean interstellar field \citep{Marconi_1998}. Because of its proximity \mbox{(414 $\pm$ 7 pc}, \citealt{Menten_2007}) and nearly edge-on orientation, 
the Orion Bar provides an excellent template to determine the chemical content and
also to investigate the structure and dynamics of strongly FUV-irradiated molecular gas \citep[e.g.][]{Tielens_1993,Hogerheijde_1995,Cuadrado_2015a,Cuadrado_2016,Goicoechea_2016}.

Multi-wavelength observations towards different positions of the Orion Bar have been historically used in the development of PDR models \citep[e.g.][]{Tielens_1985b, Tielens_1985a} and today they are still used as a template to understand the unresolved emission from sources as different as the nuclei of distant starburst galaxies \citep[e.g.][]{Fuente_2008} or the illuminated surfaces of protoplanetary disks \citep[e.g.][]{Agundez_2008a}. 
The transition from ionised to neutral gas in the Orion Bar has been extensively mapped, generally at low angular resolution,
in various atomic and molecular tracers \citep[see e.g.][]{Tielens_1993, Hogerheijde_1995, vanderWerf_1996,Walmsley_2000, Leurini_2010, Ossenkopf_2013}. 
The detailed analysis of these observations suggested an inhomogeneous density distribution. 
The most commonly accepted scenario is that an extended gas component, with mean gas densities
of \mbox{10$^{4}$ $-$ 10$^{5}$ cm$^{-3}$}, causes the chemical stratification seen perpendicular to the dissociation 
front as the FUV field is attenuated \citep{Hogerheijde_1995, Jansen_1995, Simon_1997, Wiel_2009, Habart_2010, van_der_Tak_2013}.
 In addition, another component of clumpy material with higher densities (\mbox{$\gtrsim$ 10$^6$ cm$^{-3}$}) and more shielded from FUV radiation is embedded in the interclump gas \citep[][]{Burton_1990, Parmar_1991, Stoerzer_1995, YoungOwl_2000, Lis_2003, Batrla_2003, Andree-Labsch_2017}. 
 Previous observations of H$_{2}$CO and CH$_{3}$OH in the Orion Bar have shown that the two molecules trace these two different environments: CH$_{3}$OH, the denser and cooler clumps seen deeper inside the Bar, and H$_{2}$CO, the warmer interclump medium directly exposed to the strong FUV-field \citep{Leurini_2006, Leurini_2010}.

\begin{figure*}
\centering
\includegraphics[scale=0.45,angle=0]{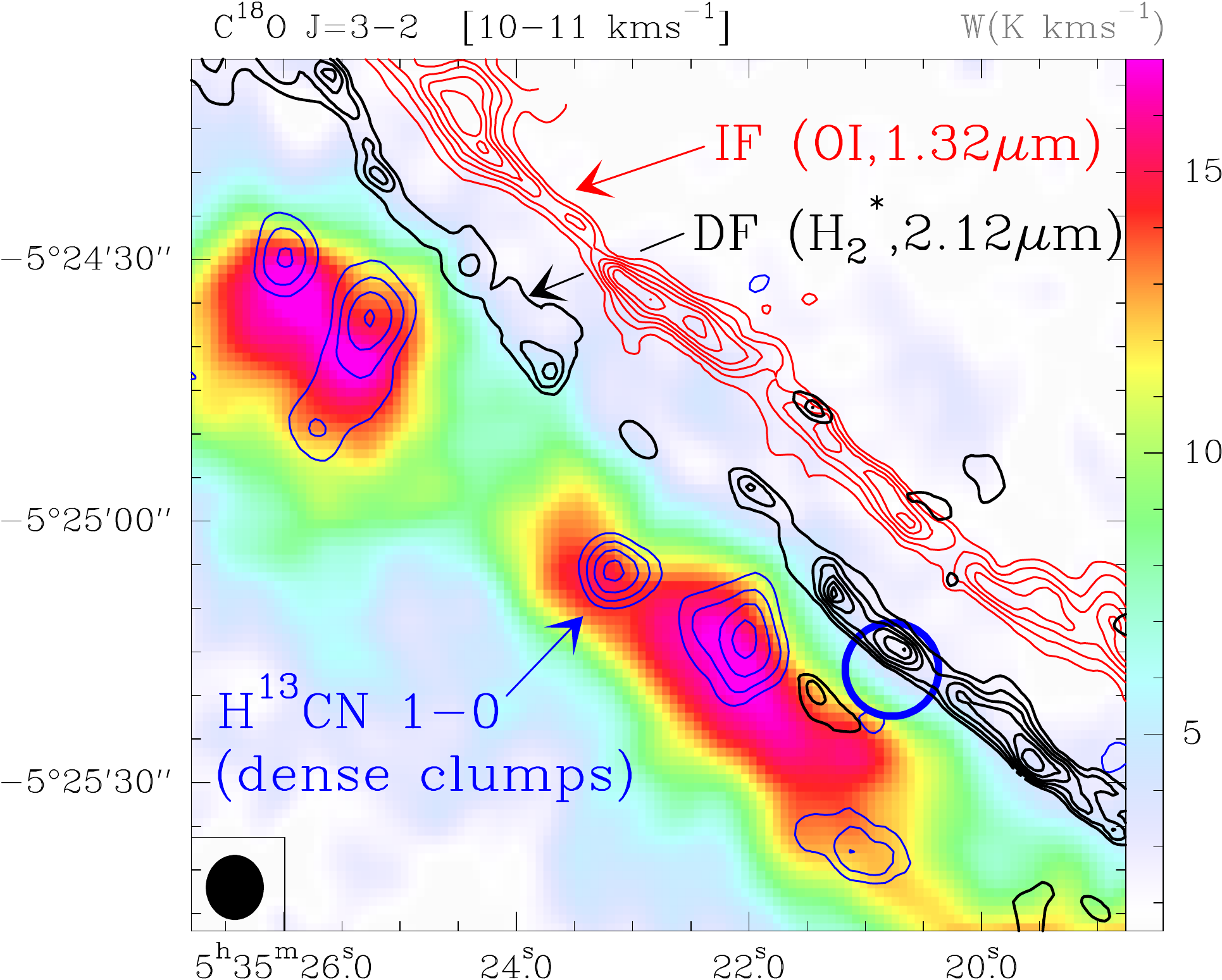} \hspace{0.5cm}
\includegraphics[scale=0.45,angle=0]{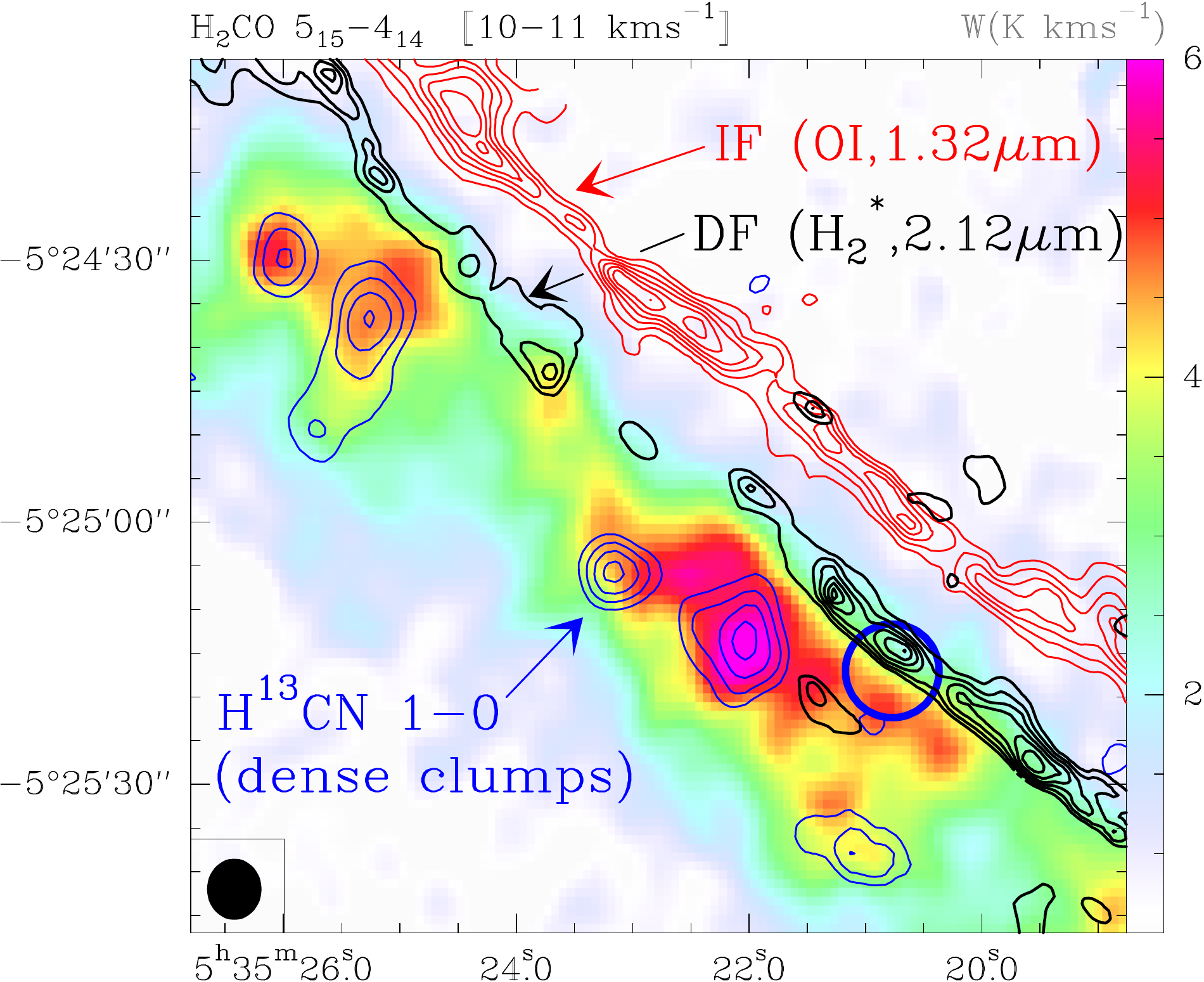}
\caption{C$^{18}$O \mbox{$J$ = 3 $\rightarrow$ 2} (left panel) and H$_2$CO \mbox{$J_{K_{\rm a},K_{\rm c}}$ = 5$_{1,5} \rightarrow$ 4$_{1,4}$} (right panel) line integrated intensity maps (W in \mbox{K km s$^{-1}$)} in the \mbox{10 $-$ 11\,km~s$^{-1}$} velocity interval observed with the IRAM~30~m telescope at $\sim$329\,GHz and $\sim$351\,GHz, respectively (colour scale).
Black contours are the H$_2^*$ \mbox{$v$ = 1 $\rightarrow$ 0} $S$(1) emission delineating the H$_2$ dissociation front (DF, \citealt{Walmsley_2000}). The red contours represent the \OI\, fluorescent line at 1.32~$\upmu$m \citep{Walmsley_2000} marking the position of the ionisation front (IF) that separates the \HII\,region and the neutral cloud. The blue contours represent the H$^{13}$CN \mbox{$J$ = 1 $\rightarrow$ 0} emission tracing dense molecular clumps inside the Bar \citep{Lis_2003}. The target position of the Orion Bar survey, close to the DF, is indicated with a blue circle. The IRAM~30~m beam at 1~mm is plotted in the bottom left corner (black circle).
}
\label{fig:Maps}
\end{figure*}

We have performed a complete millimetre line survey using the IRAM~30~m telescope towards the edge of Orion Bar, a high FUV-illuminated position close to what \citet{Stoerzer_1995} call the ``CO$^{+}$ peak'', near the H$_2$ dissociation front (see Fig.~\ref{fig:Maps}).
This position shows a distinctive chemistry that can only be understood due to the presence of a strong flux of FUV photons (e.g. compared to that in the more shielded clumps deeper inside the Bar): enhanced abundances of simple reactive ions (e.g. CH$^+$, CO$^+$, and HOC$^+$) and small hydrocarbon ions (e.g. $l$-C$_{3}$H$^+$) that are only abundant in the presence of C$^+$;  vibrationally excited H$_2$; and high gas temperatures \citep[e.g.][]{Nagy_2013, Guzman_2015, Cuadrado_2015a}. Indeed, high-angular resolution  ALMA-ACA images do show that reactive ions such as SH$^+$ or HOC$^+$ do not emit from the more shielded clumps \citep{Goicoechea_2017}. Therefore, their chemistry is different to that of the PDR edge observed in this work. Our survey covers $\sim$220 GHz of bandwidth, between 80 GHz and 360 GHz. These observations have been complemented with several 8.5$''$ resolution maps of different molecules at 0.9~mm to put our line survey position in the context of the Bar large-scale emission. In this paper we focus on the detection of rotational lines from COMs and related precursors.

The paper is organised as follows. In Sect.~\ref{Obs} we describe the line survey and the mapping observations. In Sect.~\ref{Results} we report the observational features of the detected organic molecules, while in Sect.~\ref{Spatial_distribution} we present the C$^{18}$O and H$_2$CO integrated line intensity maps at 0.9~mm.
The data analysis is explained in Sect.~\ref{Analysis}. In Sect.~\ref{Discussion} we discuss the results, and finally in Sect.~\ref{Summary} we summarise the main conclusions.


\section{Observations and data reduction} \label{Obs}

\begin{table}
\centering 
\caption{Observed frequency ranges and telescope parameters.}
\label{Table_efficiences}     
\begin{tabular}{c c c c c c@{\vrule height 10pt depth 5pt width 0pt}} 
\hline\hline      
 Rec.\tablefootmark{a}   &   Backend &   Freq.\tablefootmark{b} & $\updelta v$\tablefootmark{c} &  $\mathrm{\eta_{_{MB}}}$\tablefootmark{d}    &  HPBW\tablefootmark{e} \\

   &   &  [GHz] & [km s$^{-1}$]   &           &   [arcsec]   \\
\hline   

E0                            &  FFTS    & \ \ 80$-$117  &   0.75$-$0.51   &    0.87$-$0.82    &    31$-$21    \\  \hline   
E1                            &  WILMA   & 128$-$176     &    4.7$-$3.4    &    0.80$-$0.74    &    19$-$14    \\  \hline
E2                            &  FFTS    & 202$-$275     &   0.30$-$0.22   &    0.70$-$0.56    &    12$-$9 \ \    \\  \hline
\multirow{2}{*}[-0.1cm]{E3}   &  FFTS    & 275$-$305     &   0.22$-$0.20   &    0.56$-$0.50    &     9$-$8      \\  
                              &  FFTS    & 328$-$359     &   0.18$-$0.17   &    0.46$-$0.40    &     8$-$7      \\  
                                                                 
\hline

\end{tabular}
\tablefoot{
\tablefoottext{a}{Emir receiver.} 
\tablefoottext{b}{Observed frequency range.}
\tablefoottext{c}{Spectral resolution in velocity units ($\updelta v$) in the observed frequency ranges.}
\tablefoottext{d}{Antenna efficiencies.}
\tablefoottext{e}{The half power beam width can be well fitted by \mbox{HPBW[arcsec]$\approx$2460/Frequency[GHz]}.}}
\end{table}

\subsection{Line survey} \label{Survey}

We performed a complete millimetre line survey towards the edge of Orion Bar 
 using the IRAM \mbox{30 m} telescope. The target position is at \mbox{$\mathrm{\alpha_{2000}=05^{h}\,35^{m}\,20.8^{s}\,}$}, 
\mbox{$\mathrm{\delta_{2000}=-\,05^{\circ}25'17.0''}$}, close to the dissociation front. 
This position is at \mbox{$(\Delta\upalpha,\Delta\updelta) = (3'',-3'')$} from the ``CO$^{+}$ peak'' of \citet{Stoerzer_1995}.

Our observations cover a total of \mbox{217 GHz} along 3, 2, 1, and \mbox{0.9 mm} bands of the EMIR receivers using the WILMA (\mbox{2 MHz} spectral resolution) and FTS (\mbox{200 kHz} spectral resolution) backends. The observing procedure was position switching (PSW) with the reference position located at an offset \mbox{($-$600$''$, 0$''$)} to avoid the extended molecular emission from the Orion Molecular Cloud (OMC-1). The antenna temperature, $T^{*}_{_{\rm A}}$, was converted to the main beam temperature, $T_{_{\rm MB}}$, through the \mbox{$T_{_{\rm MB}}$ = $T^{*}_{_{\rm A}}/ \eta_{_{\rm MB}}$} relation, where $\mathrm{\eta_{_{MB}}}$ is the antenna efficiency. All intensities in tables and figures are in main beam temperature. A local standard of rest (LSR) of \mbox{10.7 km s$^{-1}$} has been used in the line survey target position in the Orion Bar dissociation front.  Table~\ref{Table_efficiences} shows an overview of the frequency ranges observed with each backend, the spectral resolution in velocity units ($\updelta v$) in the observed frequency ranges, as well as the variation in the telescope efficiencies, $\mathrm{\eta_{_{MB}}}$, and the half power beam width (HPBW) across the covered frequency range.
 
In Appendix~\ref{errorbeam} we study the possible contamination of bright molecular line
emission from the Orion BN/KL region (located at a distance of $\sim$2$'$)
through the beam side lobes \citep[several arcmin, see][]{Greve_1998}.
 Although the contribution to the detected power ranges from $\sim$12\% (3\,mm) to $\sim$30\% (0.8\,mm), most of
the emission from the hot core region arises at different velocities ($\sim$8~km\,s$^{-1}$) than those of the Orion Bar \mbox{($\sim$10 $-$ 11~km\,s$^{-1}$)} and thus can be easily separated.

Data reduction and spectral analysis were done using the CLASS software of the GILDAS package\footnote{http://www.iram.fr/IRAMFR/GILDAS/} developed by IRAM. Weighted spectra were averaged and calibrated, and a polynomial baseline of low order (typically second or third order) was subtracted from each \mbox{$\sim$200 MHz} wide spectrum. Finally, Gaussian profiles were fitted to all the detected lines (Appendix~\ref{Tables}). The rms noise of our observations obtained after integration during \mbox{$\sim$4 h} ranges between 4 mK and 20 mK per resolution channel.

\subsection{Maps} \label{maps}

The 0.9\,mm line emission from different molecules was mapped with the IRAM~30~m telescope
in January 2014 under excellent winter conditions ($<$1\,mm of precipitable water vapour). The E3 receiver and the FTS backend at 200\,kHz spectral resolution were used.
On-the-fly (OTF) scans were obtained along and perpendicular to the Bar over a \mbox{170$''\times170''$}
region, with an OFF position at \mbox{($-$600$''$,0$''$)} relative
to the map centre at 
\mbox{$\mathrm{\alpha_{2000}=05^{h}\,35^{m}\,20.1^{s}}$}, 
\mbox{$\mathrm{\delta_{2000}=-\,05^{\circ}25'07.0''}$}, which is slightly different from that of the line survey (see above).
 The HPBW at this frequency is \mbox{$\sim$7$''$}. Data processing consisted in a linear baseline subtraction in each observed spectra. The resulting spectra were gridded to a data cube through convolution with a Gaussian kernel providing a final resolution of \mbox{8.5$''$}. The total integration time was approximately 6~h, and the achieved rms noise is $\sim$1~K per $\sim$0.2~km~s$^{-1}$ channel. Figure~\ref{fig:Maps} shows the C$^{18}$O \mbox{$J$ = 3 $\rightarrow$ 2} and  H$_2$CO \mbox{$J_{K_{\rm a},K_{\rm c}}$ = 5$_{1,5} \rightarrow$ 4$_{1,4}$} integrated line intensity maps in the range \mbox{$v_{\rm LSR}$ = 10 $-$ 11\,km~s$^{-1}$} in which the Orion Bar shows prominent emission.
We note that \citet{Wiel_2009} also presented a smaller H$_2$CO \mbox{$J_{K_{\rm a},K_{\rm c}}$ = 5$_{1,5} \rightarrow$ 4$_{1,4}$} map at lower angular resolution (a factor of $\sim$2).

\section{Results: complex organic molecule detections} \label{Results}

We have identified more than 250 lines from COMs and related organic precursors. The detected lines are attributed to ten different molecules with up to seven atoms: HCO, HNCO, H$_{2}$CO, H$_{2}$CS, CH$_{2}$NH, H$_{2}$CCO, HC$_{3}$N, CH$_{3}$OH, CH$_{3}$CN, and CH$_{3}$CHO.
We also detect several lines of the isotopologue H$_{2}^{13}$CO. 
We note that among the organic molecules with up to seven atoms, we also identified several lines of formic acid, trans- and cis-HCOOH, in the Orion Bar PDR. \citet{Cuadrado_2016} have recently reported the first detection of the cis conformer of HCOOH in the interstellar medium as well as a detailed analysis of the detected lines of both HCOOH conformers in the 3~mm spectral band.
Line profiles peak at \mbox{$v_{_{\rm LSR}}$ = 10 $-$ 11\,km~s$^{-1}$}, the velocity of the Bar, thus confirming their PDR origin. 
The detected organic molecules and their dipole moments are listed in Table~\ref{Table_detected_COMs}.

\begin{table}
\centering 
\caption{Dipole moments ($\mu$), number of detected lines of complex organic molecules and related organic precursors, and their corresponding Figure and Table numbers.}
\label{Table_detected_COMs}     
\begin{tabular}{l c l l c c@{\vrule height 9pt depth 5pt width 0pt}} 
\hline\hline      
Molecule         &  No. of  &  \multicolumn{2}{c}{$\mathrm{\mu}$}    &  Fig. & Table   \\
                &  lines &  \multicolumn{2}{c}{[Debyes]}          &       &    \\

\hline  
HCO              &    16      &  $\mu_{a}$=1.363         & $\mu_{b}$=0.700$^a$  &  \ref{fig:HCO_lines}      &   \ref{Table_HCO}      \\ 
H$_{2}$CO        &    15      &  $\mu_{a}$=2.332$^b$     &                      &  \ref{fig:H2CO_lines}     &   \ref{Table_H2CO}      \\ 
H$_{2}^{13}$CO   &     8      &  $\mu_{a}$=2.332$^b$     &                      &  \ref{fig:H2-13CO_lines}  &   \ref{Table_H2-13CO}      \\ 
H$_{2}$CS        &    26      &  $\mu_{a}$=1.649$^b$     &                      &  \ref{fig:H2CS_lines}     &   \ref{Table_H2CS}      \\ 
HNCO             &     6      &  $\mu_{a}$=1.602         & $\mu_{b}$=1.350$^c$  &  \ref{fig:HNCO_lines}     &   \ref{Table_HNCO}      \\ 
CH$_{2}$NH       &     6      &  $\mu_{a}$=1.340         & $\mu_{b}$=1.446$^d$  &  \ref{fig:CH2NH_lines}    &   \ref{Table_CH2NH}       \\ 
H$_{2}$CCO       &     30     &  $\mu_{a}$=1.422$^b$     &                      &  \ref{fig:H2CCO_lines}    &   \ref{Table_H2CCO}      \\ 
HC$_{3}$N        &     11     &  $\mu_{a}$=3.732$^e$     &                      &  \ref{fig:HC3N_lines}     &   \ref{Table_HC3N}      \\ 
CH$_{3}$CN       &     44     &  $\mu_{a}$=3.922$^f$     &                      &  \ref{fig:CH3CN_lines}    &   \ref{Table_CH3CN}      \\ 
CH$_{3}$OH       &     55     &  $\mu_{a}$=0.896         & $\mu_{b}$=1.412$^g$  &  \ref{fig:CH3OH_lines}    &   \ref{Table_CH3OH}      \\ 
CH$_{3}$CHO      &     36     &  $\mu_{a}$=2.423         & $\mu_{b}$=1.260$^h$  &  \ref{fig:CH3CHO_lines}   &   \ref{Table_CH3CHO}      \\ 
                                                                                           
\hline
 
\end{tabular}
\tablebib{$^{(a)}$~\citet{Blake_1984}; $^{(b)}$~\citet{Fabricant_1977}; $^{(c)}$~\citet{Hocking_1975}; $^{(d)}$~\citet{Allegrini_1979}; $^{(e)}$~\citet{DeLeon_1985}; 
$^{(f)}$~\citet{Gadhi_1995}; $^{(g)}$~\citet{Sastry_1981}; $^{(h)}$~\citet{Kleiner_1996}.}

\end{table}

Previous studies of the Orion Bar have already reported the presence of HCO, H$_{2}$CO, and CH$_3$OH in several positions close to the ``CO$^{+}$ peak'': (i) \citet{Schilke_2001} detected one hyperfine component of HCO \mbox{$N$ = 1 $\rightarrow$ 0} using the IRAM~30~m telescope; (ii) \citet{Hogerheijde_1995} detected nine lines of H$_{2}$CO and three lines of CH$_3$OH in several positions along the Bar using the JCMT, IRAM~30~m, and CSO telescopes; 
(iii) \citet{Nagy_2017} detected ten submm H$_{2}$CO lines with Herschel/HIFI. In addition, H$_{2}$CO, H$_{2}^{13}$CO, HDCO, H$_{2}$CS, and CH$_3$OH have been detected towards the colder, denser, and more FUV-shielded clumps: (iv) \citet{Leurini_2006} detected seven lines of H$_{2}$CO, four lines of H$_{2}$CS, and ten lines of CH$_3$OH using APEX between 279~GHz and 361.5~GHz towards clump $\#$1 of \citet{Lis_2003}; (v) \citet{Parise_2009} also detected two isotopologues, one line of H$_{2}^{13}$CO and one line of HDCO, towards clump $\#$3 of \citet{Lis_2003} using the IRAM~30~m telescope.
To our knowledge, our work presents the first detection
 of CH$_{2}$NH, H$_{2}$CCO, HC$_{3}$N, CH$_{3}$CN, and CH$_{3}$CHO in the Orion Bar PDR.

Line assignment was carried out using J.~Cernicharo's spectral catalogue \citep[MADEX,][]{Cernicharo_2012}, and the JPL \citep{Pickett_1998}\footnote{http://spec.jpl.nasa.gov/} and 
CDMS \citep{Muller_2001,Muller_2005}\footnote{http://www.astro.uni-koeln.de/cdms/}
public databases.  
Although some lines have peak intensities lower than 3$\sigma$ noise level, we included them in our results as tentative detections because other transitions from the same molecule are well detected in this survey, and these 3$\sigma$ limits are in agreement with our excitation models. Given their intricate spectroscopy, below we summarise the main spectroscopic information of the \mbox{detected species.}

\begin{figure}[!t]
\centering
\hspace{-0.5cm}
\includegraphics[scale=0.5,angle=0]{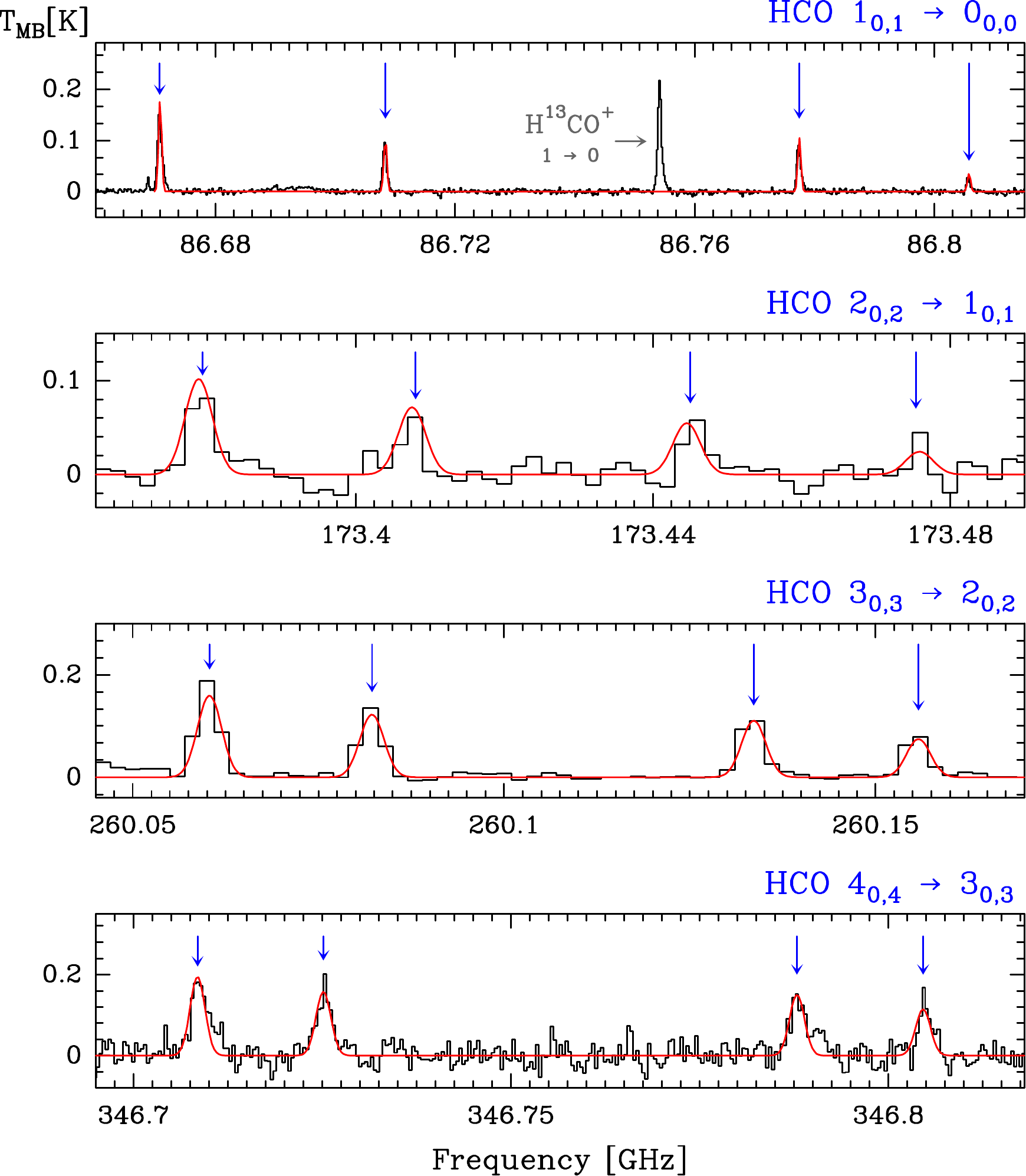}\\
\caption{Detected HCO hyperfine structure (HFS) lines of the \mbox{$N$ = 1 $\rightarrow$ 0}, \mbox{2 $\rightarrow$ 1}, \mbox{3 $\rightarrow$ 2}, and
\mbox{4 $\rightarrow$ 3} rotational transitions (black histogram spectra). A single excitation temperature model is shown overlaid in red (see Sect.~\ref{DR}). HFS lines are indicated by the blue arrows. The other spectral features appearing in the selected windows are labelled with their corresponding identification.}
\label{fig:HCO_lines}
\end{figure}

\subsection{Formyl radical: HCO} \label{HCO}

The formyl radical, HCO, 
is a light asymmetric rotor with one unpaired electron and one non-zero nuclear spin.
The quantum numbers designating the energy levels are $N$, $J$, and $F$. Spin doubling \mbox{($J = N + S$)} is produced by the coupling between the rotational angular momentum, $N$, and the unpaired electron spin, $S$, while the hyperfine structure \mbox{($F = J + I$)} is due to the coupling of the angular momentum, $J$, and the spin of the hydrogen nucleus, $I$. Each rotational level $N_{K_{\rm a},K_{\rm c}}$ is split into a doublet by the spin-rotation interaction, the levels of which are further split into doublets by magnetic hyperfine interactions, with the final energy levels labelled by the quantum number \mbox{$F = J$ $\pm$ 1/2} \citep[e.g.][]{Bowater_1971,Saito_1972,Austin_1974,Blake_1984}. 
$a$-type \mbox{($\Delta$$K_{a}$ = 0,} $\pm$2...)
 and $b$-type \mbox{($\Delta$$K_{a}$ = $\pm$1,} $\pm$3...)
 transitions are allowed, with a stronger dipole moment for the $a$-type transitions (see Table~\ref{Table_detected_COMs}). 

We identified a total of 16 lines of HCO. They consist of four sets of rotational transitions 
corresponding to the hyperfine splitting of the \mbox{$N$ = 1 $\rightarrow$ 0} to \mbox{4 $\rightarrow$ 3} transitions.
All transitions observed here are $a$-type.
The quantum numbers 
of the detected HCO transitions, their spectroscopic parameters, and the results from fitting the line profiles with Gaussians are listed in Table~\ref{Table_HCO}. In Fig.~\ref{fig:HCO_lines} we present the spectra of the  
\mbox{$N$ = 1 $\rightarrow$ 0} to \mbox{4 $\rightarrow$ 3} rotational lines.

\begin{figure}
\centering
\hspace{-0.7cm}
\includegraphics[scale=0.56,angle=0]{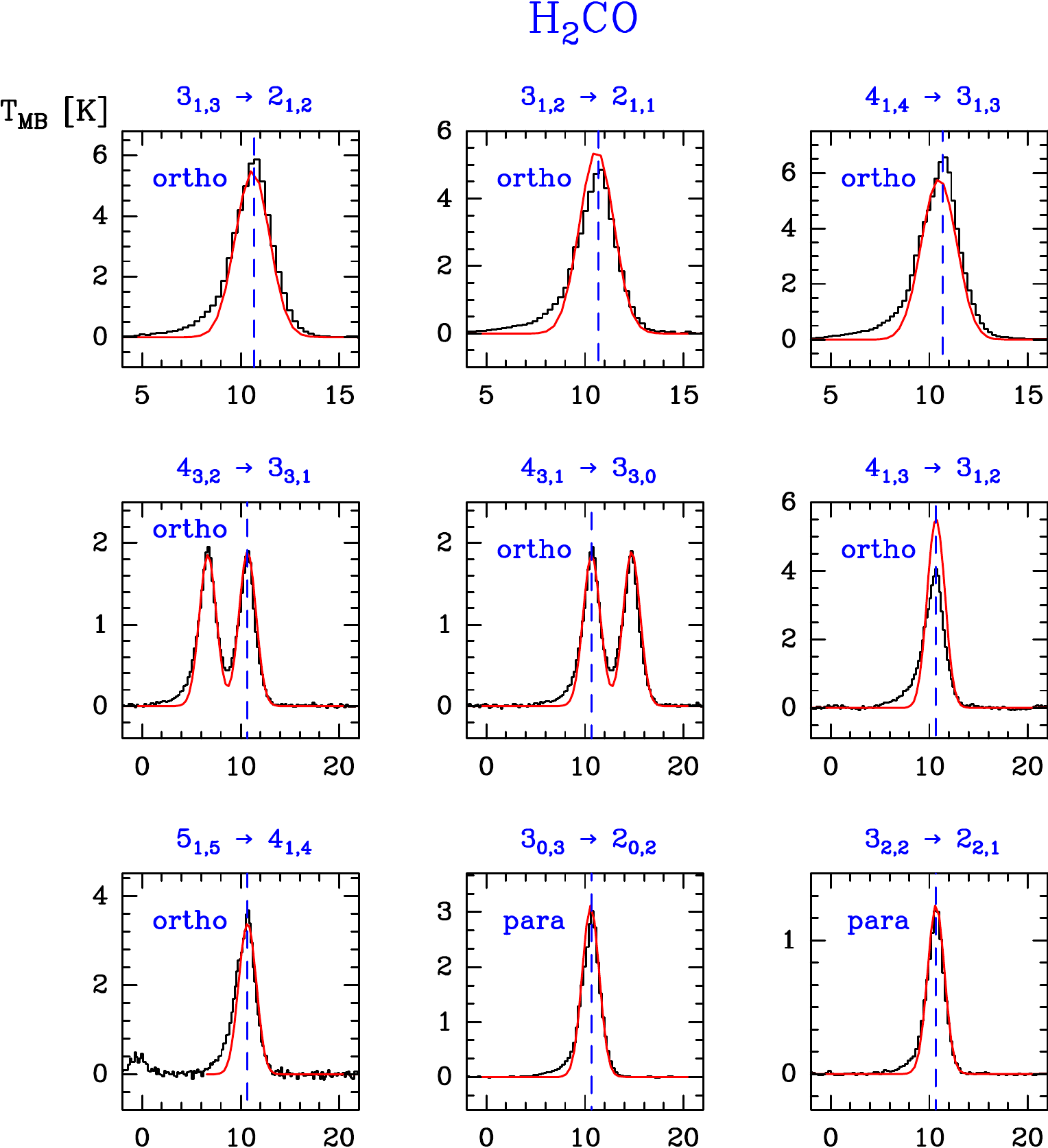} \\
\vspace{0.4cm}\includegraphics[scale=0.56,angle=0]{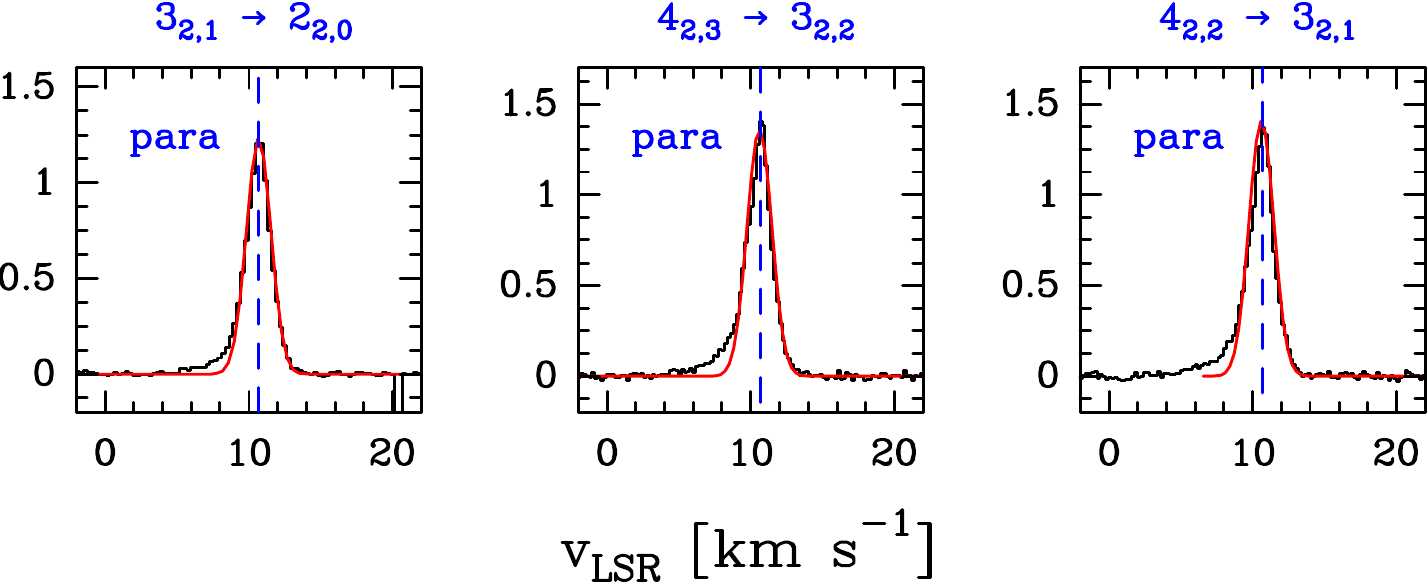}
\caption{Example of ortho- and para-H$_{2}$CO lines (black histogram spectra). A non-LTE LVG model (\mbox{$T_{\rm k}$ $\simeq$ 200 K},
\mbox{$n$(H$_{2}$) $\simeq$ 1 $\times$ 10$^{6}$ cm$^{-3}$}, 
\mbox{$N$(o-H$_{2}$CO) = 4.0 $\times$ 10$^{13}$ cm$^{-2}$}, and 
\mbox{$N$(p-H$_{2}$CO) = 1.8 $\times$ 10$^{13}$ cm$^{-2}$}) is shown overlaid in red (see Sect.~\ref{Non-LTE}). The dashed lines indicate the LSR velocity \mbox{(10.7 km s$^{-1}$)} of the Orion Bar PDR.}
\label{fig:H2CO_lines}
\end{figure} 

 \begin{figure}
\includegraphics[scale=0.56,angle=0]{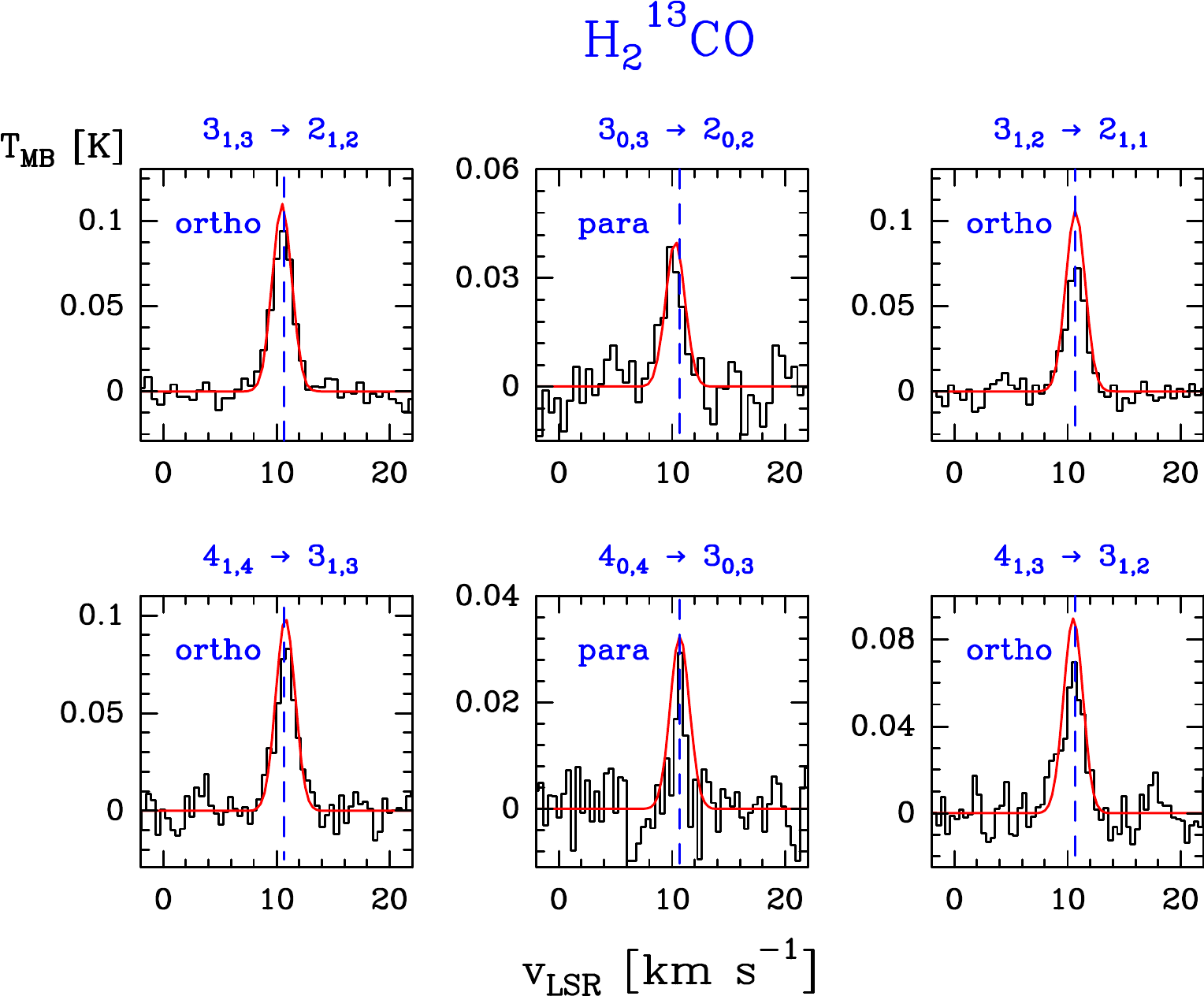}\\
\caption{Example of ortho- and para-H$_{2}^{13}$CO lines (black histogram spectra). A non-LTE LVG model (\mbox{$T_{\rm k}$ $\simeq$ 200 K},
\mbox{$n$(H$_{2}$) $\simeq$ 1 $\times$ 10$^{6}$ cm$^{-3}$}, 
\mbox{$N$(o-H$_{2}^{13}$CO) = 7.3 $\times$ 10$^{11}$ cm$^{-2}$}, and 
\mbox{$N$(p-H$_{2}^{13}$CO) = 2.3 $\times$ 10$^{11}$ cm$^{-2}$}) is shown overlaid in red (see Sect.~\ref{Non-LTE}).}
\label{fig:H2-13CO_lines}
\end{figure}

\subsection{Formaldehyde: H$_{2}$CO} \label{H2CO}

Formaldehyde, H$_{2}$CO, was the first polyatomic organic molecule found in the ISM \citep{Snyder_1969}. It is a slightly asymmetric prolate rotor molecule with $K$-type doublets and an ortho-para symmetry because of the two indiscernible off-axis hydrogen atoms. 
The ortho states are those with $K_{\rm a}$ odd, and the para states with $K_{\rm a}$ even. The nuclear spin weights are 3 and 1 for \mbox{o-H$_{2}$CO} and \mbox{p-H$_{2}$CO}, respectively. The $K$-type doublets are the result of the slight asymmetry in the H$_{2}$CO produced by the light H atoms. There are no line doublets in the \mbox{$K_{\rm a}$ = 0} para state \citep[e.g.][]{Bocquet_1996,Brunken_2003,Eliet_2012}.

We detected nine lines of the ortho species (in the \mbox{$K_{\rm a}$ = 1} and 3 ladders) with \mbox{$E_{\rm u}/k$ $\leq$ 125.8~K}, and six lines for the para species (in the \mbox{$K_{\rm a}$ = 0} and 2 ladders) with \mbox{$E_{\rm u}/k$ $\leq$ 82.1~K}. Line profiles are shown in Fig.~\ref{fig:H2CO_lines}. Table~\ref{Table_H2CO} gives the observed line parameters.

We also detected eight lines of H$_{2}^{13}$CO, consisting of six lines of ortho species and two lines of the para species, in the \mbox{$K_{\rm a}$ = 0} and 1 ladders, respectively (\mbox{$E_{\rm u}/k$ $\leq$ 34.0~K}). Line profiles are shown in Fig.~\ref{fig:H2-13CO_lines}. Table~\ref{Table_H2-13CO} gives the observed line parameters. No lines of deuterated formaldehyde were detected.

\subsection{Thioformaldehyde: H$_{2}$CS} \label{H2CS}

Thioformaldehyde, H$_{2}$CS, has a formaldehyde-like structure. It is a slightly asymmetric prolate rotor, with only $a$-type transitions and ortho-para symmetry \citep[e.g.][]{Johnson_1971,Beers_1972,Maeda_2008}.

\begin{figure}
\centering
\includegraphics[scale=0.55,angle=0]{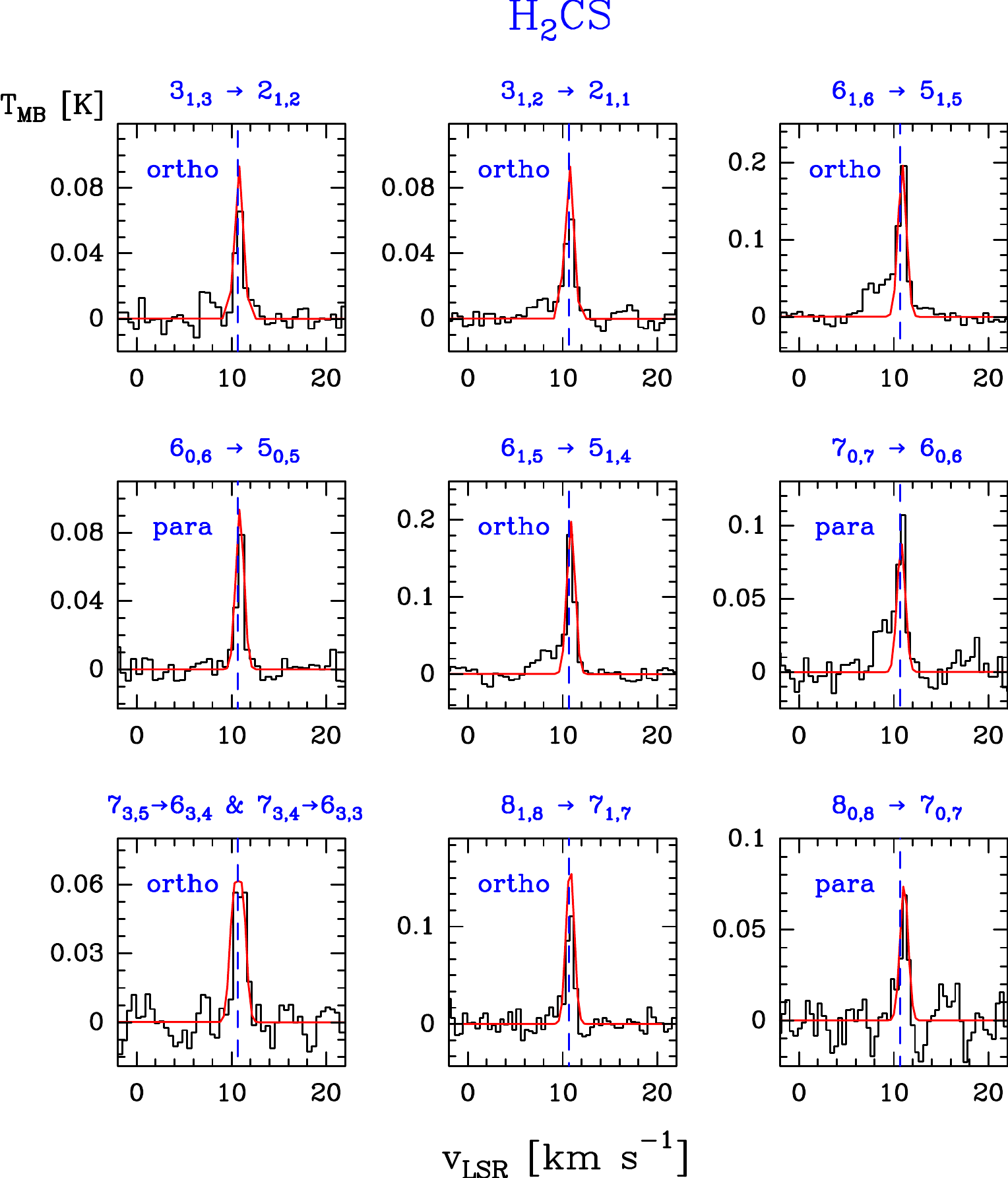}\\
\caption{Example of ortho- and para-H$_{2}$CS lines (black histogram spectra). Single excitation temperature models are shown overlaid in red (see Sect.~\ref{DR}).}
\label{fig:H2CS_lines}
\end{figure}

We detected 15 lines of ortho species (in the \mbox{$K_{\rm a}$ = 1} and 3 ladders) with \mbox{$E_{\rm u}/k$ $\leq$ 149.8~K}, and 11 lines of para species (in the \mbox{$K_{\rm a}$ = 0} and 2 ladders) with \mbox{$E_{\rm u}/k$ $\leq$ 112.0~K}. Figure~\ref{fig:H2CS_lines} shows a selection of ortho and para detected lines. Table~\ref{Table_H2CS} gives the observed line parameters.

\subsection{Isocyanic acid: HNCO} \label{HNCO}

Isocyanic acid, H-N=C=O, is the simplest molecule containing carbon, hydrogen, oxygen, and nitrogen. It is a slightly asymmetric prolate rotor with $a$- and $b$-type transitions. Its rotational levels are designated as $J_{K_{\rm a},K_{\rm c}}$ \citep[e.g.][]{Kukolich_1971,Hocking_1975,Lapinov_2007}.

\begin{figure}
\centering
\hspace{-0.5cm}
\includegraphics[scale=0.55,angle=0]{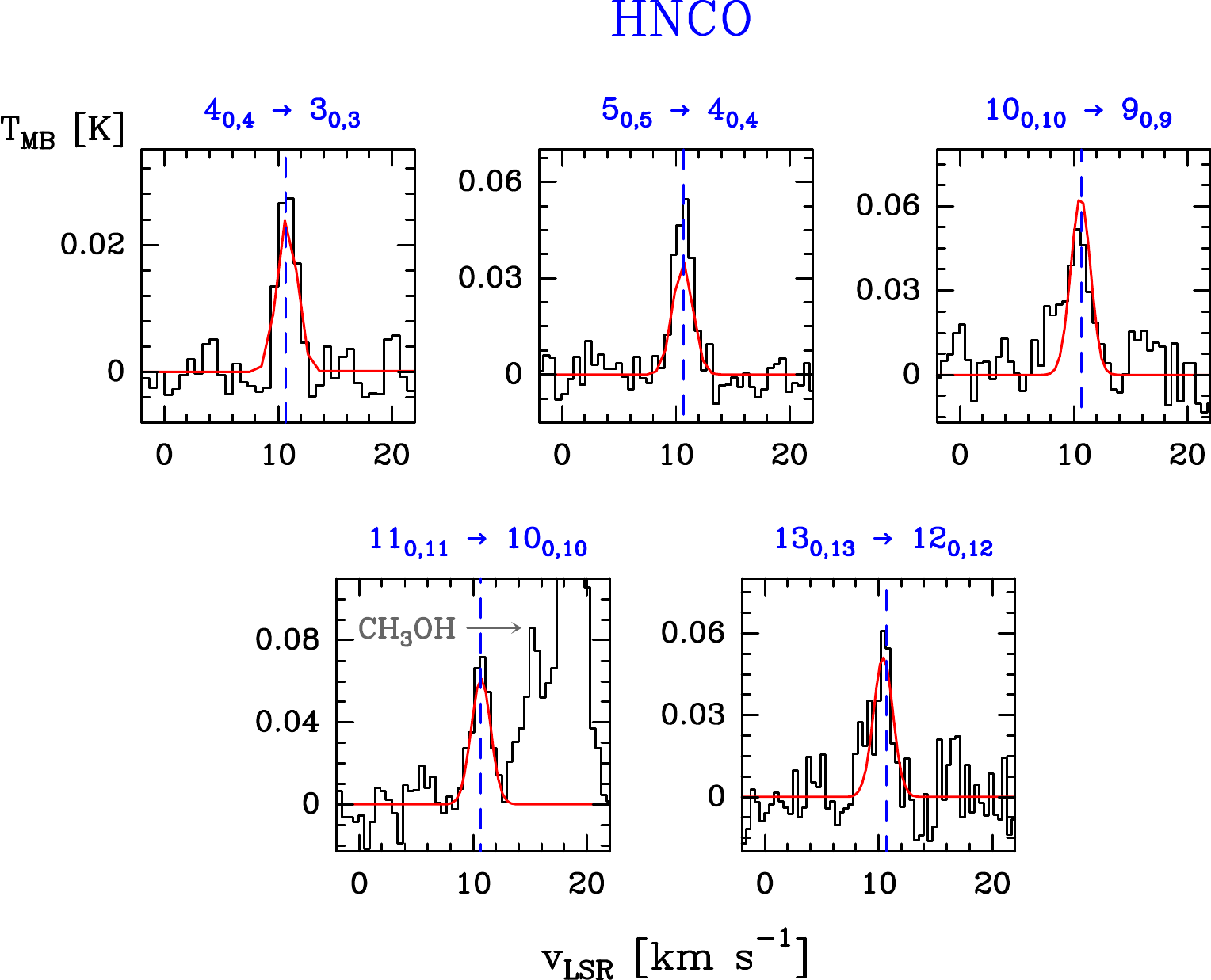}\\
\caption{Example of HNCO detected lines (black histogram spectra). A single excitation temperature model is shown overlaid in red (see Sect.~\ref{DR}).}
\label{fig:HNCO_lines}
\end{figure}

We detect six $a$-type transitions inside the \mbox{$K_{a}$ = 0} rotational ladder (\mbox{$E_{\rm u}/k$ $\leq$ 96.0~K}). The observed lines do not show evidence for hyperfine splittings. Line profiles are shown in Fig.~\ref{fig:HNCO_lines}. Table~\ref{Table_HNCO} gives the observed line parameters.
 
\begin{figure}
\includegraphics[scale=0.55,angle=0]{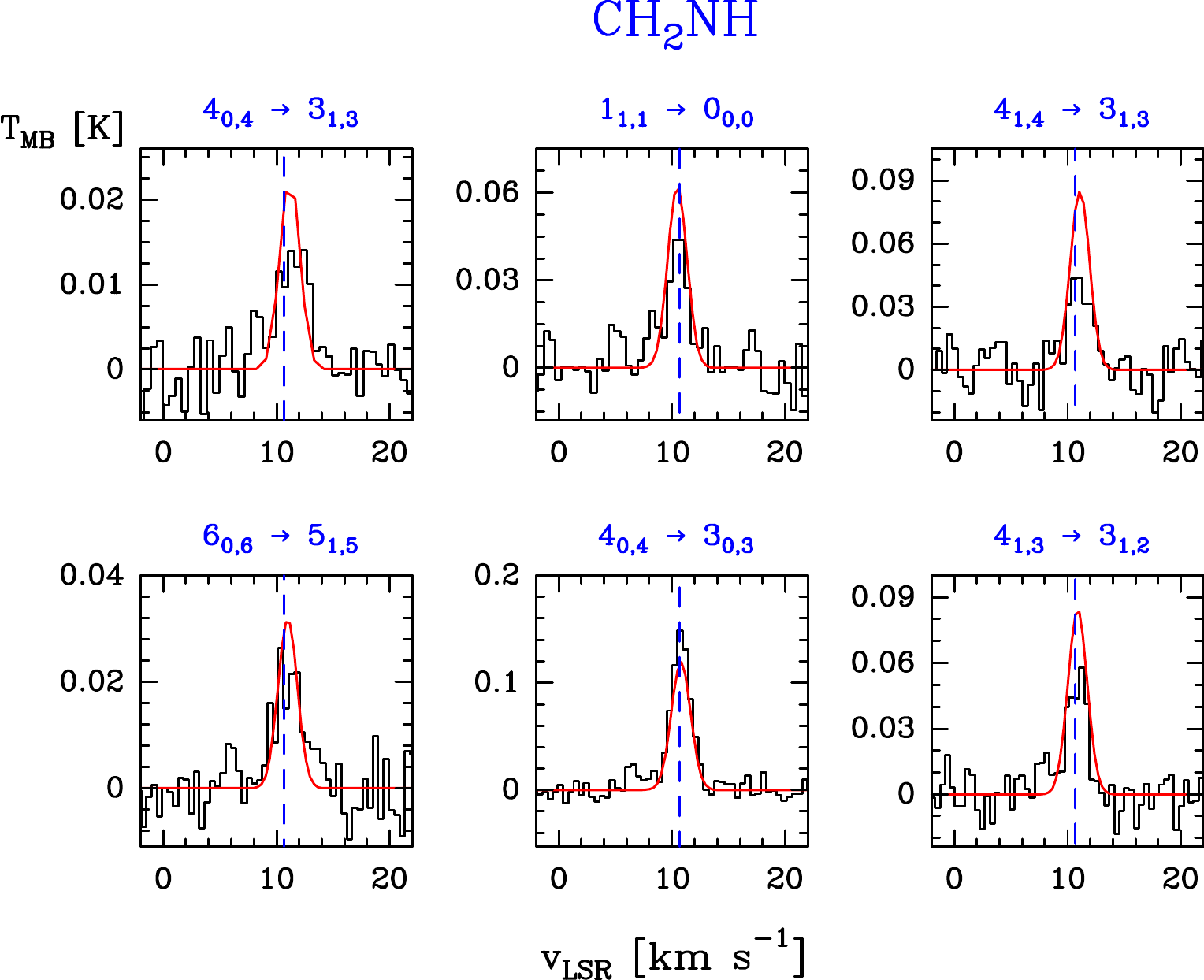}\\ 
\caption{CH$_{2}$NH detected lines (black histogram spectra). A single excitation temperature model is shown overlaid in red (see Sect.~\ref{DR}).}
\label{fig:CH2NH_lines}
\end{figure}

\subsection{Methanimine: CH$_{2}$NH} \label{CH2NH}

Methanimine is the simplest molecule containing a carbon-nitrogen double bond (H$_{2}$C=NH). It is a planar molecule and a nearly prolate asymmetric rotor with components of the electric dipole moment along both the $a$ and $b$ molecular axes \citep[e.g.][]{Kirchhoff_1973,Dore_2010,Dore_2012}.

We detected six lines of CH$_{2}$NH consisting of (i) three \mbox{$a$-type} transitions (in the \mbox{$K_{\rm a}$ = 0} and 1 ladders) and (ii) three \mbox{$b$-type} transitions, with \mbox{$E_{\rm u}/k$ $\leq$ 64.1~K}. The observed lines do not show evidence for hyperfine splittings. The line profiles are shown in Fig.~\ref{fig:CH2NH_lines}. Table~\ref{Table_CH2NH} gives the observed line parameters.

\subsection{Ketene: H$_{2}$CCO} \label{H2CCO}

\begin{figure}
\includegraphics[scale=0.55,angle=0]{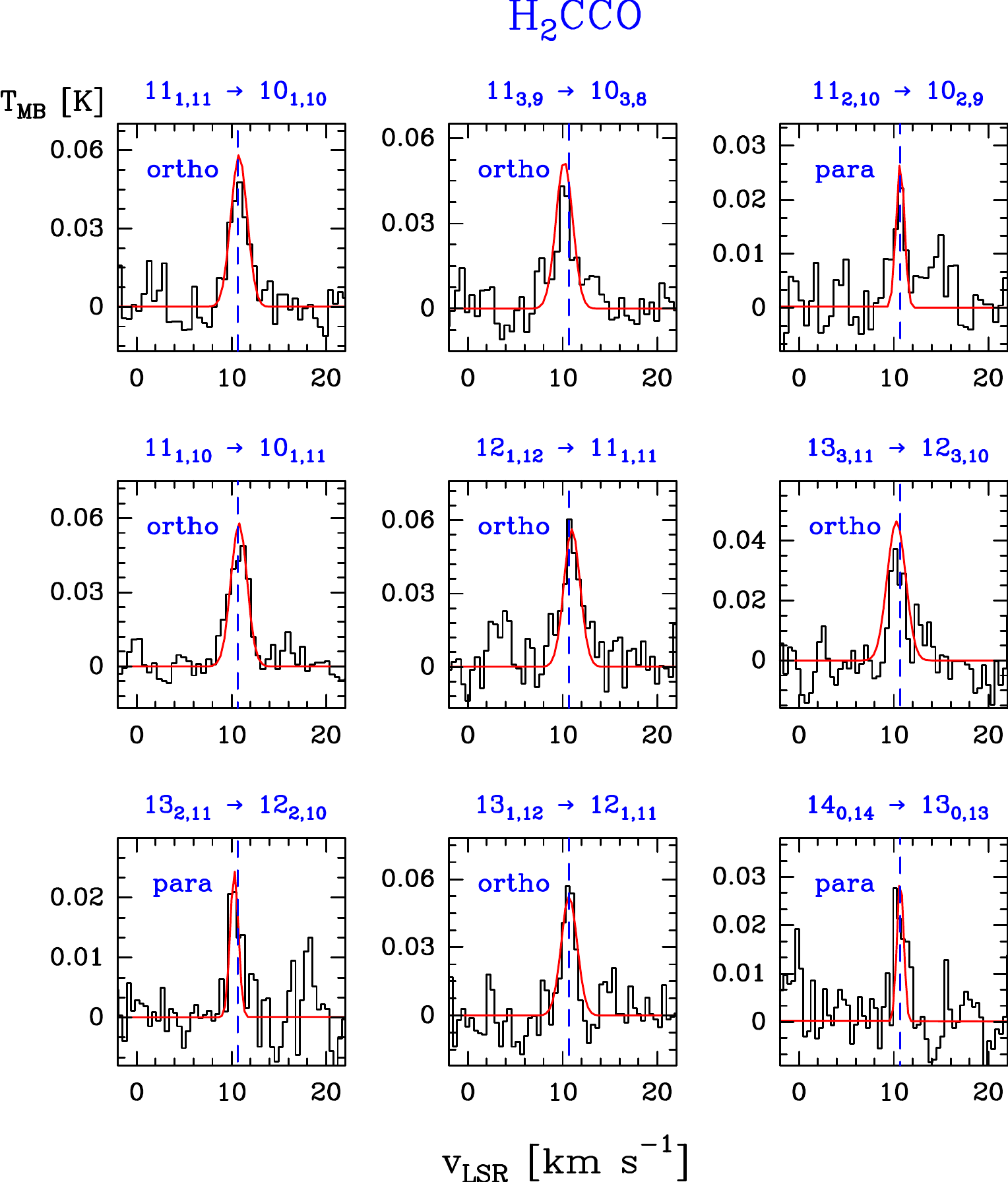} 
\caption{Example of ortho- and para-H$_{2}$CCO lines (black histogram spectra). Single excitation temperature models are shown overlaid in red (see Sect.~\ref{DR}).}
\label{fig:H2CCO_lines}
\end{figure}

Ketene, H$_{2}$C=C=O, is a slightly asymmetric prolate rotor with only $a$-type transitions and ortho-para symmetry \citep[e.g.][]{Johnson_1952,Fabricant_1977,Brown_1990,Johns_1992}.
We detected 20 lines of ortho species (in the \mbox{$K_{\rm a}$ = 1} and 3 ladders) with \mbox{$E_{\rm u}/k$ $\leq$ 205.2~K}, and ten lines for the para species (in the \mbox{$K_{\rm a}$ = 0} and 2 ladders) with \mbox{$E_{\rm u}/k$ $\leq$ 154~K}. 
Examples of the line profiles of ortho and para lines are shown in Fig.~\ref{fig:H2CCO_lines}. Table~\ref{Table_H2CCO} gives the observed line parameters.

\subsection{Cyanoacetylene: HC$_{3}$N} \label{HC3N}

Cyanoacetylene, H--C$\equiv$C--C$\equiv$N, is a linear molecule with a large dipole moment of \mbox{$\mu_{a}$ = 3.732 D} \citep{DeLeon_1985}. It is the simplest example of cyanopolyynes, HC$_{n}$N (with \mbox{$n$ = 3,} 5, 7...; see e.g. \citealt{Turner_1971,Avery_1976,Cernicharo_1986}). 
The rotational transitions show hyperfine splittings owing to the interaction of the electric-quadrupole moment of the  nitrogen nuclear spin with the electronic-charge distribution \citep[e.g.][]{Zafra_1971,Mbosei_2000,Chen_1991,Yamada_1995,Thorwirth_2000}.
 This hyperfine splitting
is only fully resolved in the lowest rotational transitions \citep[e.g.][]{Turner_1971,Dickinson_1972}, therefore, HC$_{3}$N is described here by a simple rotational spectrum, with \mbox{$J$+1 $\rightarrow$ $J$} transitions. 

\begin{figure}
\centering
\vspace{0.2cm}
\includegraphics[scale=0.55,angle=0]{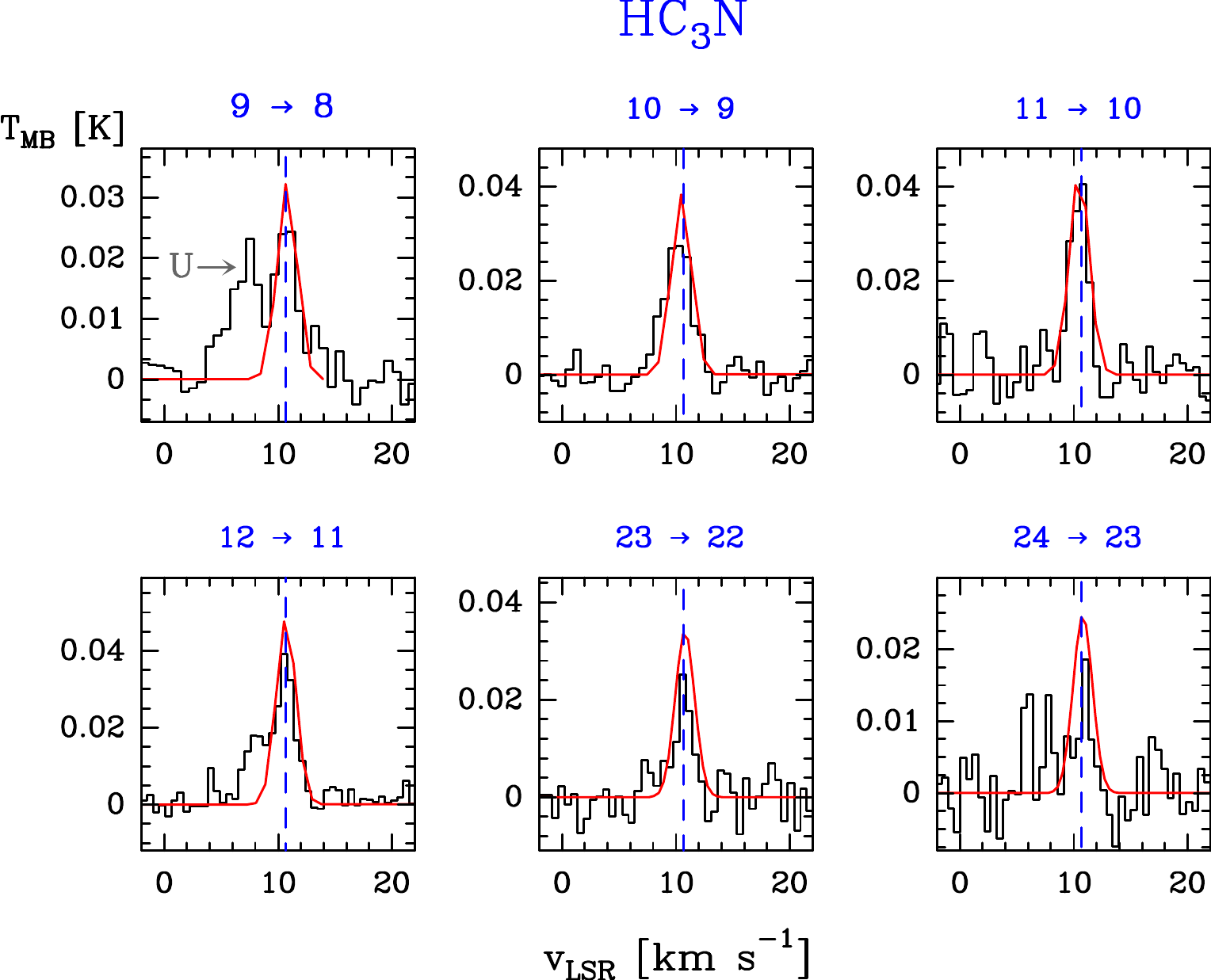}
\caption{Example of HC$_{3}$N detected lines (black histogram spectra). 
A non-LTE LVG model (\mbox{$T_{\rm k}$ $\simeq$ 80 K},
\mbox{$n$(H$_{2}$) $\simeq$ 1 $\times$ 10$^{6}$ cm$^{-3}$}, 
\mbox{$N$(HC$_{3}$N) = 4.0 $\times$ 10$^{11}$ cm$^{-2}$}) is shown overlaid in red (see Sect.~\ref{Non-LTE}).}
\label{fig:HC3N_lines}
\end{figure}

We detected 11 lines of HC$_{3}$N, from \mbox{$J$ = 9 $\rightarrow$ 8} to \mbox{24 $\rightarrow$ 23}, with \mbox{$E_{\rm u}/k$ $\leq$ 131.0~K}. Figure~\ref{fig:HC3N_lines} shows a selection of detected lines and Table~\ref{Table_HC3N} gives the observed line parameters.

\begin{figure}
\centering
\vspace{0.5cm}
\includegraphics[scale=0.45,angle=0]{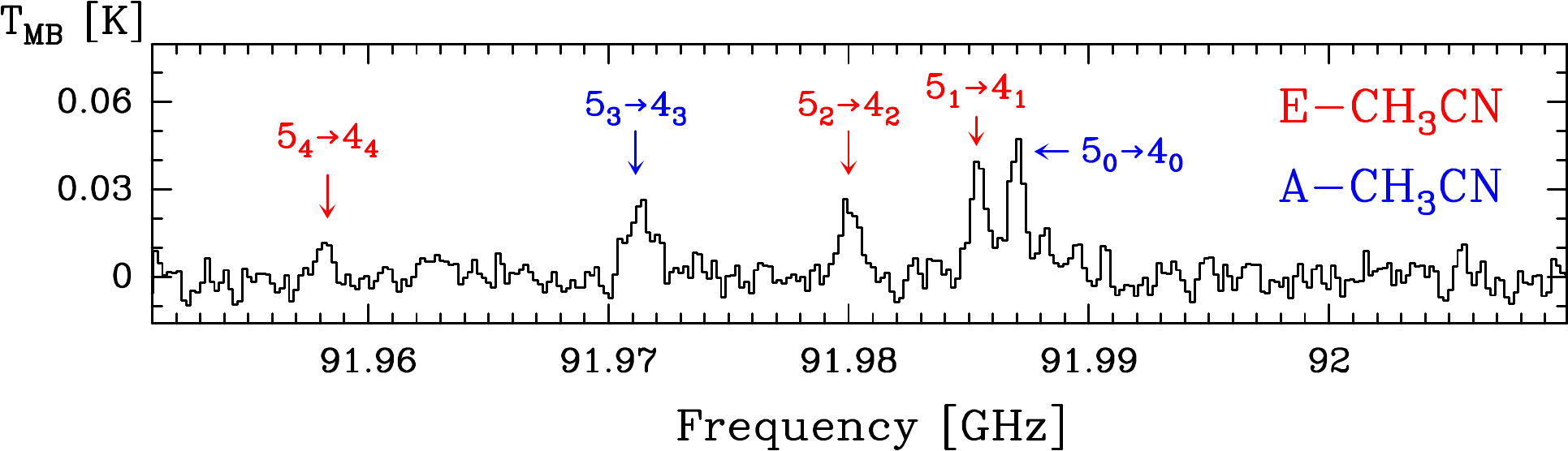}
\caption{Observed emission lines from the \mbox{$J$ = 5 $\rightarrow$ 4} transitions of \mbox{CH$_{3}$CN}. Transitions with different $K$ but the same $J$ occur in narrow frequency regions despite they have different level energies.}
\label{fig:CH3CN_5-4}
\end{figure}

\begin{figure*}
\includegraphics[scale=0.5,angle=0]{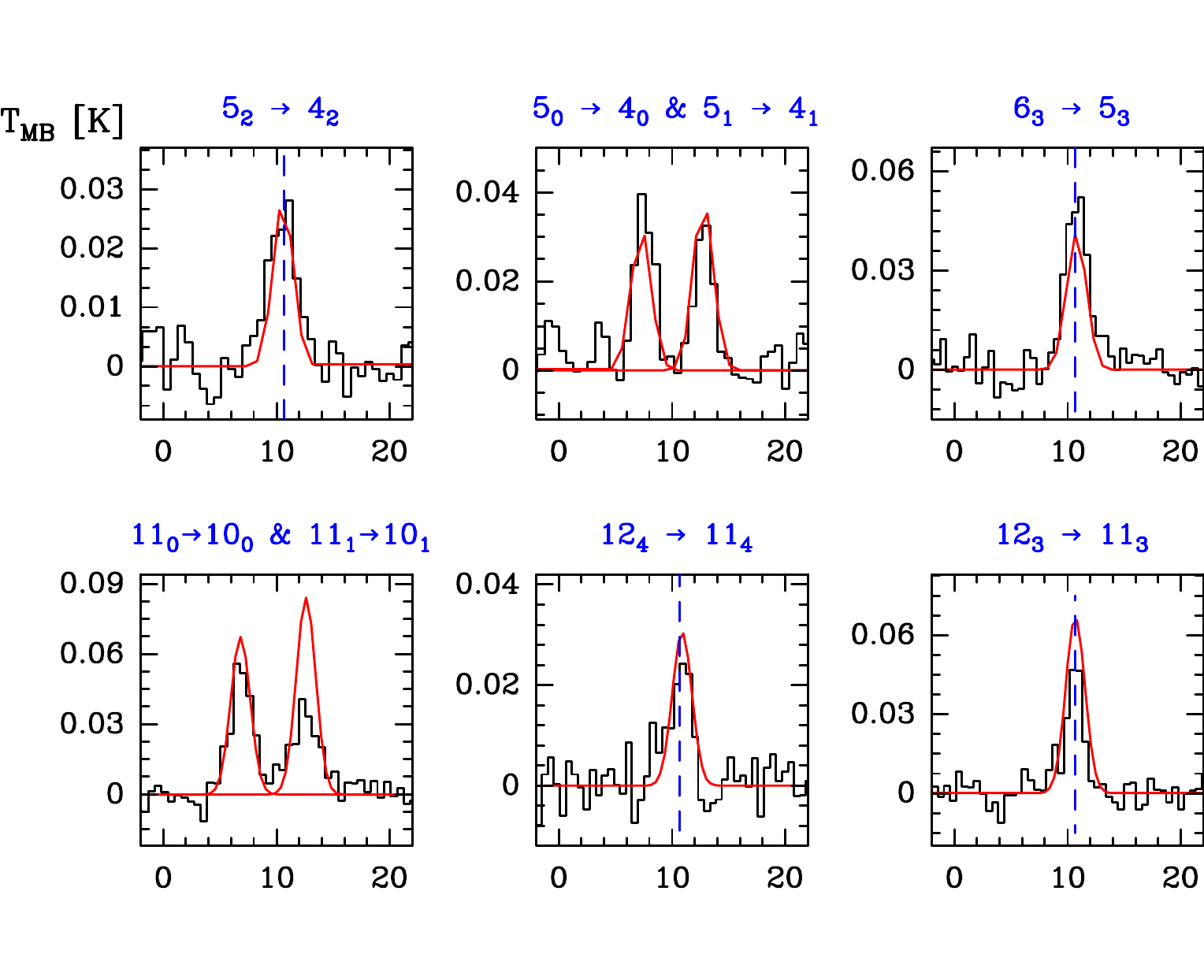} 
\hspace*{0.2cm}\includegraphics[scale=0.5,angle=0]{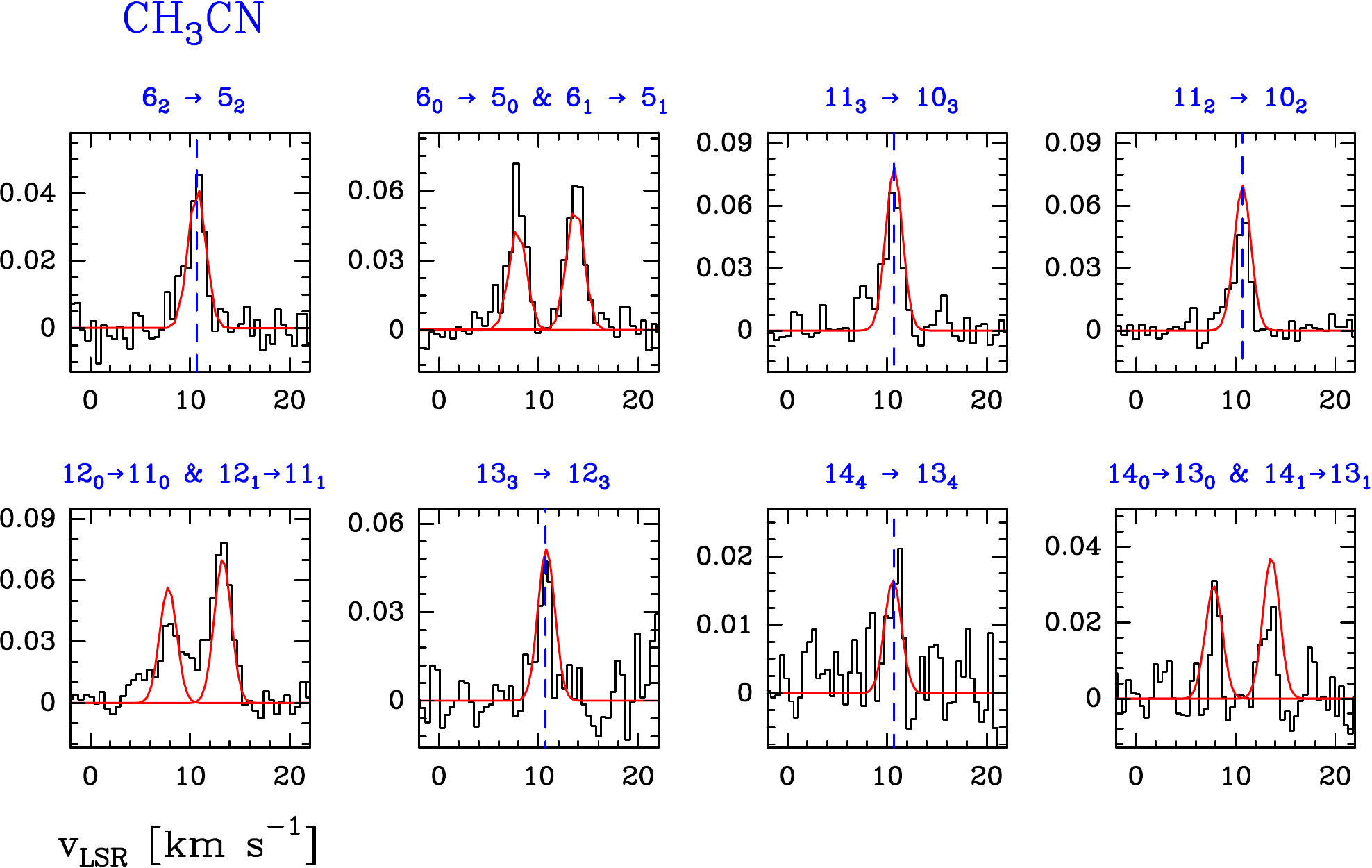}
\caption{Example of A- and \mbox{E-CH$_{3}$CN} lines (black histogram spectra). 
A non-LTE LVG model (\mbox{$T_{\rm k}$ $\simeq$ 150 K},
\mbox{$n$(H$_{2}$) $\simeq$ 1 $\times$ 10$^{6}$ cm$^{-3}$}, 
\mbox{$N$(A-CH$_{3}$CN) = 5 $\times$ 10$^{11}$ cm$^{-2}$}, and 
\mbox{$N$(E-CH$_{3}$CN) = 6 $\times$ 10$^{11}$ cm$^{-2}$}) is shown overlaid in red (see Sect.~\ref{Non-LTE}).}
\label{fig:CH3CN_lines}
\end{figure*}

\begin{figure*}
\includegraphics[scale=0.5,angle=0]{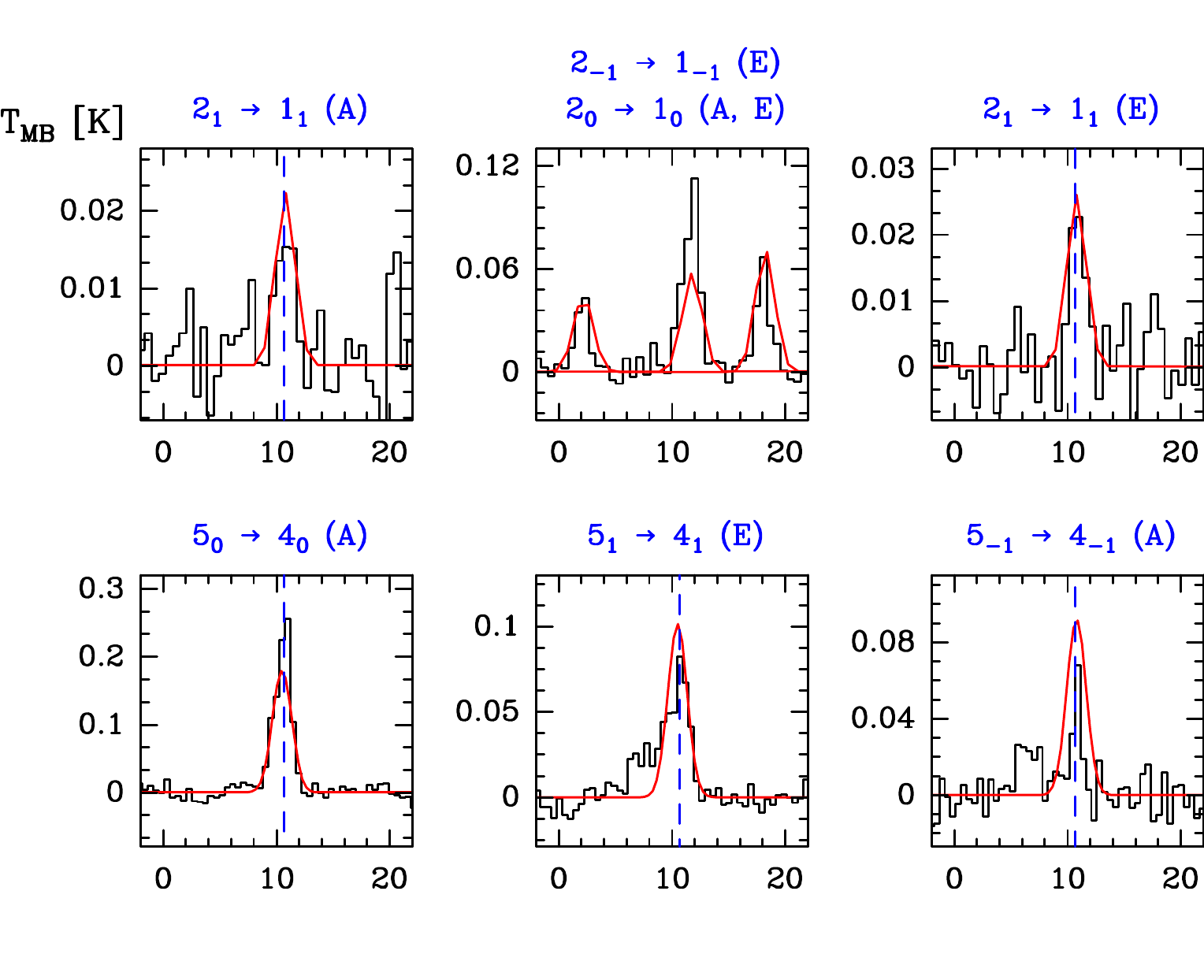} 
\hspace*{0.2cm}\includegraphics[scale=0.5,angle=0]{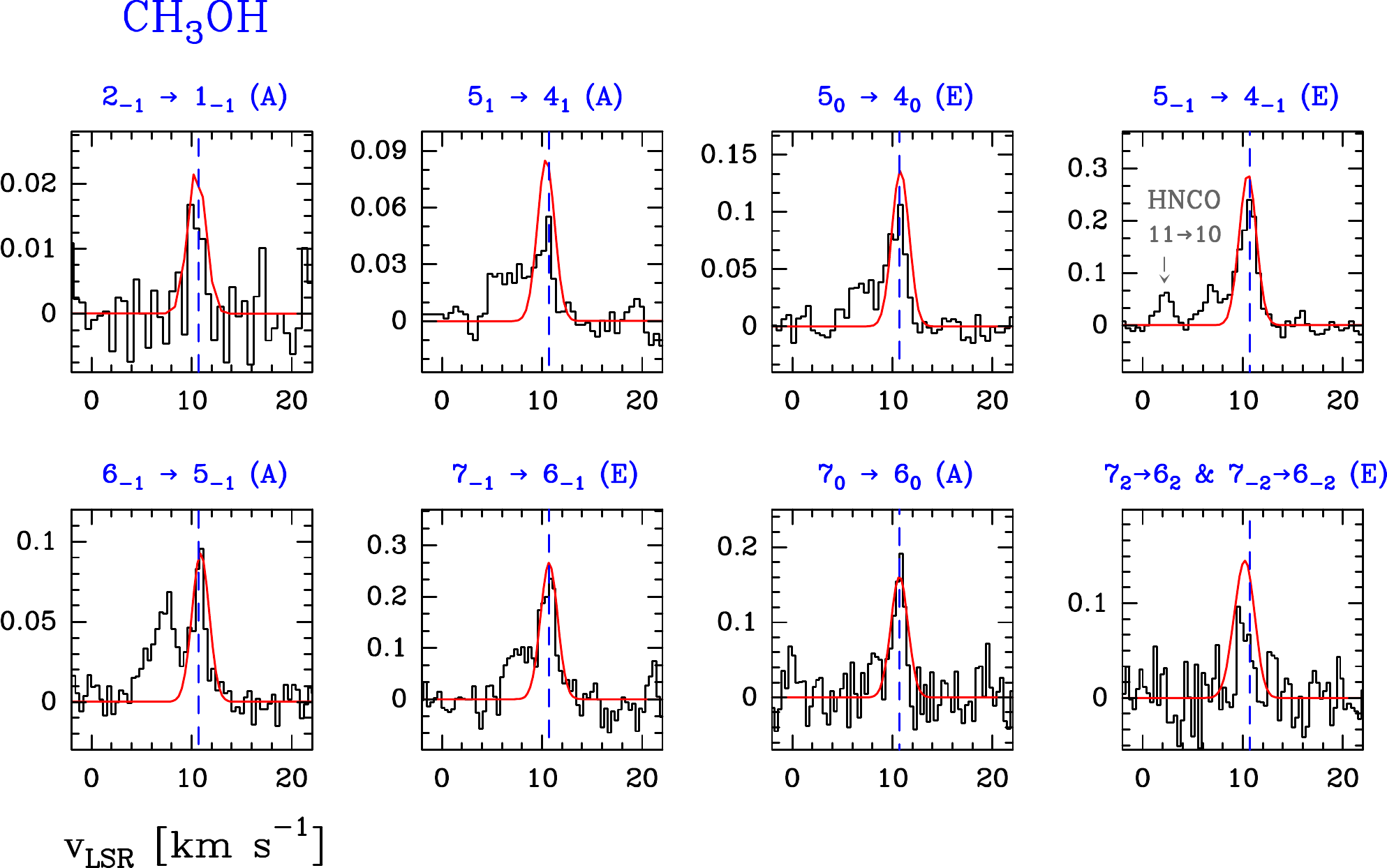}
\caption{Example of A- and \mbox{E-CH$_{3}$OH} lines (black histogram spectra). 
A non-LTE LVG model (\mbox{$T_{\rm k}$ $\simeq$ 40 K},
\mbox{$n$(H$_{2}$) $\simeq$ 3 $\times$ 10$^{7}$ cm$^{-3}$}, 
\mbox{$N$(A-CH$_{3}$OH) = 1.2 $\times$ 1$0^{13}$ cm$^{-2}$}, and 
\mbox{$N$(E-CH$_{3}$OH) = 1.9 $\times$ 10$^{13}$ cm$^{-2}$}) is shown overlaid in red (see Sect.~\ref{Non-LTE}).}
\label{fig:CH3OH_lines}
\end{figure*}

\begin{figure*}
\includegraphics[scale=0.5,angle=0]{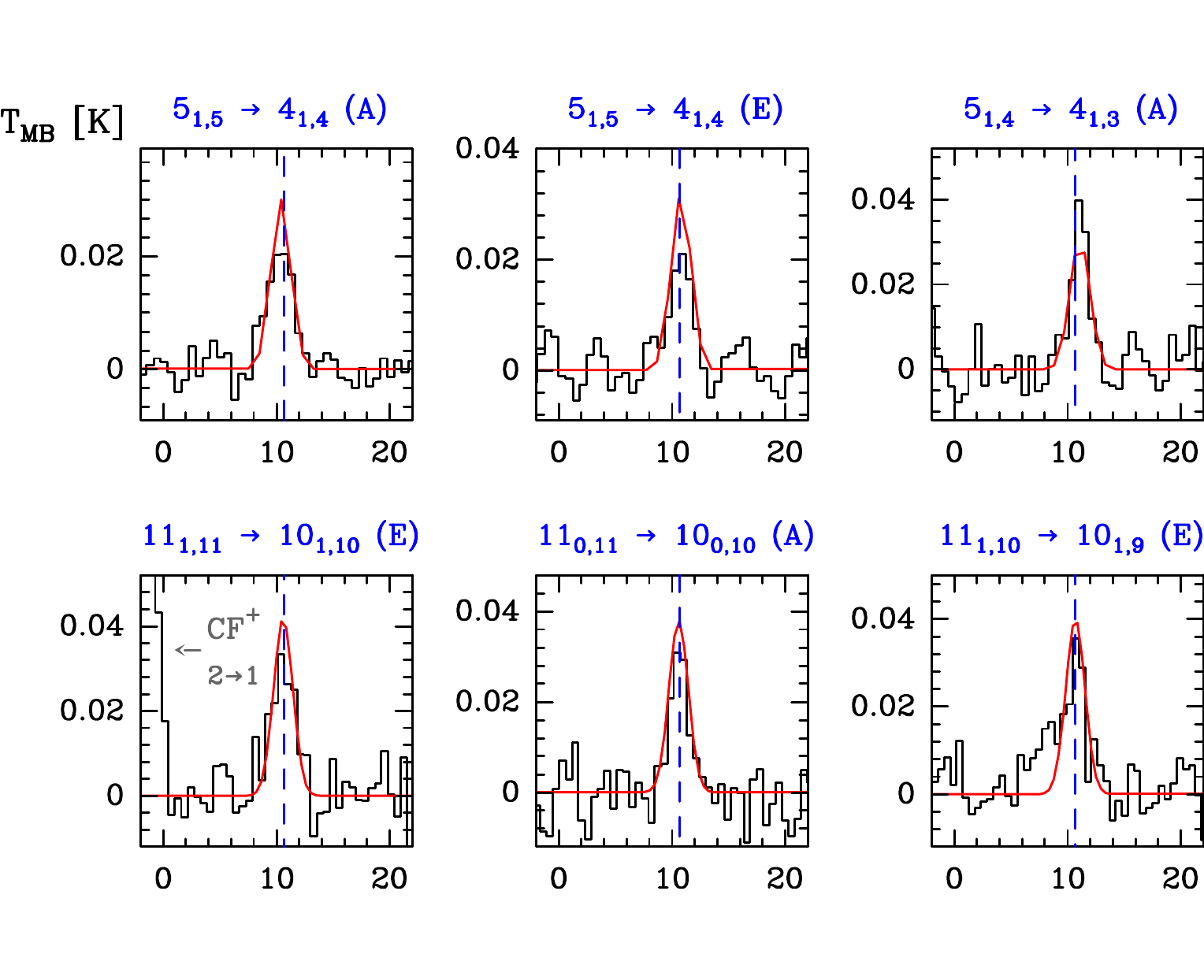} 
\hspace*{0.2cm}\includegraphics[scale=0.5,angle=0]{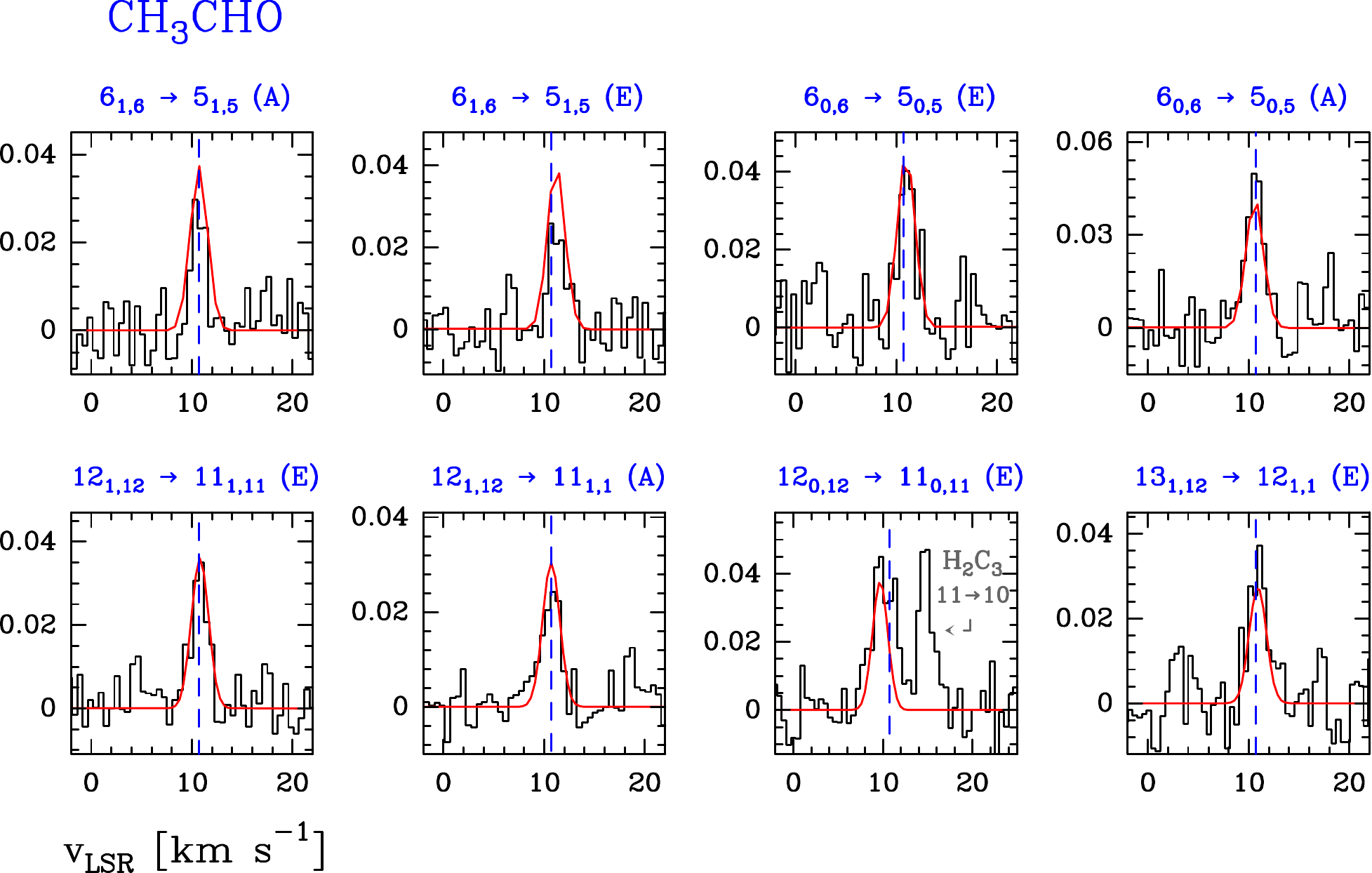}
\caption{Examples of A- and E-CH$_{3}$CHO lines (black histogram spectra). Single excitation temperature models are shown overlaid in red (see Sect.~\ref{DR}).}
\label{fig:CH3CHO_lines}
\end{figure*}

\subsection{Methyl cyanide: CH$_{3}$CN} \label{CH3CN}

Methyl cyanide, CH$_{3}$CN, is a strongly prolate symmetric rotor, therefore, the permitted radiative transitions are all \mbox{$\Delta$$K$ = 0}.
 The internal rotation of the methyl group
gives rise to two non-interacting torsional substates, denoted A and E. 
The A state levels are described by \mbox{$K$ = 3$n$}, and those of E state by \mbox{$K$ = 3$n$ $\pm$ 1}, with \mbox{$n$ $\geq$ 0} \citep[e.g.][]{Kukolich_1973,Kukolich_1982,Boucher_1977,Anttila_1993,Simeckova_2004,Cazzoli_2006,Muller_2009}. Transitions with different $K$ but the same $J$ occur in narrow frequency regions 
but they have quite different energies (see Fig.~\ref{fig:CH3CN_5-4}).

We detected 18 lines of A-CH$_{3}$CN (in the \mbox{$K$ = 0} and 3 ladders) with \mbox{$E_{\rm u}/k$ $\leq$ 184.4~K}, 23 lines for the E-CH$_{3}$CN (in the \mbox{$K$ = 1}, 2, and 4 ladders) with \mbox{$E_{\rm u}/k$ $\leq$ 199.0~K}, and three lines corresponding to several fully overlapped A-E transitions. 
Examples of line profiles of A- and \mbox{E-CH$_{3}$CN} lines are shown in Fig.~\ref{fig:CH3CN_lines}. Table~\ref{Table_CH3CN} gives the observed line parameters. We did not detect the CH$_{3}$NC isomer (see Sect.~\ref{Undetected_COMs}).

\subsection{Methanol: CH$_{3}$OH} \label{CH3OH}

Methanol, CH$_{3}$OH, is the simplest alcohol molecule. It is a slightly asymmetric rotor showing hindered internal rotation. Its energy levels are classified as a symmetric rotor with quantum number $J_{K}$ and a threefold barrier that causes two states in the molecule, A and E; the latter being doubly degenerate \citep[e.g.][]{Lees_1968,Lees_1973,Xu_2008}. 

We detected 20 lines of A-CH$_{3}$OH (in the \mbox{$|$K$|$ = 0} and 1 ladders) with \mbox{$E_{\rm u}/k$ $\leq$ 104.4~K}, 33 lines for the E-CH$_{3}$OH (in the \mbox{$|$K$|$ = 1}, 2, 3, and 4 ladders) with \mbox{$E_{\rm u}/k$ $\leq$ 114.8~K}, and two lines corresponding to several fully overlapped A-E transitions.
Examples of line profiles of A- and \mbox{E-CH$_{3}$OH} lines are shown in Fig.~\ref{fig:CH3OH_lines}. Table~\ref{Table_CH3OH} gives the observed line parameters.

\subsection{Acetaldehyde: CH$_{3}$CHO} \label{CH3CHO}

Acetaldehyde, CH$_{3}$CHO, is an asymmetric-top molecule with A and E symmetry states and $a$- and $b$-type transitions (e.g. \citealt{Kleiner_1996}, and references therein).

We detected 18 lines of \mbox{A-CH$_{3}$CHO}, 15 lines of \mbox{E-CH$_{3}$CHO}, and three lines corresponding to a fully overlapped A-E transition ($a$-type transitions with \mbox{$E_{\rm u}/k$ $\leq$ 109.7~K}). Figure~\ref{fig:CH3CHO_lines} shows a selection of detected lines and Table~\ref{Table_CH3CHO} gives the observed line parameters.

\section{Spatial distribution of C$^{18}$O and H$_2$CO} \label{Spatial_distribution}

Figure~\ref{fig:Maps} shows the spatial distribution of the rotationally excited C$^{18}$O \mbox{$J$ = 3 $\rightarrow$ 2} (\mbox{$E_{\rm u}/k$ $\simeq$ 32~K}, \mbox{$A_{\rm ul}$ $\simeq$ 2 $\times$ 10$^{-6}$\,s$^{-1}$}) and H$_2$CO \mbox{$J_{K_{\rm a},K_{\rm c}}$ = 5$_{1,5} \rightarrow$ 4$_{1,4}$} (\mbox{$E_{\rm u}/k$ $\simeq$ 47~K}, \mbox{$A_{\rm ul}$ $\simeq$ 1 $\times$ 10$^{-3}$\,s$^{-1}$}) lines along the Bar. The molecular dissociation front (DF) is traced by the vibrationally excited
H$_{2}$ \mbox{$\nu$ = 1 $\rightarrow$ 0} $S$(1) line emission (H$_{2}^{*}$; black contours in Fig.~\ref{fig:Maps};  \citealt{Walmsley_2000}). The red contours represent the \OI\, fluorescent line at 1.32~$\upmu$m \citep{Walmsley_2000} marking the position of the ionisation front (IF) that separates the \HII\, region and the neutral cloud. The blue contours represent the H$^{13}$CN \mbox{$J$ = 1 $\rightarrow$ 0} emission tracing dense molecular clumps inside the Bar \citep{Lis_2003}.

\begin{figure}
\centering
\includegraphics[scale=0.6,angle=0]{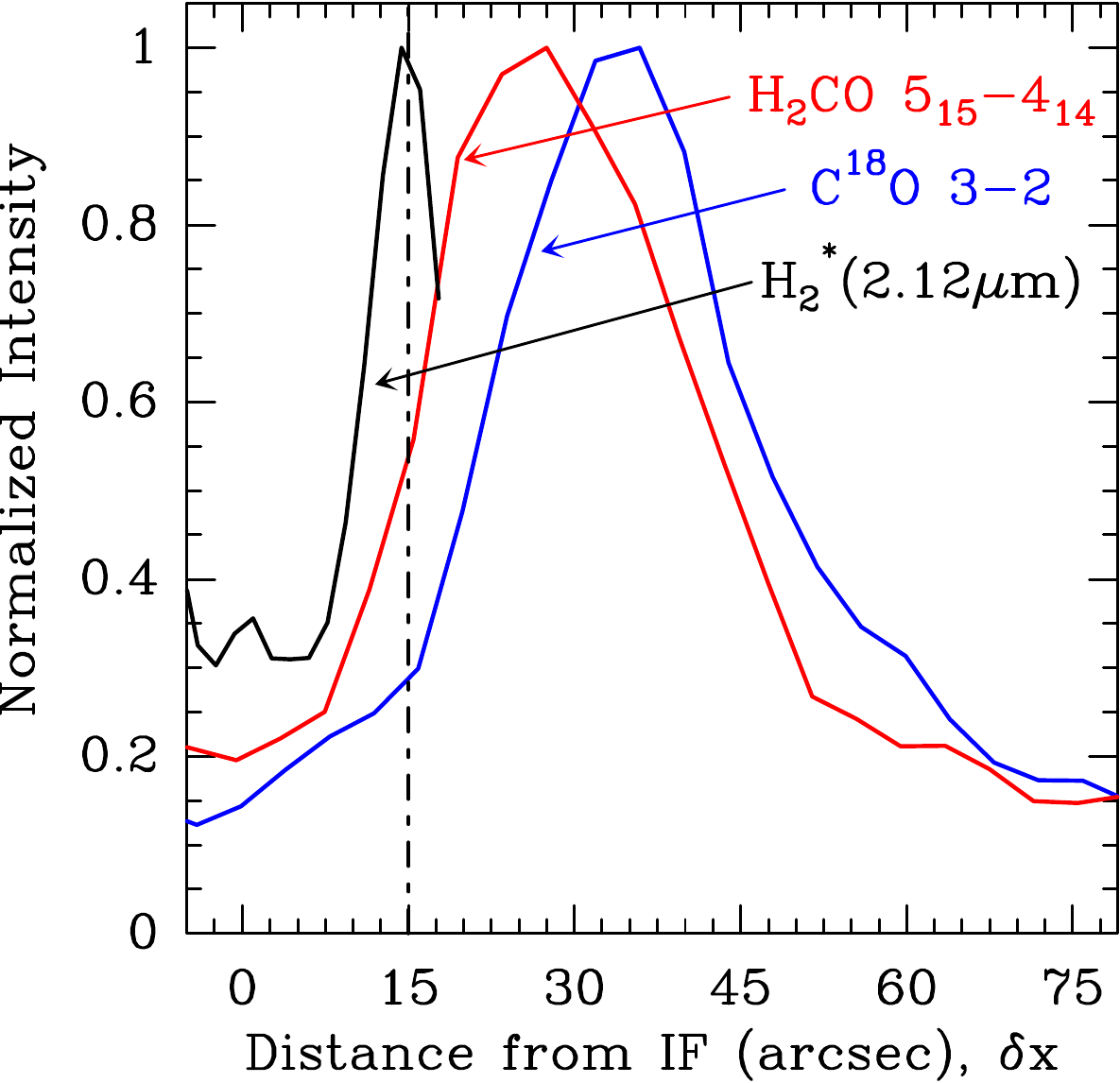} 
\caption{Normalised intensity crosscuts for C$^{18}$O \mbox{$J$ = 3 $\rightarrow$ 2} (blue curve), H$_2$CO \mbox{$J_{K_{\rm a},K_{\rm c}}$ = 5$_{1,5} \rightarrow$ 4$_{1,4}$} (red curve), and H$_{2}^{*}$ (black curve) as a function of distance from the IF (in arcsec). The vertical dot-dashed line indicates the H$_2$ dissociation front. FUV radiation comes from left.}
\label{fig:Cross-cuts}
\end{figure}

The emission from the Orion Bar can be distinguished from the extended OMC-1 cloud component by the emission LSR velocity. While OMC-1 is brighter in the \mbox{8 $-$ 10 km s$^{-1}$} velocity range, the Orion Bar emits predominantly in the \mbox{10 $-$ 11 km s$^{-1}$} range. The C$^{18}$O and H$_2$CO line emissions shown in Fig.~\ref{fig:Maps} are integrated in this latter interval. In these maps, H$_2$CO peaks at the position of the dense \mbox{($n$(H$_2$) $\simeq$ 6 $\times$ 10$^6$\,cm$^{-3}$)} and lukewarm \mbox{($T_{\rm k}$ $\simeq$ 50\,K)} clumps inside the Bar \citep{Lis_2003}. In addition, our maps reveal extended H$_2$CO emission along the PDR and close to the DF. This conclusion was initially inferred by \citet{Leurini_2010} from $\sim$11$''$ resolution observations of lower-excitation H$_2$CO lines.

To investigate the chemical stratification as a function of FUV flux attenuation, we built averaged line intensity crosscuts perpendicular to the Bar. The maps were rotated by 50$^{\circ}$ to bring the incident FUV radiation from left, and then, line intensities in the mapped area were averaged parallel to the IF. In Fig.~\ref{fig:Cross-cuts} we plot the resulting normalised intensity cuts as a function of distance to the IF position. Distances are in arcseconds and increase with depth into the molecular cloud. Although these cuts do not represent true abundance variations (different lines have different excitation conditions), to first order they can be used to study the chemical stratification in the PDR. The H$_{2}^{*}$ emission peak is at $\sim$15$''$ (vertical dot-dashed line) from the IF, and shows a narrow emission profile. The H$_2$CO \mbox{$J_{K_{\rm a},K_{\rm c}}$ = 5$_{1,5} \rightarrow$ 4$_{1,4}$} emission peak appears at $\sim$27$''$, followed by C$^{18}$O  \mbox{$J$ = 3 $\rightarrow$ 2} that peaks at $\sim$35$''$.

We note that the different position of the emission peaks of the  C$^{18}$O \mbox{$J$ = 3 $\rightarrow$ 2} and H$_2$CO \mbox{$J_{K_{\rm a},K_{\rm c}}$ = 5$_{1,5} \rightarrow$ 4$_{1,4}$} lines shown in  Fig.~\ref{fig:Cross-cuts} could
 be related to a sharp thermal gradient in the region, with the gas temperature increasing from $\sim$50\,K inside the dense clumps seen in the PDR interior \citep{Lis_2003}, to $\sim$300\,K near the DF. The  H$_2$CO \mbox{$J_{K_{\rm a},K_{\rm c}}$ = 5$_{1,5} \rightarrow$ 4$_{1,4}$} line requires higher excitation conditions (higher temperature/density) and thus it would naturally peak closer to the DF than  the  C$^{18}$O \mbox{$J$ = 3 $\rightarrow$ 2} line, even for
constant  H$_2$CO and  C$^{18}$O abundances throughout the PDR. 
Still, the main point is that observations show that H$_2$CO is present in the warm interclump gas and close to the PDR edge where the FUV flux is strong.

\section{Analysis} \label{Analysis}

\subsection{Line parameter fitting procedure} \label{Fitting_procedure}

Detected lines show simple Gaussian line profiles centred at the Orion Bar LSR velocity and thus
Gaussian profiles were fitted to all detected lines. 
For the resolved and unblended COM lines detected with significant $S/N$, the
mean line width is \mbox{$\Delta$$v$ = 1.8~km\,s$^{-1}$},
 with a standard deviation of 
\mbox{$\sigma$ = 0.6~km\,s$^{-1}$}.

The spectroscopic and observational line parameters of the detected lines are given in \mbox{Tables~\ref{Table_HCO}-\ref{Table_CH3CHO}}. 
We provide line frequencies (in MHz), 
energy of the upper level of each transition (\mbox{$E_{\rm u}/k$} in K), 
Einstein coefficient for spontaneous emission (\mbox{$A_{\rm ul}$} in \mbox{s$^{-1}$}), 
intrinsic line strength \mbox{($S_{\rm ul}$)}, 
and the level degeneracy \mbox{($g_{\rm u}$)} 
from the MADEX spectral catalogue, and the JPL and CDMS molecular databases. 
The velocity-integrated intensity \mbox{($\int T_{_{\rm MB}}$d$v$} in \mbox{mK km s$^{-1}$)}, 
LSR velocity ($v_{_{\rm LSR}}$ in km s$^{-1}$), 
FWHM (full width at half maximum) line width \mbox{($\Delta$$v$}  
in \mbox{km s$^{-1}$)}, and the line peak temperature ($T_{_{\rm MB}}$ in \mbox{mK)} 
were obtained from Gaussian fits. Parentheses indicate the uncertainty. When two or more transitions were found to be overlapping, the total profile was fitted. Fully overlapping transitions are marked with connecting symbols in the tables.

 \begin{figure*}[!ht]
\centering
\includegraphics[scale=0.41,angle=0]{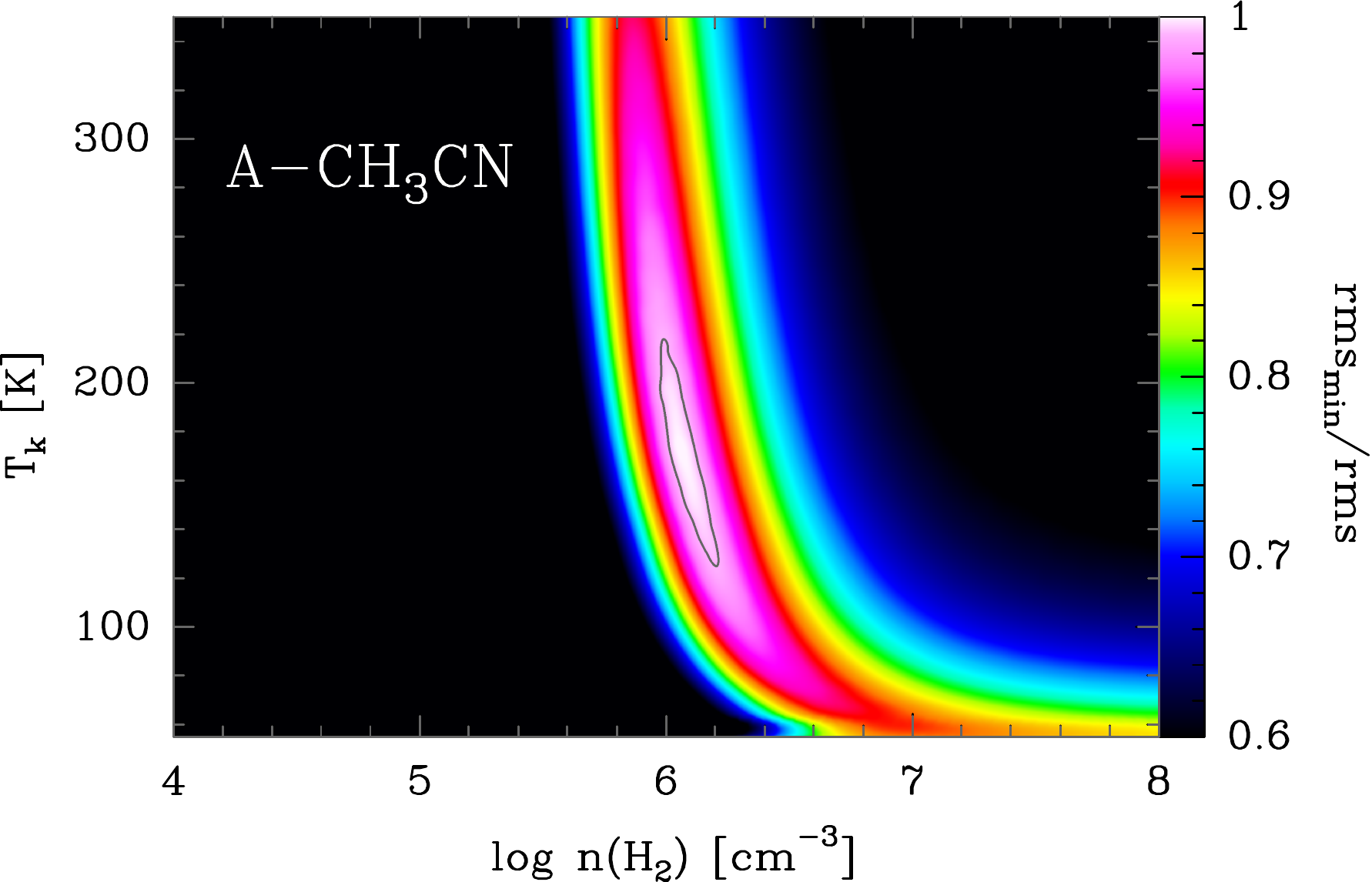} \hspace{1cm} 
\includegraphics[scale=0.41,angle=0]{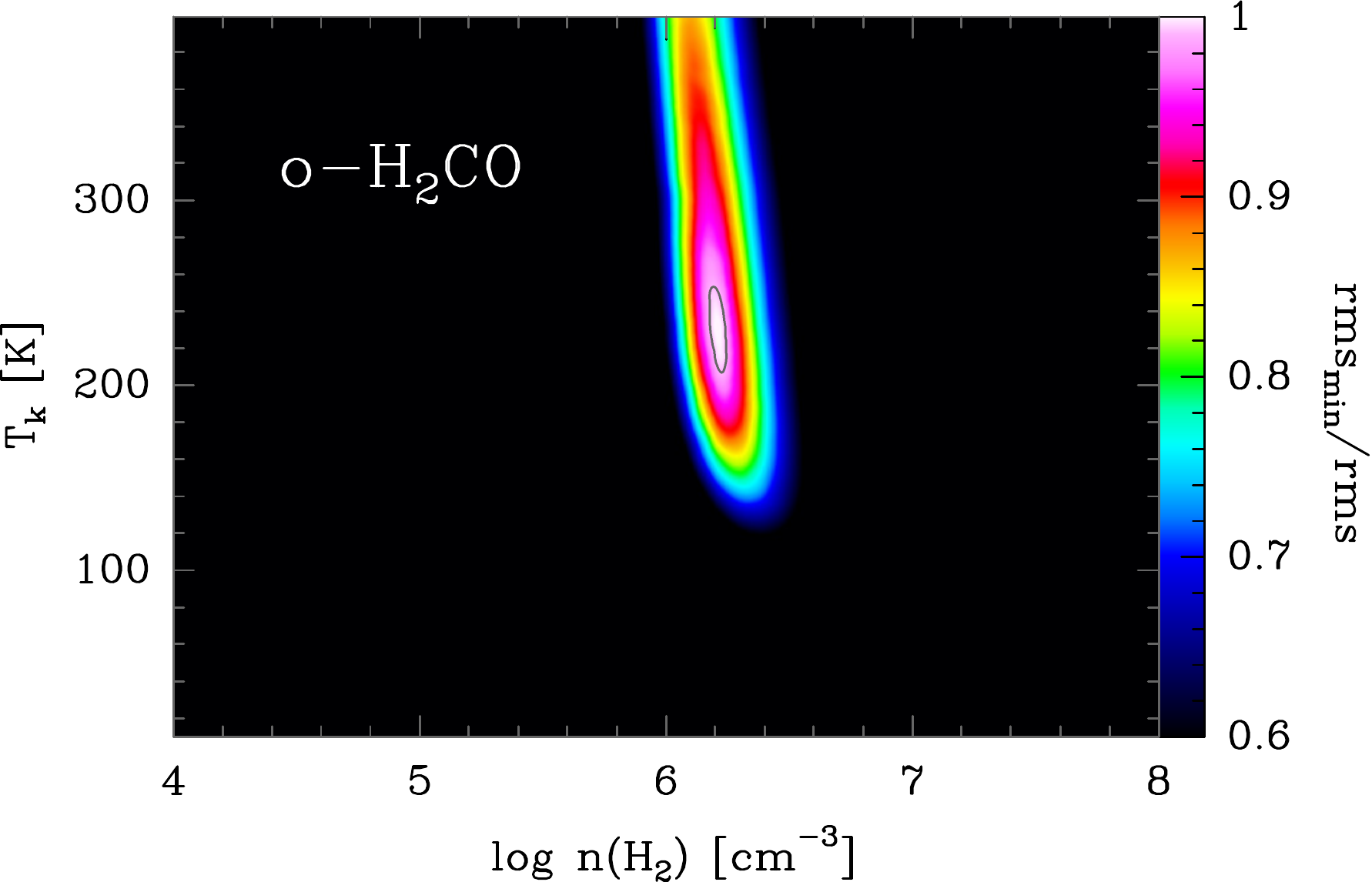}\\ \vspace{0.5cm} 
\includegraphics[scale=0.41,angle=0]{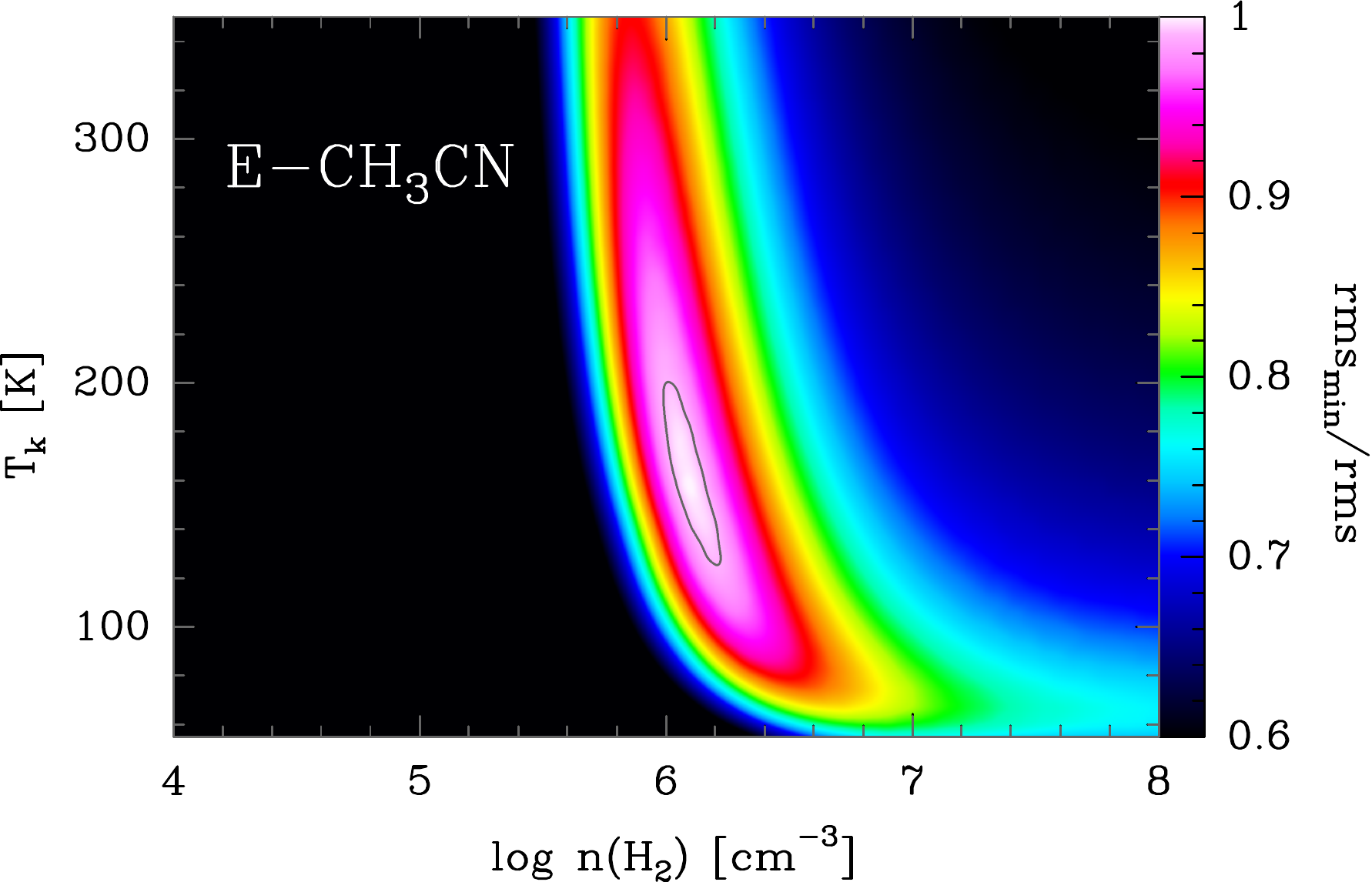} \hspace{1cm} 
\includegraphics[scale=0.41,angle=0]{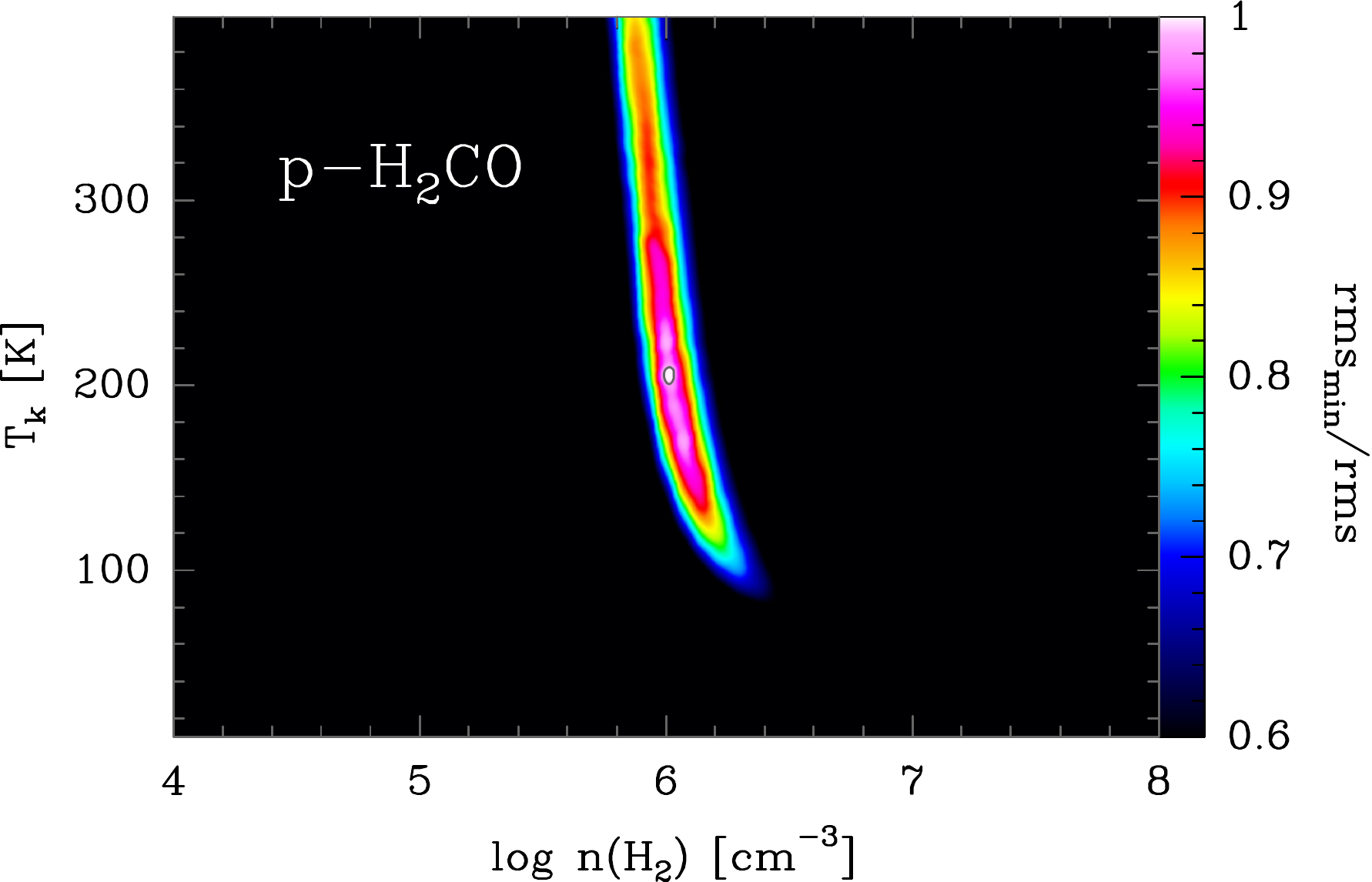}\\ \vspace{0.5cm} 
\includegraphics[scale=0.41,angle=0]{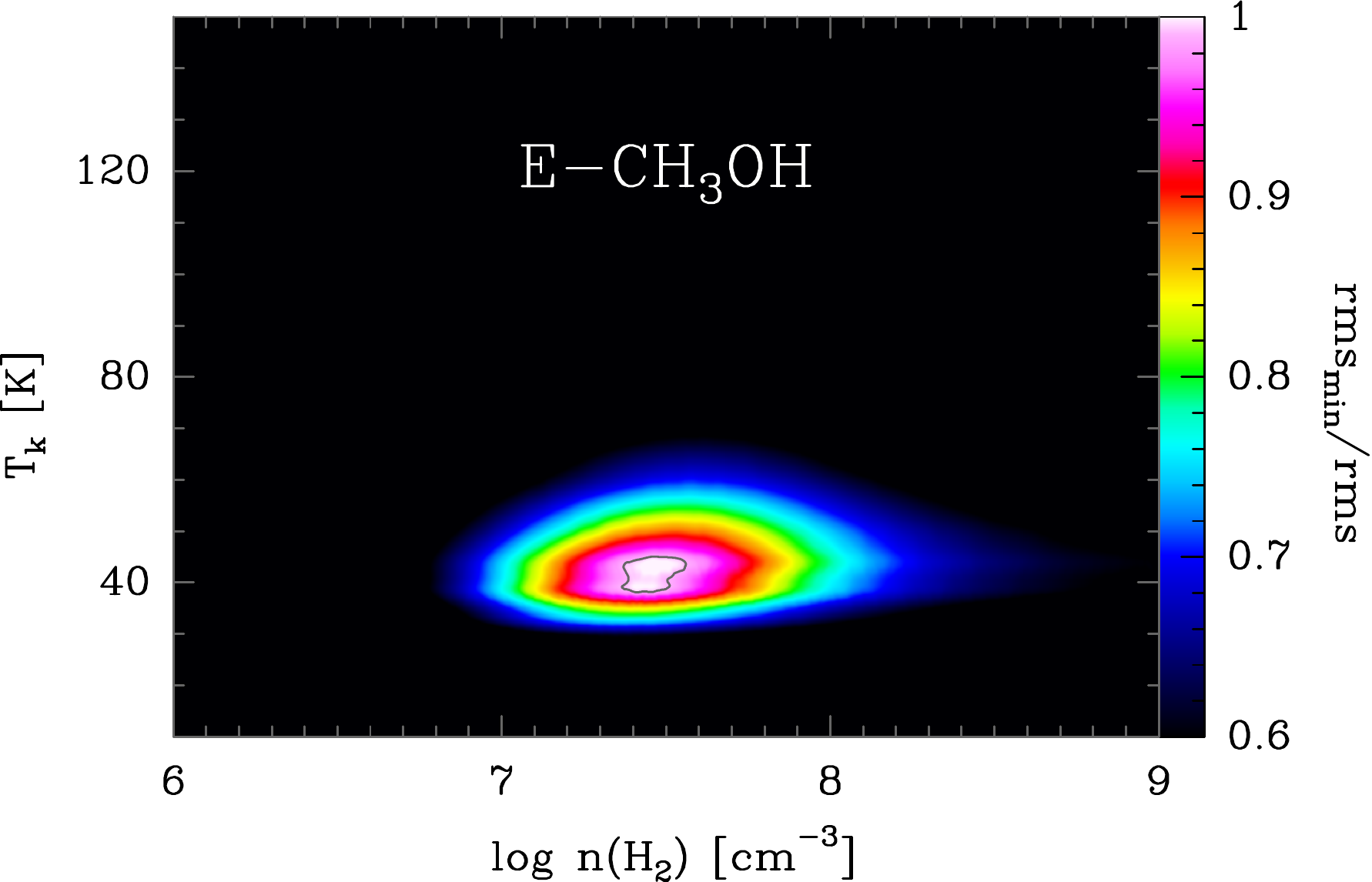}\hspace{1cm} 
\includegraphics[scale=0.41,angle=0]{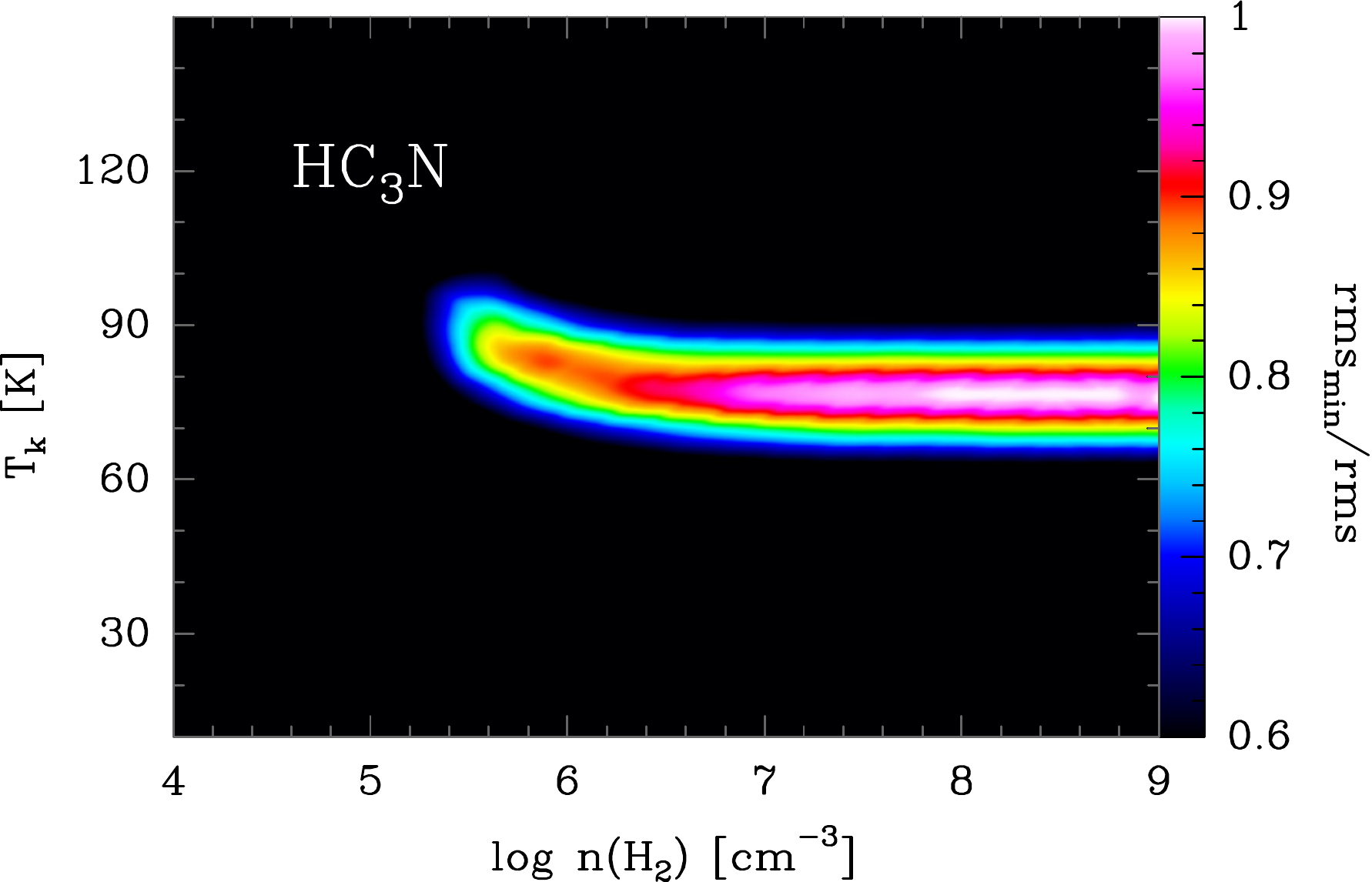}\\ \vspace{0.4cm} 
\caption{Grid of LVG excitation and radiative transfer models for 
CH$_{3}$CN, H$_{2}$CO, CH$_{3}$OH, and HC$_{3}$N lines towards the PDR position
assuming uniform beam filling and using column densities shown in Table~\ref{Table_resultsLVG}. The best fits are those with the higher rms$_{min}$ / rms value.}
\label{fig:rms}
\end{figure*}

\subsection{Non-LTE LVG excitation analysis: physical conditions} \label{Non-LTE}

We estimate the beam-averaged physical conditions towards the line survey position from non-LTE (local thermodynamic equilibrium) excitation models for molecules with known inelastic collisional rate coefficients (\mbox{H$_2$CO}, \mbox{HC$_{3}$N}, \mbox{CH$_{3}$CN}, and \mbox{CH$_{3}$OH})\footnote{We used \mbox{[H$_2$CO--H$_2$]} collisional rates for ortho and para H$_2$CO \citep{Green_1991}, \mbox{[A/E-CH$_{3}$OH--H$_2$]} for A- and \mbox{E-CH$_{3}$OH} \citep{Rabli_2010}, \mbox{[A/E-CH$_{3}$CN--H$_2$]} for A- and \mbox{E-CH$_{3}$CN} \citep{Green_1986}, and \mbox{[HC$_{3}$N--H$_2$]} for HC$_{3}$N \citep{Wernli_2007}.}.
In particular, we run a large grid of non-LTE excitation and radiative transfer models in which the statistical equilibrium equations are explicitly solved in the LVG (large velocity gradient) approximation using MADEX \citep[][see also Appendix~A in \citealt{Cuadrado_2015a}]{Cernicharo_2012}.

We tested a broad range of column density ($N$), H$_{2}$ density (\mbox{$n$(H$_{2}$) = 10$^{3}$ $-$ 10$^{9}$ cm$^{-3}$}), and gas temperature \mbox{($T_{\rm k}$ = 10 $-$ 1000 K)} values. We adopted \mbox{$\Delta$$v$ = 2 km s$^{-1}$}
 line widths.
In order to constrain the physical conditions that reproduce the observed line intensities and line profiles towards the observed position, we compared the observed \mbox{H$_2$CO}, \mbox{CH$_{3}$OH}, \mbox{CH$_{3}$CN}, and HC$_{3}$N lines to synthetic lines obtained from the grid of LVG models. 
The best fit model (assuming uniform beam filling emission) was obtained by finding the minimum root mean square (rms); we refer to \citet{Cuadrado_2015a} for a detailed explanation.
The range of column densities that we used as an input in the models is around the value derived from the rotational diagrams analysis (see Sect.~\ref{DR}). In fact, the best fit models have column densities within a factor of~2 (see Table~\ref{Table_resultsLVG}) of the inferred value from the rotational diagram analysis (we note that in these models the observed lines are optically thin).
Figure~\ref{fig:rms} represents the \mbox{rms$_{\rm min}$/rms} ratio as a function of 
$T_{\rm k}$ and $n$(H$_{2}$) for a grid of excitation models trying to fit the observed lines towards the PDR position. 

\begin{table}[!ht]
 \centering 
 \caption{Best fit LVG model parameters and comparison with the column density obtained with the rotational diagram (RD) analysis.}
 \label{Table_resultsLVG}     
  
  \begin{tabular}{l c c c c c@{\vrule height 9pt depth 5pt width 0pt}}     
 \hline\hline

  & \multicolumn{3}{c}{LVG calculations} \rule[0.15cm]{0cm}{0.2cm}\ &  &  RD analysis$^{*}$ \rule[0.2cm]{0cm}{0.2cm}\  \\ \cline{2-4} \cline{6-6}

  & $T_{\rm k}$ & $n$(H$_2$) & $N$(X)$_{\rm \, LVG}$ &  & $N$(X)$_{\rm \, RD}$ \\

  & [K] & [cm$^{-3}$] & $\mathrm{[cm^{-2}]}$  & & $\mathrm{[cm^{-2}]}$  \\

 \hline  
 
o-H$_2$CO  & $\sim$200  & $\sim$10$^{6}$ & 4.0 $\times$ 10$^{13}$  & & (4.4 $\pm$ 0.6) $\times$ 10$^{13}$ \\
p-H$_2$CO  & $\sim$200  & $\sim$10$^{6}$ & 1.8 $\times$ 10$^{13}$  & & (1.6 $\pm$ 0.1) $\times$ 10$^{13}$ \\
HC$_3$N  & $\sim$80  & $\sim$10$^{6}$ & 4.0 $\times$ 10$^{11}$  & & (4.2 $\pm$ 0.3) $\times$ 10$^{11}$ \\
A-CH$_3$CN  & $\sim$150  & $\sim$10$^{6}$ & 5.0 $\times$ 10$^{11}$  & & (5.8 $\pm$ 0.8) $\times$ 10$^{11}$ \\
E-CH$_3$CN  & $\sim$150  & $\sim$10$^{6}$ & 6.0 $\times$ 10$^{11}$  & & (5.7 $\pm$ 0.7) $\times$ 10$^{11}$ \\
E-CH$_3$OH  & $\sim$40  & $\sim$3$\times$10$^{7}$ & 1.9 $\times$ 10$^{13}$  & & (1.9 $\pm$ 0.2) $\times$ 10$^{13}$ \\

\hline

   \end{tabular}                                 
     \tablefoot{
 $^{*}$ Rotational diagram analysis assuming uniform beam filling (see Sect.~\ref{DR}).}
 \end{table}

From the excitation analysis it seems that not all COMs arise from the same PDR layers (i.e. same cloud depth). The set of beam-averaged physical conditions that best fit the \mbox{CH$_{3}$CN} and \mbox{H$_2$CO} lines lie within \mbox{$T_{\rm k}$ = 150 $-$ 250 K} and
\mbox{$n$(H$_{2}$) = (1 $-$ 3) $\times$ 10$^{6}$ cm$^{-3}$}, similar (although slightly denser gas) to that obtained for C$_{2}$H in \citet{Cuadrado_2015a}. Figures~\ref{fig:H2CO_lines} and \ref{fig:CH3CN_lines} show the best fit models overlaid over the observed lines.
The HC$_{3}$N detected lines (see Fig.~\ref{fig:HC3N_lines}) fit within \mbox{$T_{\rm k}$ = 70 $-$ 90 K} and \mbox{$n$(H$_{2}$) $>$ 1 $\times$ 10$^{6}$ cm$^{-3}$}. 
The physical conditions that fit the CH$_{3}$OH detected lines (see Fig.~\ref{fig:CH3OH_lines}) are colder and denser (\mbox{$T_{\rm k}$ = 40 $-$ 50 K} and \mbox{$n$(H$_{2}$) $\simeq$ 5 $\times$ 10$^{7}$ cm$^{-3}$}). 
The best-fit parameters (\mbox{$n$(H$_{2}$)}, $T_{\rm k}$, and $N$) of the LVG models fitting the lines in Fig.~\ref{fig:H2CO_lines}, \ref{fig:HC3N_lines}, \ref{fig:CH3CN_lines}, and  \ref{fig:CH3OH_lines} are shown in Table~\ref{Table_resultsLVG}.
 We note, however, that the best fit density value depends on the assumed emission beam filling factor\footnote{Assuming that the emission source has a 2D Gaussian shape, the beam filling factor ($\mathrm{\eta_{_{bf}}}$) is equal to \mbox{$\mathrm{\eta_{_{bf}}=\theta_{_{S}}^{\, 2}/\,(\theta_{_{S}}^{\, 2}+\theta_{_{B}}^{\, 2})}$}, with $\theta_{_{\rm B}}$the HPBW of the 30~m telescope (in arcsec) and $\theta_{_{\rm S}}$ the diameter of the Gaussian source (also in arcsec). For the semi-extended emission we assumed that \mbox{$\theta_{_{\rm S}}$ = 9$''$} (see Sect.~\ref{DR}).} for each molecule. Lower densities (by a factor of up to $\sim$10) would still
be consistent with observations if one assumes a semi-extended emission source (i.e. more compact than the beam-size at each observed frequency; see Sect.~\ref{DR}).

Higher-angular-resolution maps are needed to spatially resolve the structures
producing the molecular emission \citep{Goicoechea_2016}.
Either way,  CH$_{3}$OH and HC$_3$N  seem to arise from a different, cooler component where the FUV flux has been attenuated.
This scenario agrees with the different H$_{2}$CO and CH$_{3}$OH spatial distributions observed by \citet{Leurini_2010} with the IRAM~30~m telescope and the PdB interferometer.
In particular, Leurini et al. found that both species show different spatial distributions, with CH$_{3}$OH only tracing the denser and cooler clumps seen deeper inside the Bar, whereas H$_{2}$CO
also traces the warmer extended (interclump) gas directly exposed to the strong FUV-field.
Consistent with previous works \citep{Goicoechea_2011,Goicoechea_2016,Nagy_2013}, 
we derive
very high thermal gas pressures at the PDR edge (\mbox{$P_{\rm th}/k$ = $n_{\rm H}\,T_{\rm K}$ $\gtrsim$ 10$^{8}$~K\,cm$^{-3}$}).

\subsection{Subthermal excitation effects on the rotational diagrams of symmetric- and asymmetric-top molecules} \label{effects}

\begin{figure}[!t]
\centering
\includegraphics[scale=0.53,angle=0]{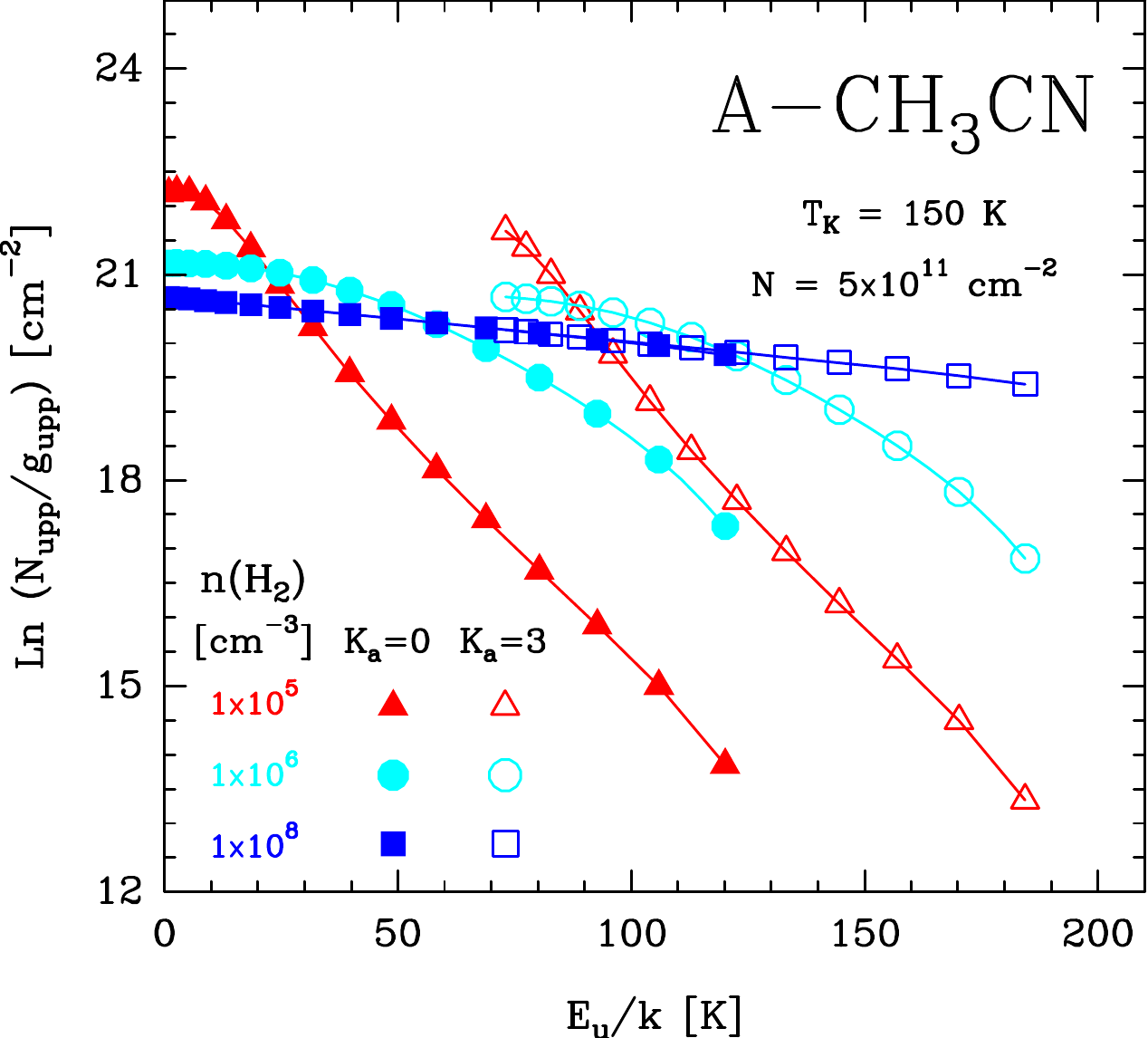}
\caption{Rotational population diagrams for \mbox{A-CH$_3$CN} computed with a non-LTE excitation code. All models adopt  the same gas temperature (\mbox{$T_{\rm k}$ = 150 K}) and column 
density (\mbox{$N_{\rm tot}$ = 5 $\times$ 10$^{11}$ cm$^{-2}$}), but  three different $n(\rm H_2)$ values. For simplicity, only rotational transitions in the 
 \mbox{$K_{\rm a}$ = 0} (filled symbols) and 3 (empty symbols) ladders are shown.}
\label{fig:A-CH3CN_DR_LVG}
\end{figure}

\begin{figure}[!t]
\centering
\includegraphics[scale=0.53,angle=0]{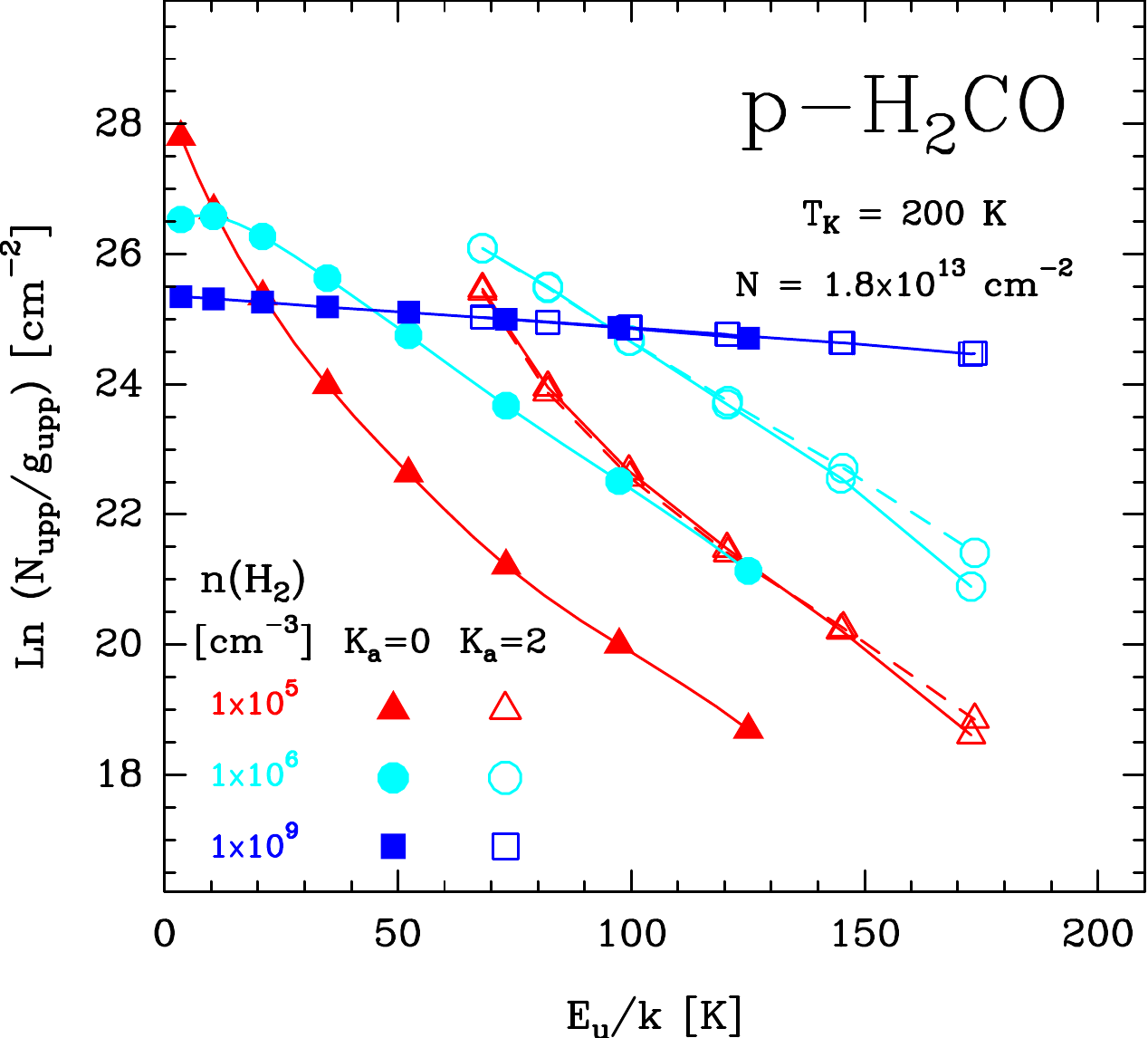}
\caption{Rotational population diagrams for \mbox{p-H$_2$CO} computed with a non-LTE excitation code. All models adopt  the same gas temperature (\mbox{$T_{\rm k}$ = 200 K}) and column 
density (\mbox{$N_{\rm tot}$ = 1.8 $\times$ 10$^{13}$ cm$^{-2}$}), but three different $n(\rm H_2)$ values. For simplicity, only rotational transitions in the \mbox{$K_{\rm a}$ = 0} (filled symbols) and 2 (empty symbols) ladders are shown. We note that at densities \mbox{$<$10$^8$ cm$^{-3}$}, the set of transitions with the \mbox{$K_{\rm a}$ = 2} number are also split into two components: 
(i) transitions with \mbox{$K_{\rm c}$ = $J$ $-$ 1} (empty symbols connected with solid lines) and 
(ii) transitions with \mbox{$K_{\rm c}$ = $J$ $-$ 2} (empty symbols connected with dashed lines).}
\label{fig:p-H2CO_DR_LVG}
\end{figure}

In this Section we show how subthermal excitation effects modify the resulting rotational population diagram of symmetric- and asymmetric-top molecules such as CH$_3$CN and H$_2$CO, respectively. In particular we run a grid of LVG models for varying gas densities, but keeping the same gas temperature and column density. Figures~\ref{fig:A-CH3CN_DR_LVG} and \ref{fig:p-H2CO_DR_LVG} show model results in the form of population diagrams for A-CH$_3$CN transitions in the \mbox{$K_{\rm a}$ = 0} and 3 rotational ladders (\mbox{$T_{\rm k}$ = 150 K} and \mbox{$N_{\rm tot}$ = 5 $\times$ 10$^{11}$ cm$^{-2}$}), and for \mbox{p-H$_{2}$CO} in the \mbox{$K_{\rm a}$ = 0} and 2 rotational ladders (\mbox{$T_{\rm k}$ = 200 K} and \mbox{$N_{\rm tot}$ = 1.8 $\times$ 10$^{13}$ cm$^{-2}$}). Only at very high gas densities, higher than the critical density for collisional excitation \mbox{($n$ $\gg$ $n_{\rm cr}$)}, do inelastic collisions drive the level populations close to LTE \mbox{($T_{\rm rot}$ $\simeq$ $T_{\rm k}$)}. In a rotational diagram, the $K_{\rm a}$ ladders merge in a single straight line with a slope equal to $-1/T_{\rm k}$. 
Owing to the high gas densities, this characteristic straight diagram is typically seen towards hot cores \citep[e.g.][]{Bisschop_2007}.
As the gas density decreases, the excitation becomes subthermal \mbox{($T_{\rm rot}$ $<$ $T_{\rm k}$)}, and
the rotational population diagram starts to show separate rotational ladders for each set of transitions with the same $K_{\rm a}$ quantum number. For very polar molecules such as CH$_3$CN and H$_2$CO
(thus high $n_{\rm cr}$), subthermal excitation happens at relatively high densities. Therefore, accurate column densities and rotational temperatures from rotational diagrams can only be obtained if the individual column densities for each rotational ladder are computed independently. This likely explains the discrepancy with the high $T_{\rm rot}$ obtained by \citet{Nagy_2017} in the rotational diagram of their submm \mbox{o-H$_2$CO} lines observed with Herschel/HIFI towards the Bar.

\begin{table*}
 \centering 
 \caption{Rotational temperatures ($T_{\rm rot}$), column densities ($N$), and abundances with respect
 to hydrogen nuclei inferred in the Orion Bar PDR.}
 \label{Table_results}     
  
  \begin{tabular}{l c c c c c c c@{\vrule height 9pt depth 5pt width 0pt}}     
 \hline\hline

  & \multicolumn{2}{c}{Uniform beam filling} \rule[0.15cm]{0cm}{0.2cm}\ &  &  \multicolumn{2}{c}{Semi-extended source} \rule[0.2cm]{0cm}{0.2cm}\ & &  \\ \cline{2-3} \cline{5-6}

  & $T_{\rm rot}$ & $N$(X) &  & $T_{\rm rot}$  & $N$(X) & Abundance$^{*}$  &  Notes \rule[0.4cm]{0cm}{0.1cm}\ \\

  & $\mathrm{[K]}$ & $\mathrm{[cm^{-2}]}$ &   & $\mathrm{[K]}$ & $\mathrm{[cm^{-2}]}$ & & \\

 \hline  
 
 \bf HCO                      &  15 $\pm$ 2     &   (1.1 $\pm$ 0.3) $\times$ 10$^{13}$   &   &    9 $\pm$ 2   &   (5.3 $\pm$ 3.5) $\times$ 10$^{13}$      &  (1.7 $-$ 8.4) $\times$ 10$^{-10}$   &   a     \\
 \hline

 o-H$_{2}$CO $K_{\rm a}$ = 1        &  17 $\pm$ 2     &   (3.7 $\pm$ 0.6) $\times$ 10$^{13}$   &   &   12 $\pm$ 1   &   (1.1 $\pm$ 0.3) $\times$ 10$^{14}$      &                                    &   a     \\
 o-H$_{2}$CO $K_{\rm a}$ = 3        &  17             &   (6.9 $\pm$ 1.0) $\times$ 10$^{12}$   &   &   12           &   (1.1 $\pm$ 0.2) $\times$ 10$^{13}$      &                                    &   b,c     \\
 p-H$_{2}$CO $K_{\rm a}$ = 0        &  16 $\pm$ 2     &   (9.8 $\pm$ 1.4) $\times$ 10$^{12}$   &   &    9 $\pm$ 1   &   (3.9 $\pm$ 0.6) $\times$ 10$^{13}$      &                                    &   a,c    \\
 p-H$_{2}$CO $K_{\rm a}$ = 2        &  18 $\pm$ 1     &   (6.2 $\pm$ 0.2) $\times$ 10$^{12}$   &   &   13 $\pm$ 1   &   (1.3 $\pm$ 0.1) $\times$ 10$^{13}$      &                                    &   a,c     \\
 
 \bf  [(o+p)-H$_{2}$CO]       &     ---         &   (6.0 $\pm$ 0.6) $\times$ 10$^{13}$   &   &     ---        &   (1.7 $\pm$ 0.3) $\times$ 10$^{14}$      &  (0.9 $-$ 2.7) $\times$ 10$^{-9}$    &   d      \\
\hline   
  
 o-H$_{2}^{13}$CO             &  15 $\pm$ 3     &   (7.3 $\pm$ 2.4) $\times$ 10$^{11}$   &   &   10 $\pm$ 2   &   (3.0 $\pm$ 1.1) $\times$ 10$^{12}$      &                                    &   a     \\
 p-H$_{2}^{13}$CO             &   9 $\pm$ 1     &   (2.3 $\pm$ 0.4) $\times$ 10$^{11}$   &   &    7 $\pm$ 1   &   (8.0 $\pm$ 1.2) $\times$ 10$^{11}$      &                                    &   a,c   \\
 
 \bf  [(o+p)-H$_{2}^{13}$CO]  &     ---         &   (9.6 $\pm$ 2.4) $\times$ 10$^{11}$   &   &     ---        &   (3.8 $\pm$ 1.1) $\times$ 10$^{12}$      &  (1.5 $-$ 6.0) $\times$ 10$^{-11}$   &   d      \\
\hline

 o-H$_{2}$CS $K_{\rm a}$ = 1        &  30 $\pm$ 3     &   (3.1 $\pm$ 0.5) $\times$ 10$^{12}$   &   &   19 $\pm$ 1   &   (1.3 $\pm$ 0.2) $\times$ 10$^{13}$      &                                     &   a     \\
 o-H$_{2}$CS $K_{\rm a}$ = 3        &  30             &   (6.7 $\pm$ 1.9) $\times$ 10$^{11}$   &   &   19           &   (2.7 $\pm$ 0.4) $\times$ 10$^{12}$      &                                     &   e,c     \\
 p-H$_{2}$CS $K_{\rm a}$ = 0        &  29 $\pm$ 6     &   (5.0 $\pm$ 1.4) $\times$ 10$^{11}$   &   &   16 $\pm$ 2   &   (2.6 $\pm$ 1.0) $\times$ 10$^{12}$      &                                     &   a     \\
 p-H$_{2}$CS $K_{\rm a}$ = 2        &  35 $\pm$ 8     &   (7.8 $\pm$ 4.4) $\times$ 10$^{11}$   &   &   15 $\pm$ 3   &   (2.6 $\pm$ 1.9) $\times$ 10$^{12}$      &                                     &   a     \\

 \bf  [(o+p)-H$_{2}$CS]       &     ---         &   (5.0 $\pm$ 0.7) $\times$ 10$^{12}$   &   &     ---        &   (2.1 $\pm$ 0.3) $\times$ 10$^{13}$      &  (0.8 $-$ 3.3) $\times$ 10$^{-10}$    &   d      \\
\hline

 \bf HNCO                     &  51 $\pm$ 7     &   (1.0 $\pm$ 0.1) $\times$ 10$^{12}$   &   &    26 $\pm$ 3  &   (5.6 $\pm$ 1.3) $\times$ 10$^{12}$      &  (1.6 $-$ 8.9) $\times$ 10$^{-11}$   &   a     \\
 \hline

 \bf   CH$_{2}$NH             &  28 $\pm$ 7     &   (1.1 $\pm$ 0.4) $\times$ 10$^{12}$   &   &    27 $\pm$ 7  &   (2.4 $\pm$ 1.0) $\times$ 10$^{12}$      &   (1.7 $-$ 3.8) $\times$ 10$^{-11}$   &   a      \\
\hline    
 o-H$_{2}$CCO  $K_{\rm a}$ = 1      &  55 $\pm$ 2     &   (3.0 $\pm$ 0.1) $\times$ 10$^{12}$   &   &    30 $\pm$ 2  &   (1.3 $\pm$ 0.2) $\times$ 10$^{13}$      &                                     &   a      \\    
 o-H$_{2}$CCO  $K_{\rm a}$ = 3      &  64 $\pm$ 6     &   (1.3 $\pm$ 0.3) $\times$ 10$^{12}$   &   &    34 $\pm$ 1  &   (4.2 $\pm$ 0.7) $\times$ 10$^{12}$      &                                     &   a      \\
 p-H$_{2}$CCO  $K_{\rm a}$ = 0      &  57 $\pm$ 4     &   (4.8 $\pm$ 0.6) $\times$ 10$^{11}$   &   &    42 $\pm$ 3  &   (1.3 $\pm$ 0.2) $\times$ 10$^{12}$      &                                     &   a      \\                                       
 p-H$_{2}$CCO  $K_{\rm a}$ = 2      &  54 $\pm$ 3     &   (1.1 $\pm$ 0.1) $\times$ 10$^{12}$   &   &    29 $\pm$ 2  &   (6.1 $\pm$ 1.8) $\times$ 10$^{12}$      &                                     &   a      \\

 \bf  [(o+p)-H$_{2}$CCO]      &     ---         &   (5.9 $\pm$ 0.3) $\times$ 10$^{12}$   &   &     ---        &   (2.5 $\pm$ 0.3) $\times$ 10$^{13}$      &   (0.9 $-$ 4.0) $\times$ 10$^{-10}$   &   d     \\
\hline   
 \bf   HC$_{3}$N              &  43 $\pm$ 2     &   (4.2 $\pm$ 0.3) $\times$ 10$^{11}$   &   &    27 $\pm$ 1  &   (3.2 $\pm$ 0.3) $\times$ 10$^{12}$      &   (0.7 $-$ 5.1) $\times$ 10$^{-11}$   &   a      \\

 \hline  
   cis-HCOOH                        &   23 $\pm$ 4    &  (4.6 $\pm$ 0.7) $\times$ 10$^{11}$      &   &    21 $\pm$ 4      &      (4.2 $\pm$ 0.6) $\times$ 10$^{12}$   &                                    &     f        \\
  trans-HCOOH $K_{\rm a}$ = 0       &   12 $\pm$ 2    &  (3.5 $\pm$ 0.5) $\times$ 10$^{11}$      &   &     6 $\pm$ 1      &      (4.1 $\pm$ 0.6) $\times$ 10$^{12}$   &                                    &     f       \\
  trans-HCOOH $K_{\rm a}$ = 1       &   12 $\pm$ 3    &  (3.3 $\pm$ 1.3) $\times$ 10$^{11}$      &   &     6 $\pm$ 1      &      (3.6 $\pm$ 2.1) $\times$ 10$^{12}$   &                                    &     f        \\
  trans-HCOOH $K_{\rm a}$ = 2       &   13 $\pm$ 3    &  (6.3 $\pm$ 2.8) $\times$ 10$^{11}$      &   &     7 $\pm$ 1      &      (5.0 $\pm$ 2.4) $\times$ 10$^{12}$   &                                    &     f        \\
  \textbf{[(cis+trans)-HCOOH]}   &   ---   &  (1.8 $\pm$ 0.3) $\times$ 10$^{12}$      &   &     ---        &      (1.7 $\pm$ 0.3) $\times$ 10$^{13}$   &    (0.3 $-$ 2.7) $\times$ 10$^{-10}$   &    g          \\
 
\hline    
 A-CH$_{3}$CN $K_{\rm a}$ = 0       &  26 $\pm$ 2     &   (2.8 $\pm$ 0.5) $\times$ 10$^{11}$   &   &    19 $\pm$ 1   &  (1.9 $\pm$ 0.6) $\times$ 10$^{12}$      &                                     &   a      \\
 A-CH$_{3}$CN $K_{\rm a}$ = 3       &  30 $\pm$ 1     &   (3.0 $\pm$ 0.6) $\times$ 10$^{11}$   &   &    21 $\pm$ 1   &  (1.5 $\pm$ 0.6) $\times$ 10$^{12}$      &                                     &   a      \\
 E-CH$_{3}$CN $K_{\rm a}$ = 1       &  28 $\pm$ 2     &   (2.9 $\pm$ 0.5) $\times$ 10$^{11}$   &   &    20 $\pm$ 1   &  (1.8 $\pm$ 0.4) $\times$ 10$^{12}$      &                                     &   a      \\
 E-CH$_{3}$CN $K_{\rm a}$ = 2       &  31 $\pm$ 1     &   (1.7 $\pm$ 0.2) $\times$ 10$^{11}$   &   &    22 $\pm$ 1   &  (1.0 $\pm$ 0.3) $\times$ 10$^{12}$      &                                     &   a      \\
 E-CH$_{3}$CN $K_{\rm a}$ = 4       &  26 $\pm$ 1     &   (1.1 $\pm$ 0.4) $\times$ 10$^{11}$   &   &    18 $\pm$ 1   &  (6.7 $\pm$ 3.2) $\times$ 10$^{11}$      &                                     &   a      \\
                  
 \bf  [(A+E)-CH$_{3}$CN]      &     ---         &   (1.2 $\pm$ 0.1) $\times$ 10$^{12}$   &   &     ---         &  (6.9 $\pm$ 1.0) $\times$ 10$^{12}$      &   (0.2 $-$ 1.1) $\times$ 10$^{-10}$   &   h    \\

   \hline
 
 A-CH$_{3}$OH $K_{\rm a}$ = 0       &  34 $\pm$ 2     &   (8.0 $\pm$ 0.6) $\times$ 10$^{12}$   &   &    18 $\pm$ 2   &  (5.3 $\pm$ 1.1) $\times$ 10$^{13}$      &                                     &   a      \\
 A-CH$_{3}$OH $K_{\rm a}$ = $\pm$1  &  26 $\pm$ 2     &   (3.5 $\pm$ 0.9) $\times$ 10$^{12}$   &   &    14 $\pm$ 1   &  (2.8 $\pm$ 0.7) $\times$ 10$^{13}$      &                                     &   a      \\
 E-CH$_{3}$OH                 &  36 $\pm$ 3     &   (1.9 $\pm$ 0.2) $\times$ 10$^{13}$   &   &    26 $\pm$ 3   &  (6.5 $\pm$ 1.4) $\times$ 10$^{13}$      &                                     &   a      \\

 \bf  [(A+E)-CH$_{3}$OH]      &     ---         &   (3.1 $\pm$ 0.2) $\times$ 10$^{13}$   &   &     ---         &  (1.5 $\pm$ 0.2) $\times$ 10$^{14}$      &   (0.5 $-$ 2.4) $\times$ 10$^{-9}$    &   h      \\

 \hline   
 A-CH$_{3}$CHO                &  31 $\pm$ 2     &   (2.3 $\pm$ 0.3) $\times$ 10$^{12}$   &   &    20 $\pm$ 1   &  (1.2 $\pm$ 0.2) $\times$ 10$^{13}$     &                                     &   a     \\
 E-CH$_{3}$CHO                &  33 $\pm$ 3     &   (2.6 $\pm$ 0.4) $\times$ 10$^{12}$   &   &    21 $\pm$ 1   &  (1.2 $\pm$ 0.2) $\times$ 10$^{13}$     &                                     &   a     \\
                                                                                                                                                                                                  
 \bf  [(A+E)-CH$_{3}$CHO]     &     ---         &   (4.9 $\pm$ 0.5) $\times$ 10$^{12}$   &   &     ---         &  (2.4 $\pm$ 0.3) $\times$ 10$^{13}$     &   (0.8 $-$ 3.8) $\times$ 10$^{-10}$   &   h     \\

\hline

   \end{tabular}                                  
                                                  
 \tablefoot{
 $^{*}$ The abundance of each species with respect to H nuclei is given by ${\frac{N(X)}{N_H}  =  \frac{N(X)}{N(H) + 2N(H_{2})} }$, with \mbox{$N$(H$_{2}$) $\simeq$ 3 $\times$ 10$^{22}$
 cm$^{-2}$} and \mbox{$N$(H) $\simeq$ 3 $\times$ 10$^{21}$ cm$^{-2}$} \citep{vanderWerf_2013}.
 \mbox{(a) $T_{\rm rot}$} and $N$ from rotational diagram analysis.
 \mbox{(b) Only} two lines detected with the same \mbox{$E_{\rm u}/k$}. $N$ calculated assuming the same $T_{\rm rot}$ that \mbox{o-H$_2$CO} \mbox{$K_{\rm a}$ = 1}.
 \mbox{(c) $\Delta N$} estimated assuming a 15$\%$ of the calculated $N$. 
 \mbox{(d) Total} $N$ calculated as the sum of the ortho and para species. 
 \mbox{(e) Only} two lines detected with upper level energies of \mbox{138.3 K} and \mbox{149.8 K}. They are too similar for an accurate estimation of $T_{\rm rot}$ and $N$ from a rotational diagram analysis. 
 $N$ calculated assuming the same $T_{\rm rot}$ that \mbox{o-H$_2$CS} \mbox{$K_{\rm a}$ = 1}.
 \mbox{(f) From} \citet{Cuadrado_2016}.
 \mbox{(g) Total} $N$ calculated as the sum of the cis and trans species. 
 \mbox{(h) Total} $N$ calculated as the sum of the A and E species.
  }
 \end{table*}

 \begin{figure*}
\centering
\includegraphics[scale=0.4, angle=0]{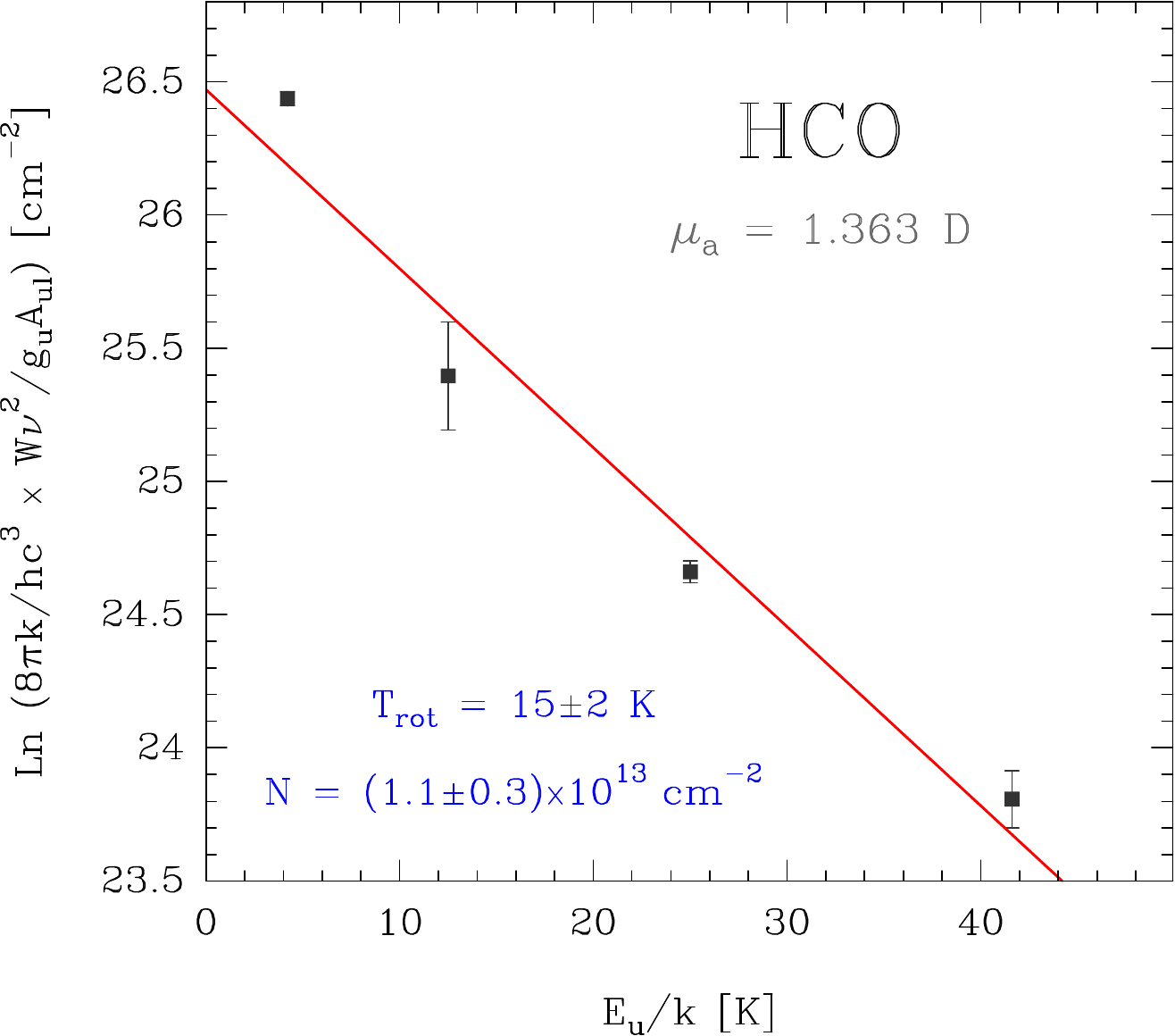} \hspace{0.5cm}
\includegraphics[scale=0.4, angle=0]{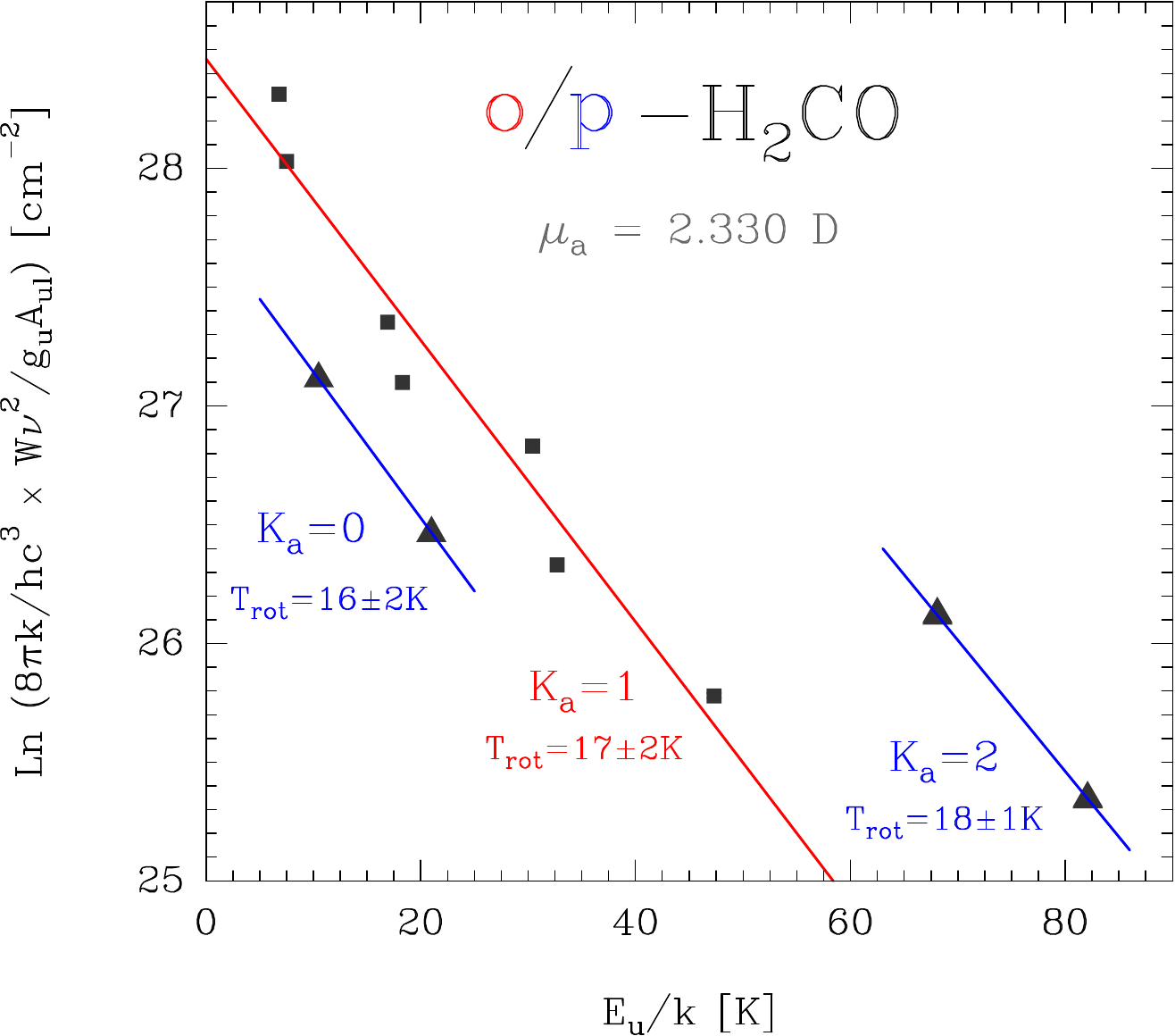} \hspace{0.5cm}
\includegraphics[scale=0.4, angle=0]{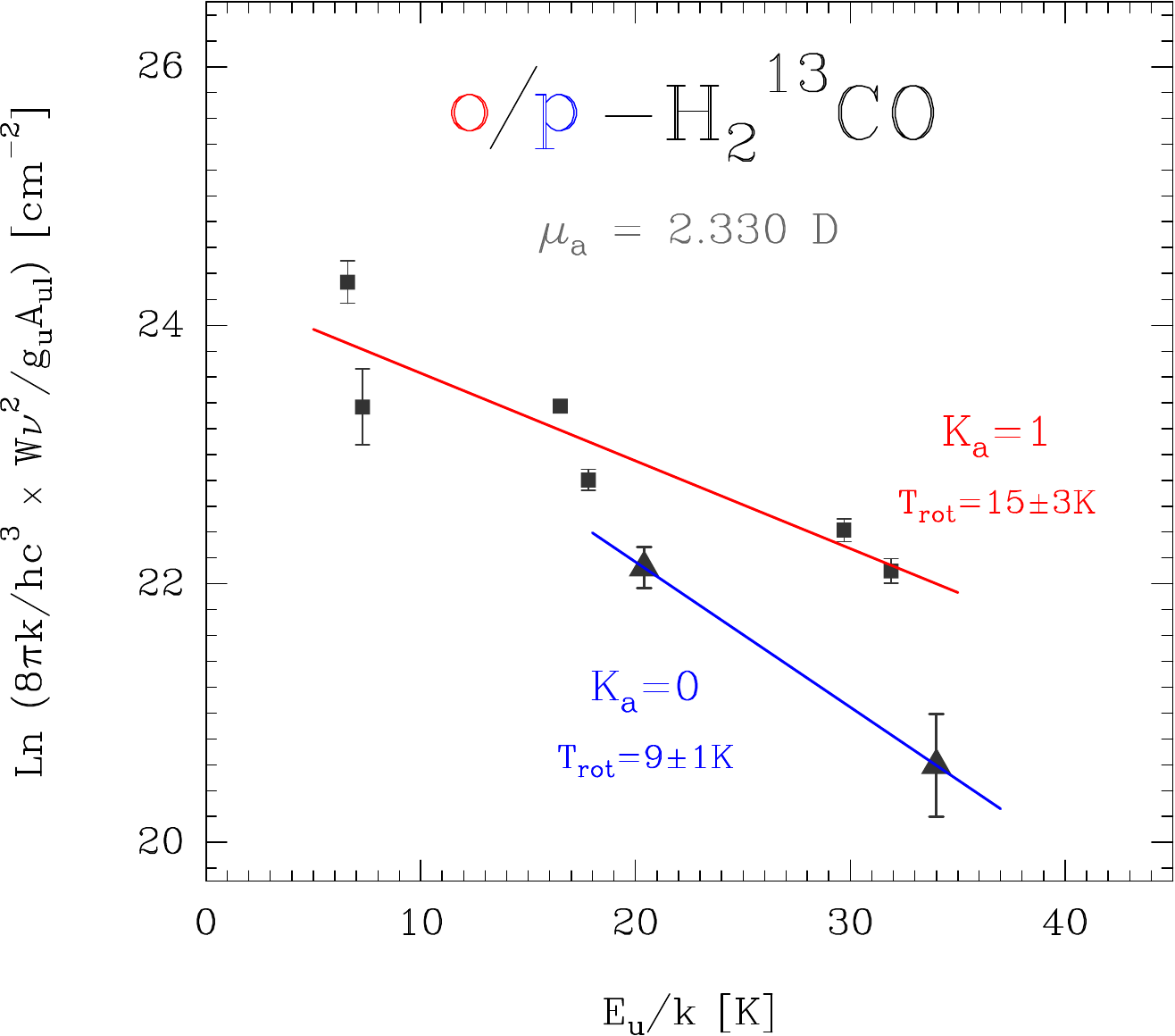} \\
\vspace{0.5cm}
\includegraphics[scale=0.4, angle=0]{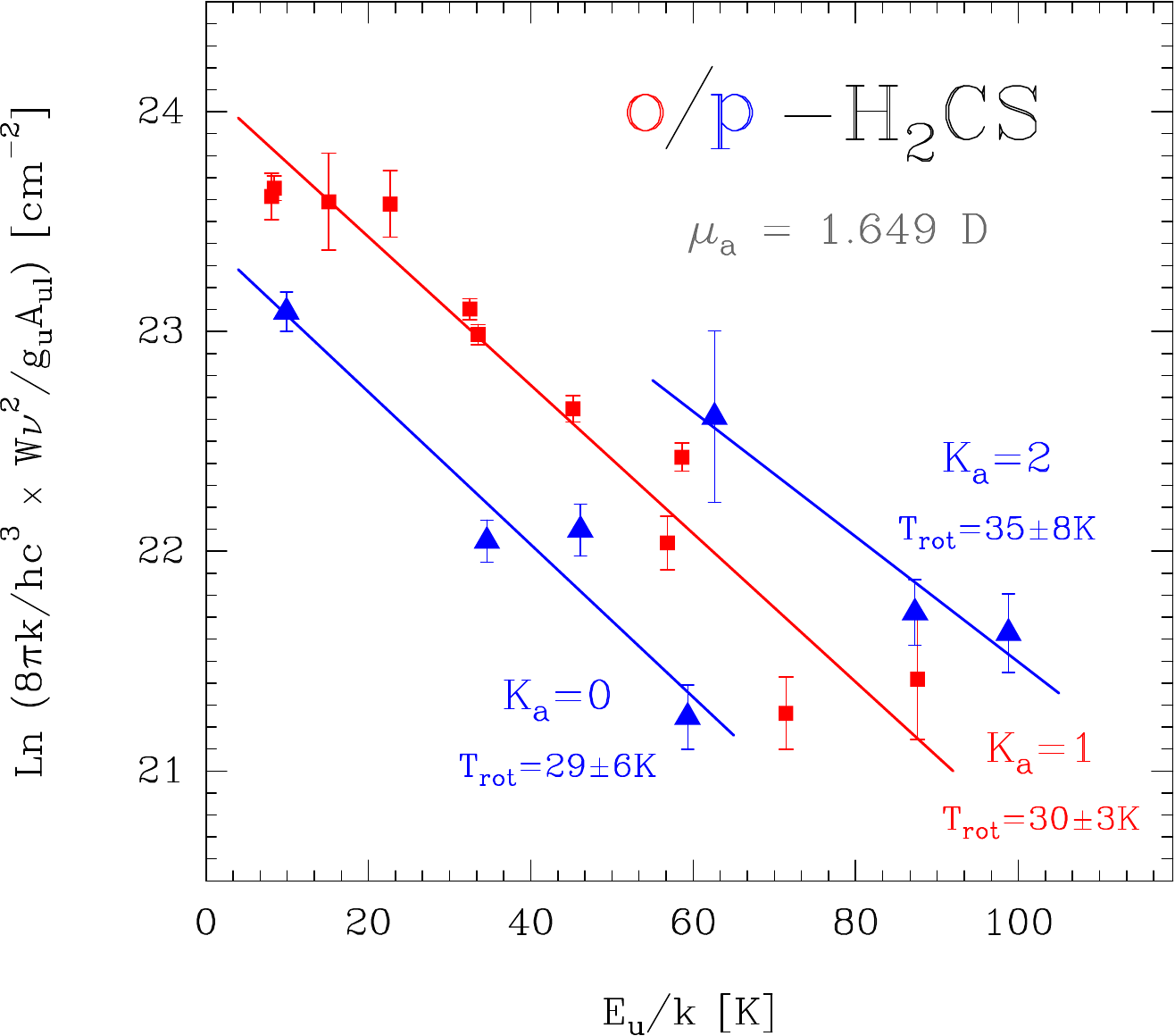}  \hspace{0.5cm}
\includegraphics[scale=0.4, angle=0]{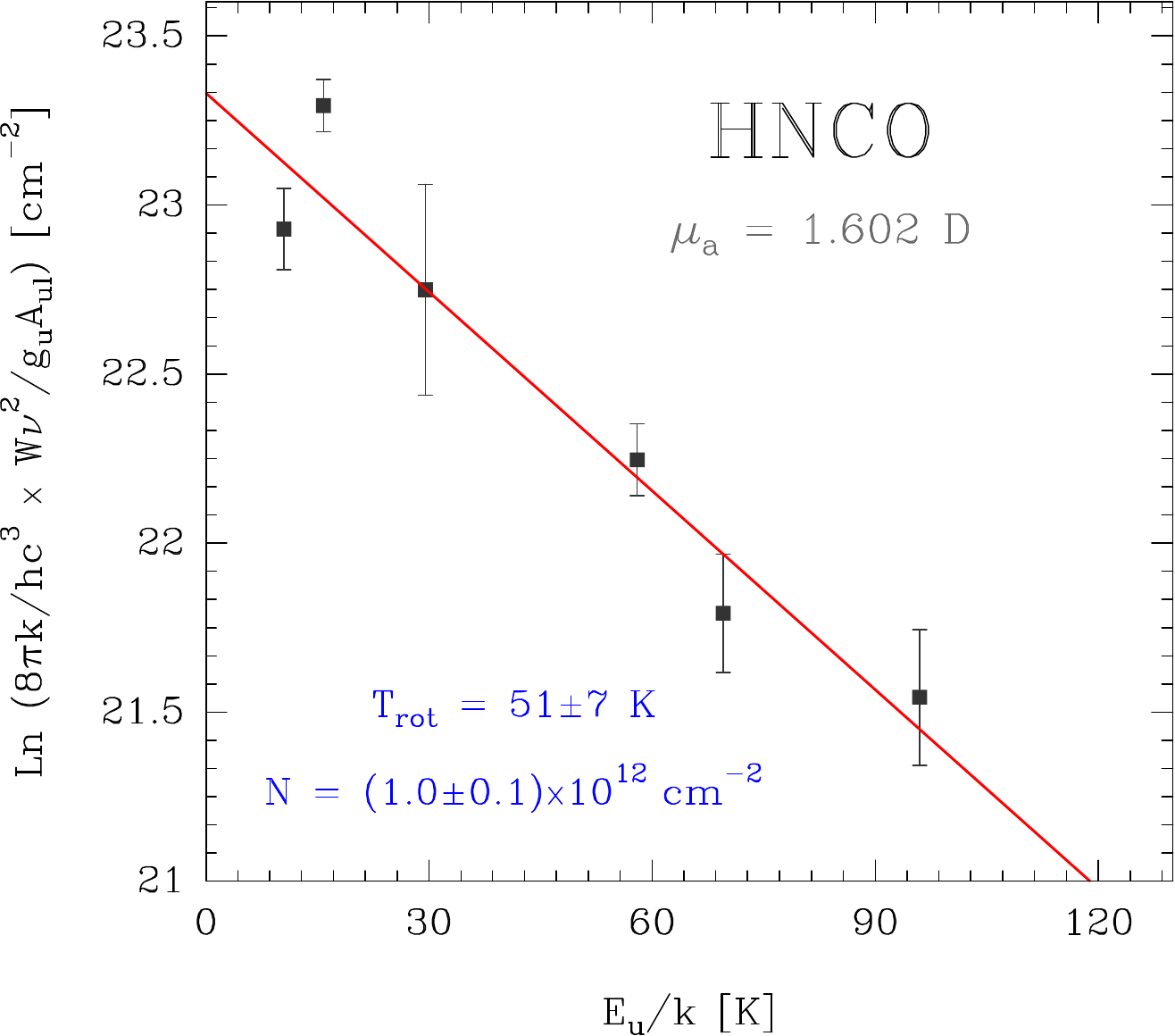}  \hspace{0.5cm}
\includegraphics[scale=0.4, angle=0]{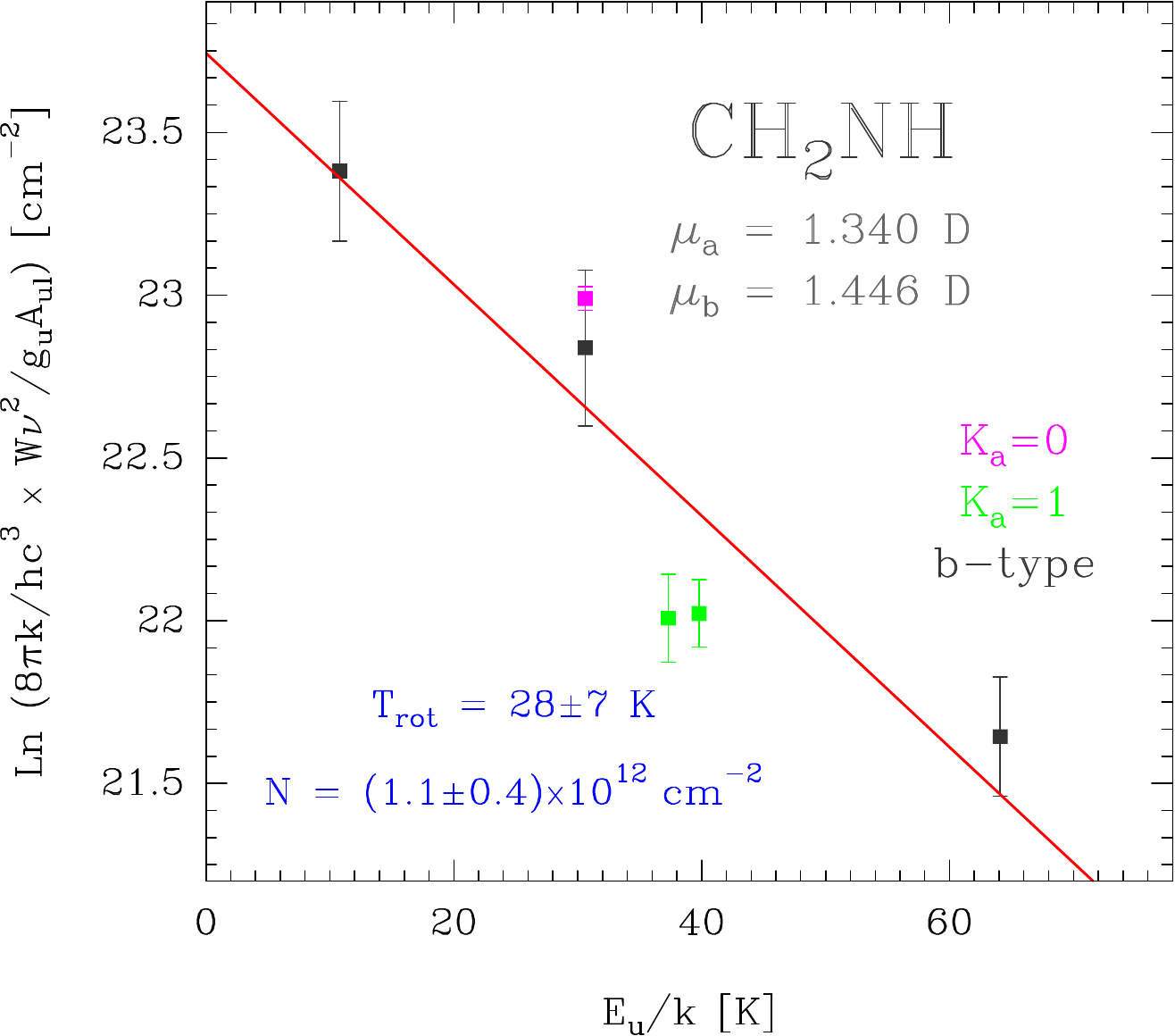} \\
\vspace{0.5cm}
\includegraphics[scale=0.4, angle=0]{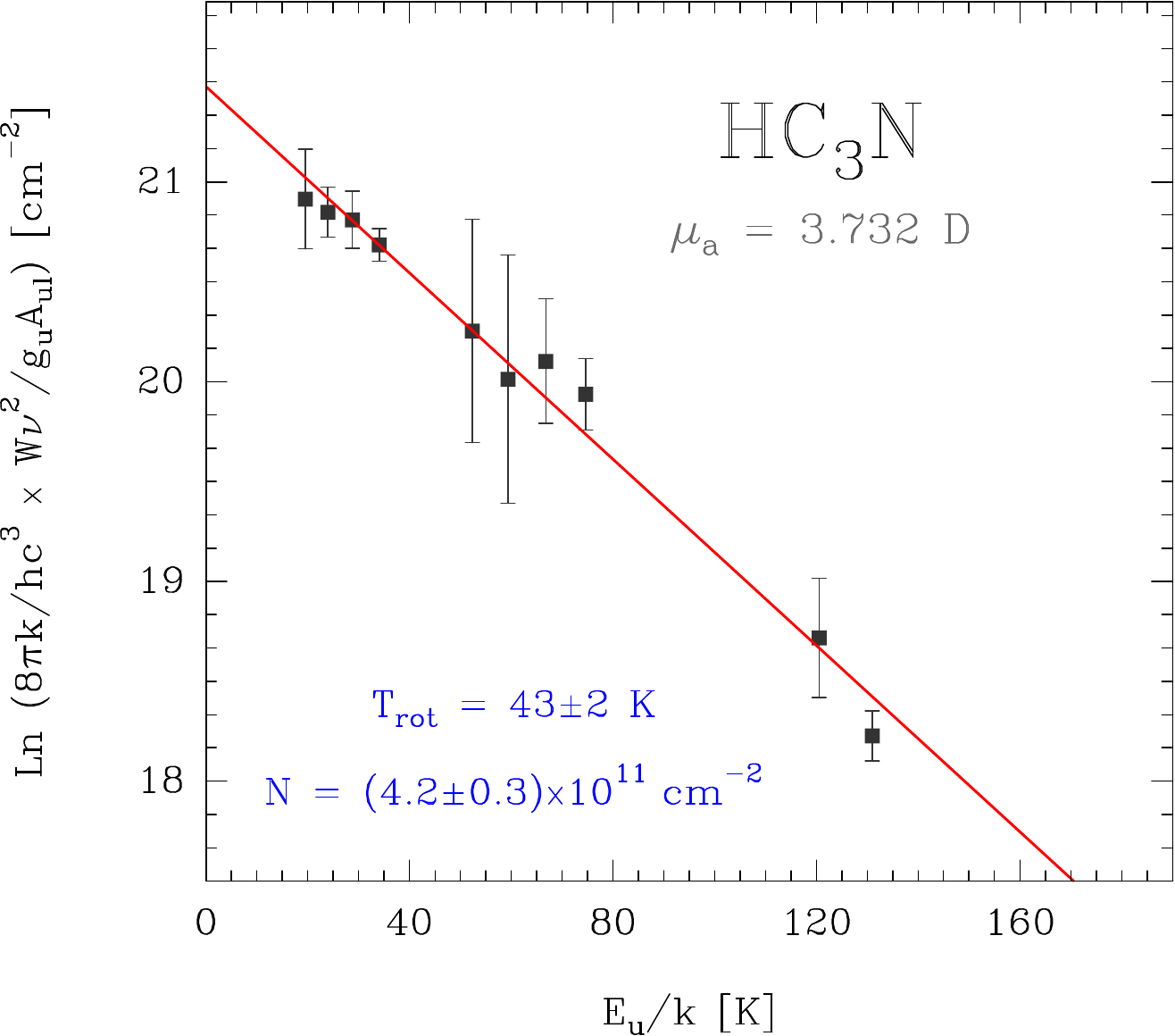} \hspace{0.5cm}
\includegraphics[scale=0.4, angle=0]{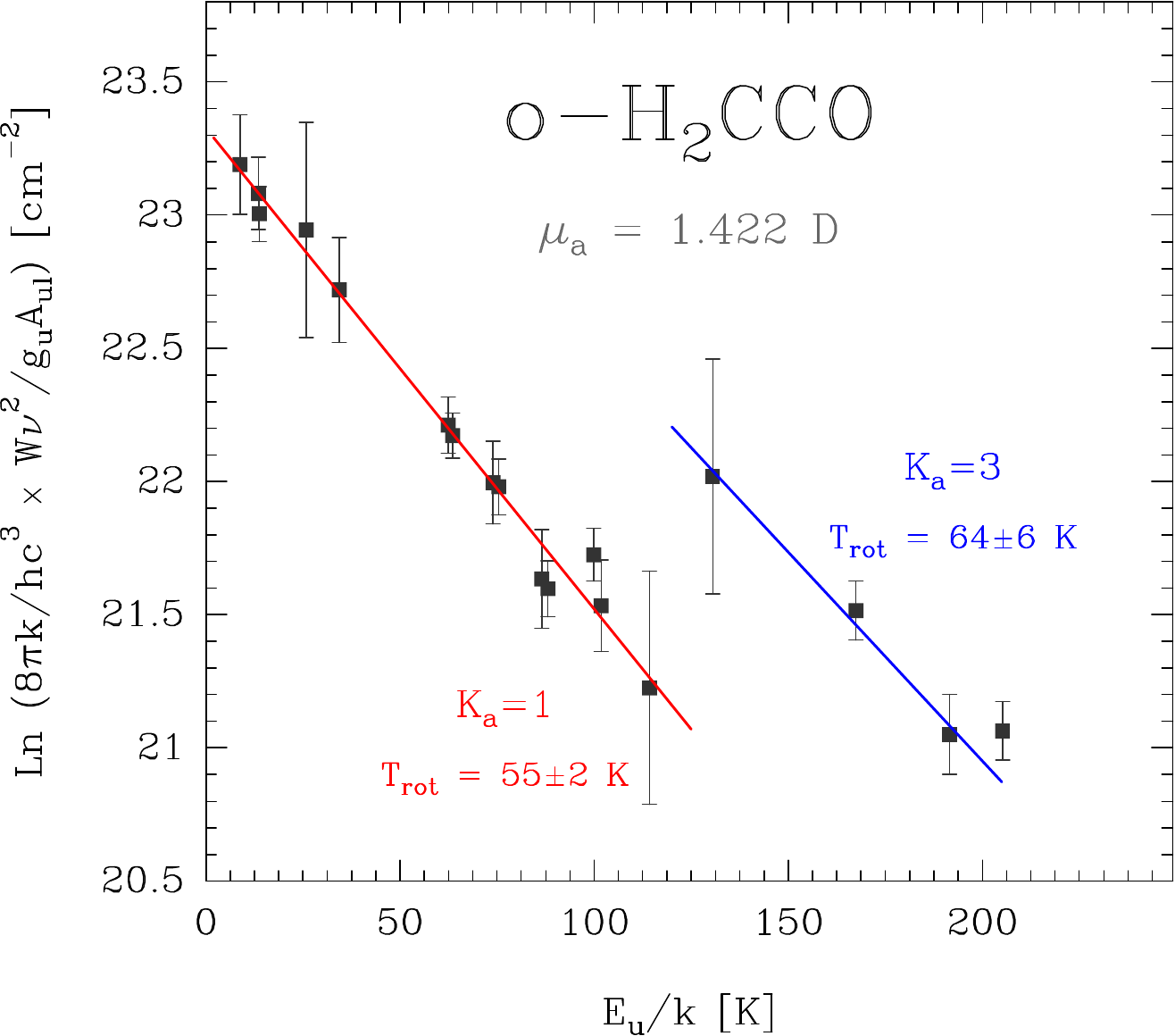}  \hspace{0.5cm}
\includegraphics[scale=0.4, angle=0]{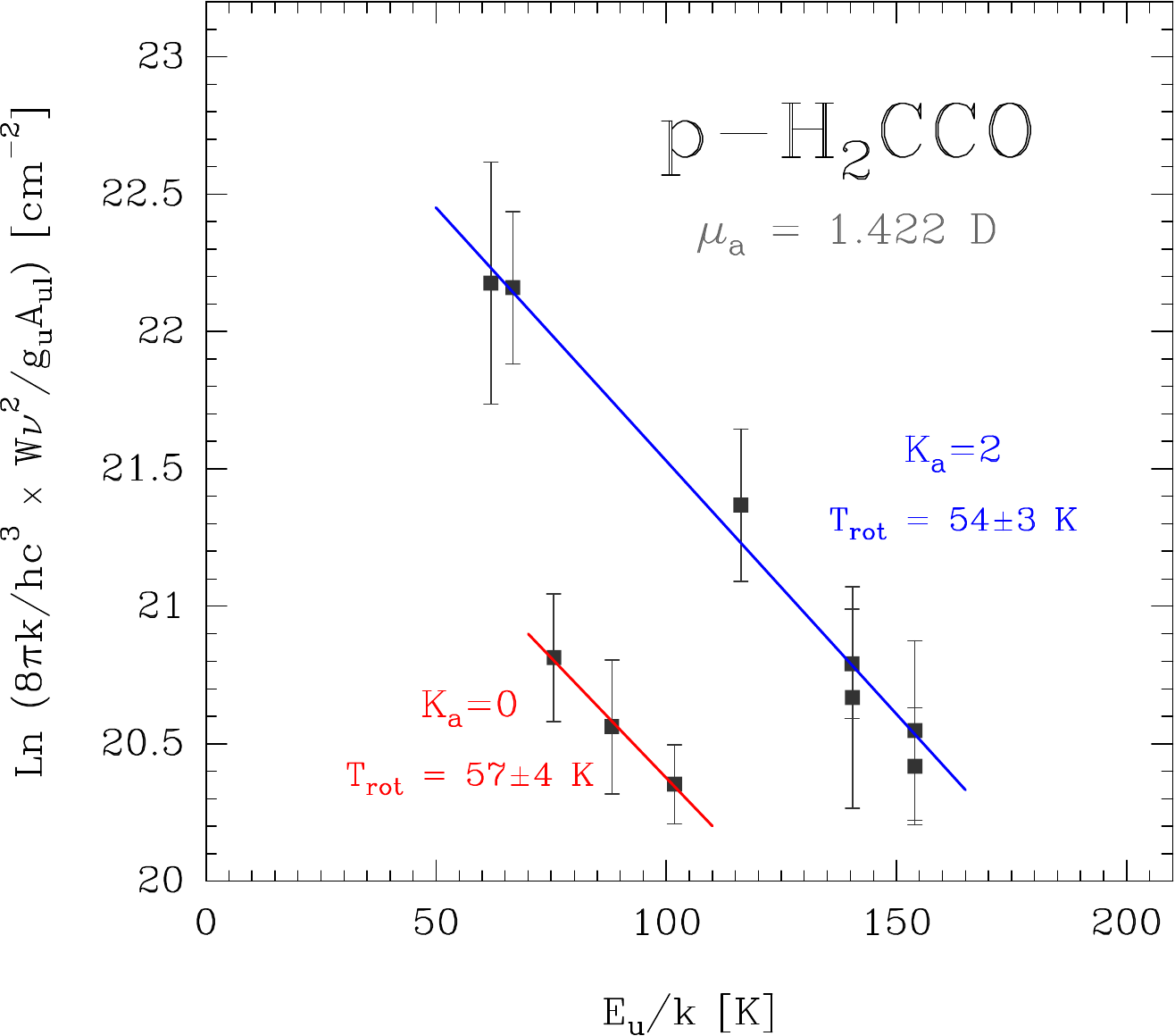} \\
\caption{Rotational diagrams of the detected complex molecules in the Orion Bar PDR (assuming uniform beam filling). Fitted values of the rotational temperature, $T_{\rm rot}$, column density, $N$, and their respective uncertainties are also indicated for each molecule.}\label{fig:DR}
\end{figure*}

\begin{figure*}
\setcounter{figure}{17}
\centering
\includegraphics[scale=0.4, angle=0]{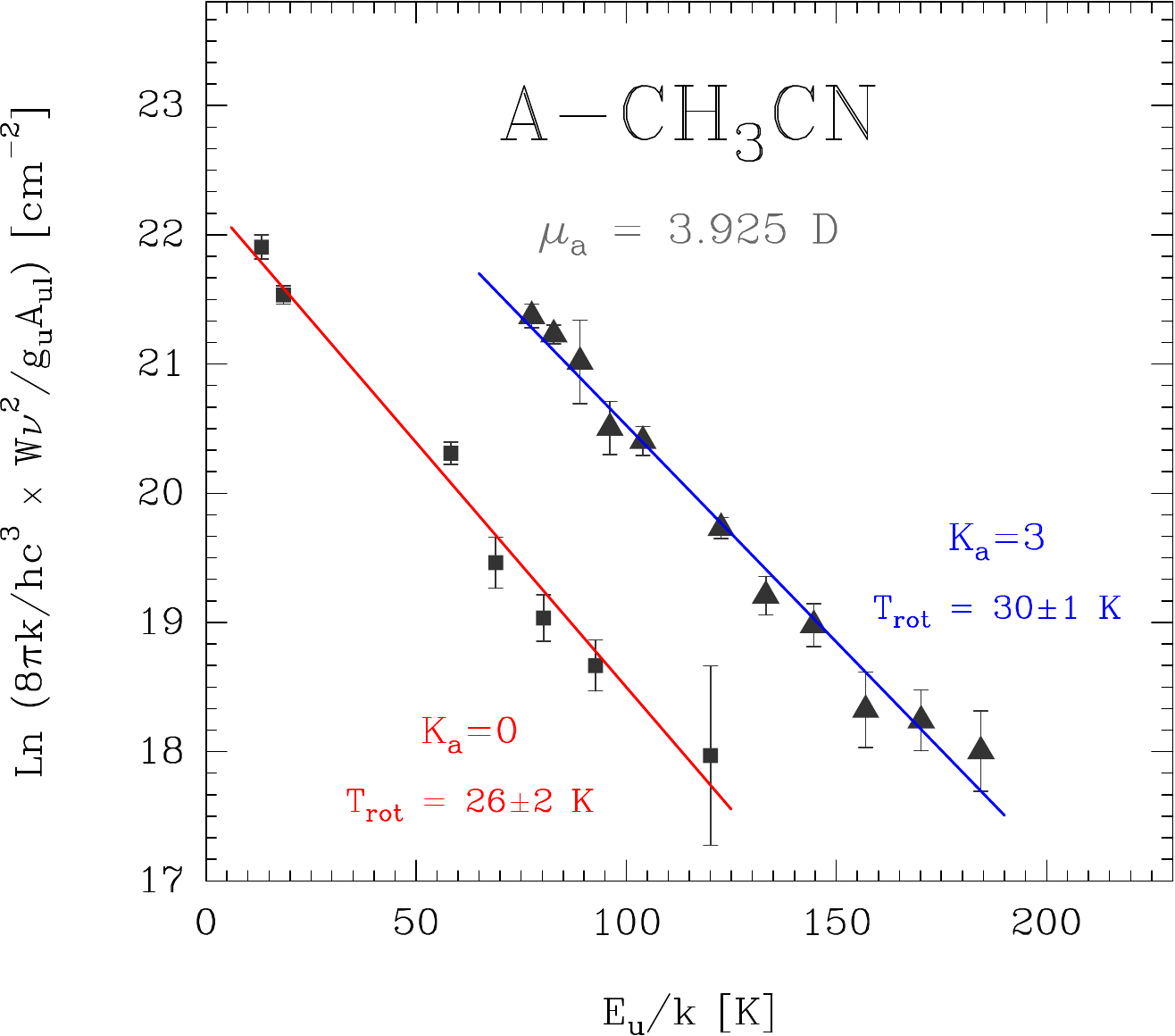}  \hspace{0.5cm}
\includegraphics[scale=0.4, angle=0]{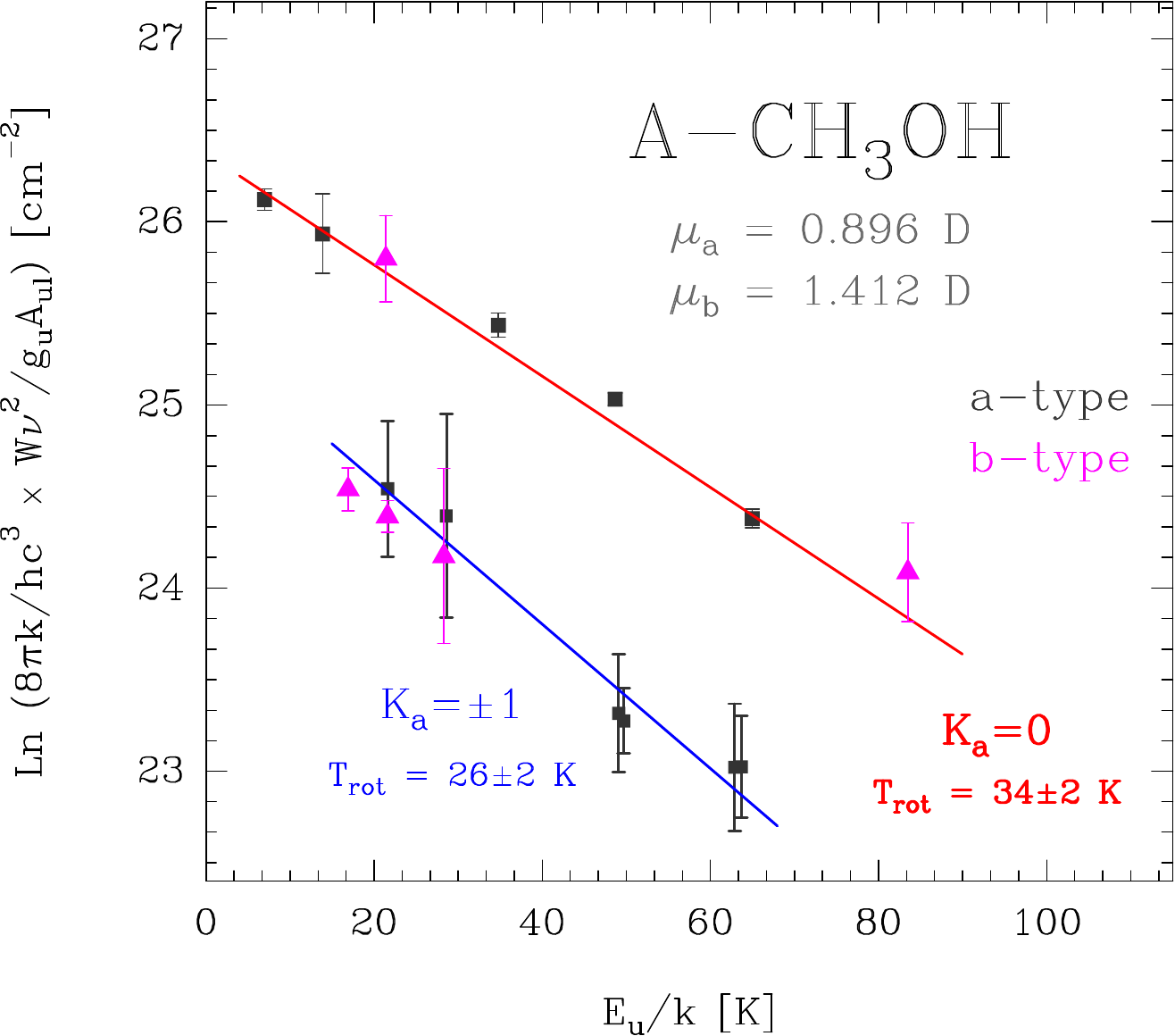} \hspace{0.5cm}
\includegraphics[scale=0.4, angle=0]{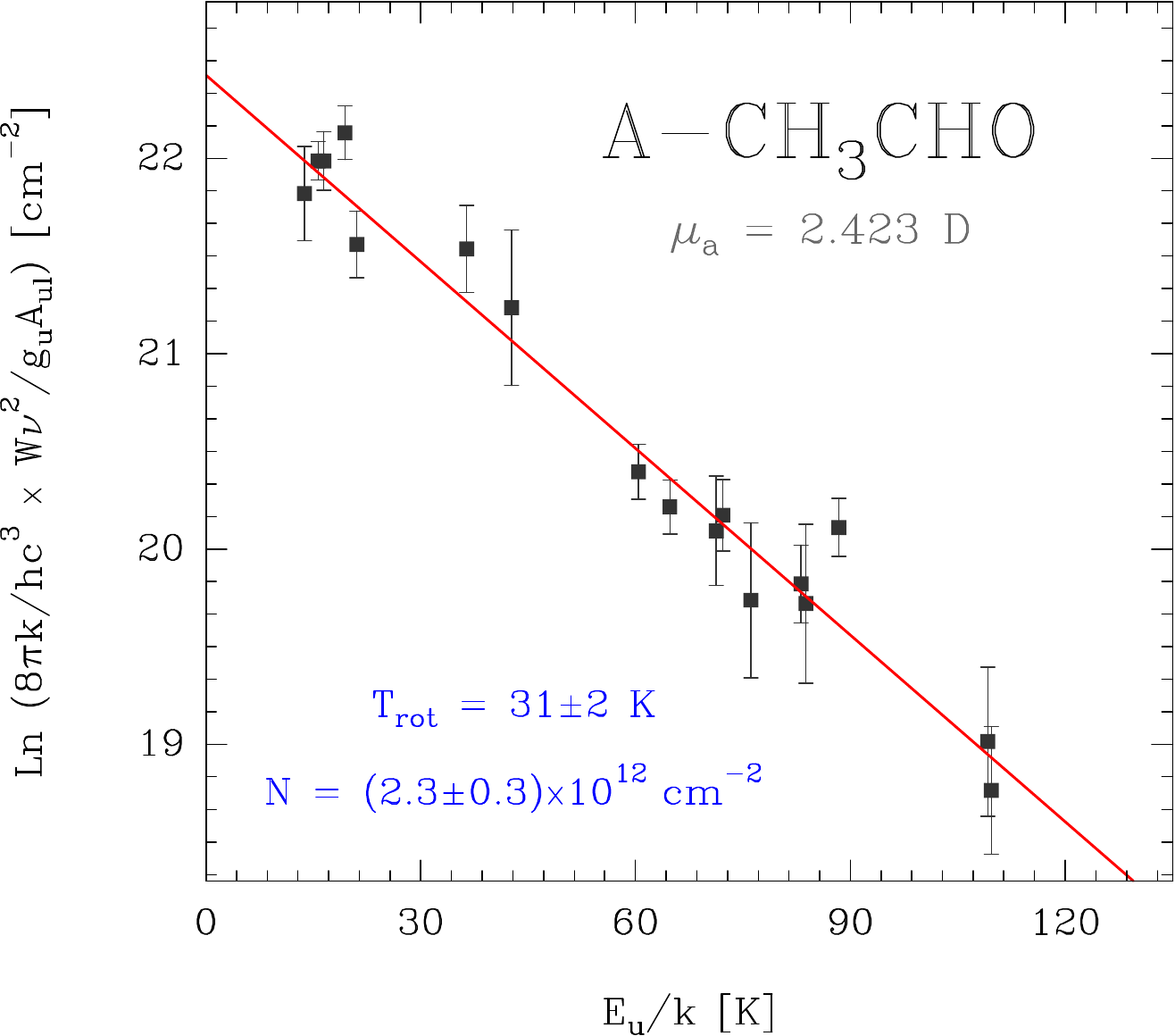}  \\
\vspace{0.5cm}
\includegraphics[scale=0.4, angle=0]{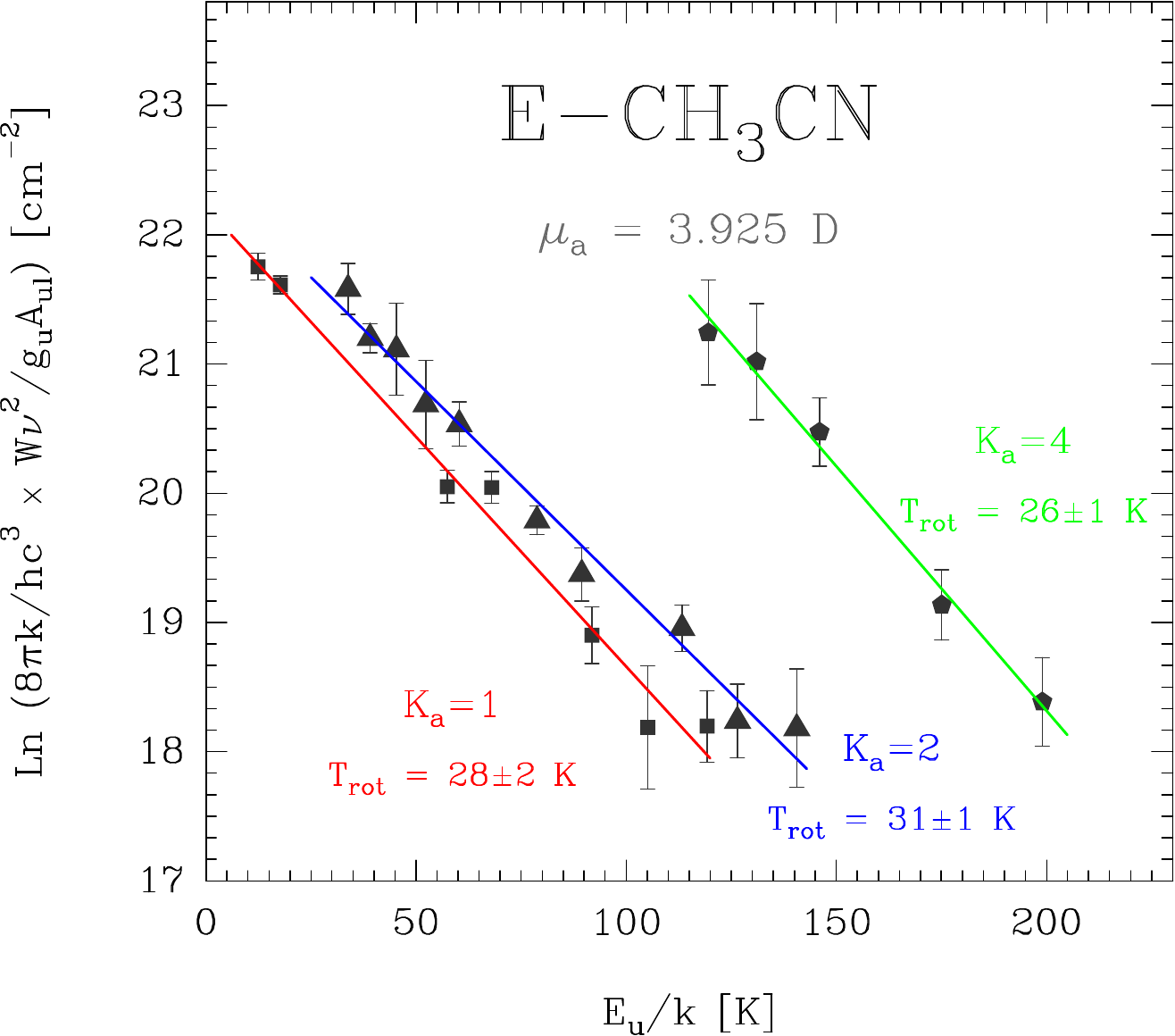} \hspace{0.5cm}
\includegraphics[scale=0.4, angle=0]{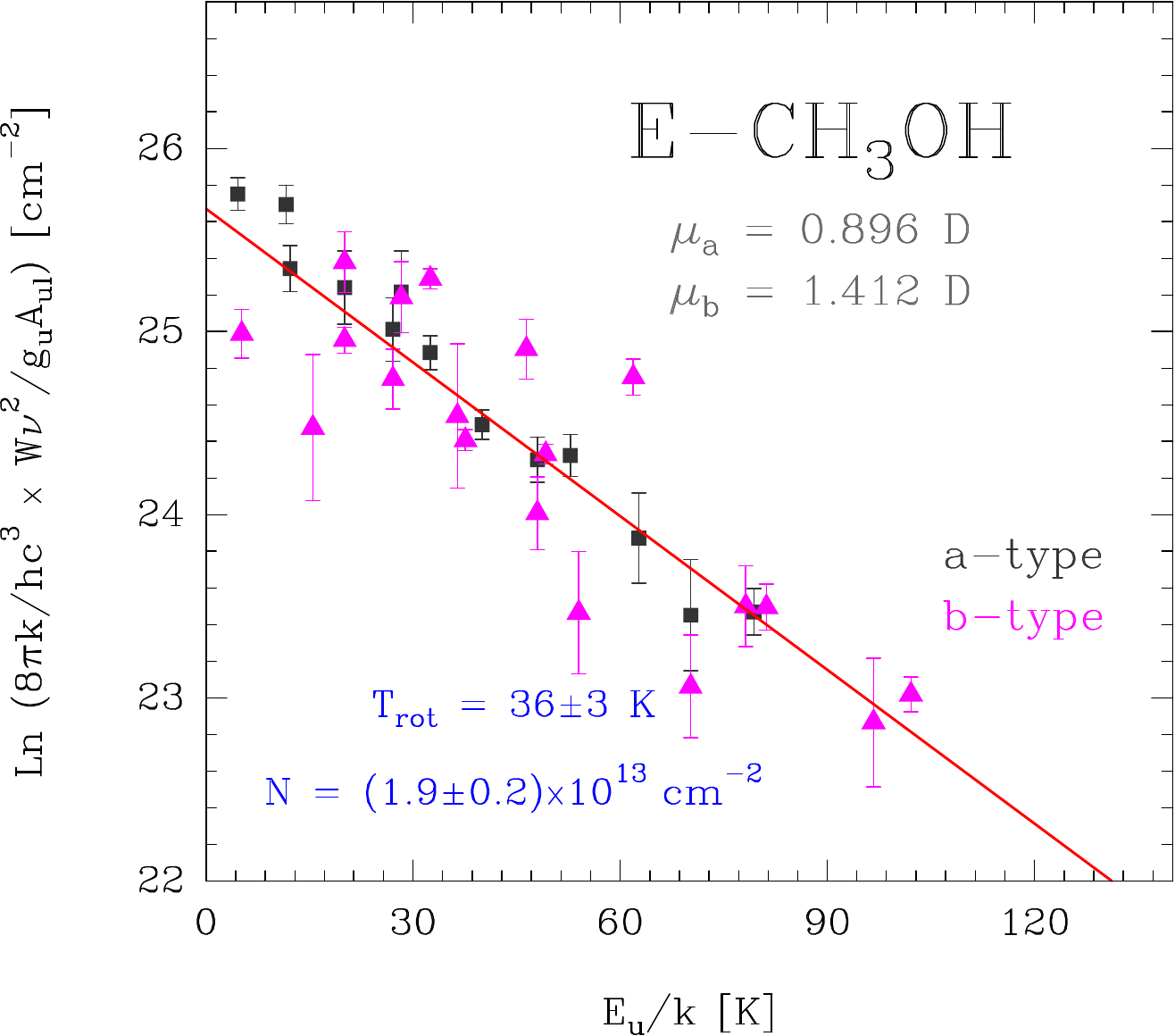}  \hspace{0.5cm}
\includegraphics[scale=0.4, angle=0]{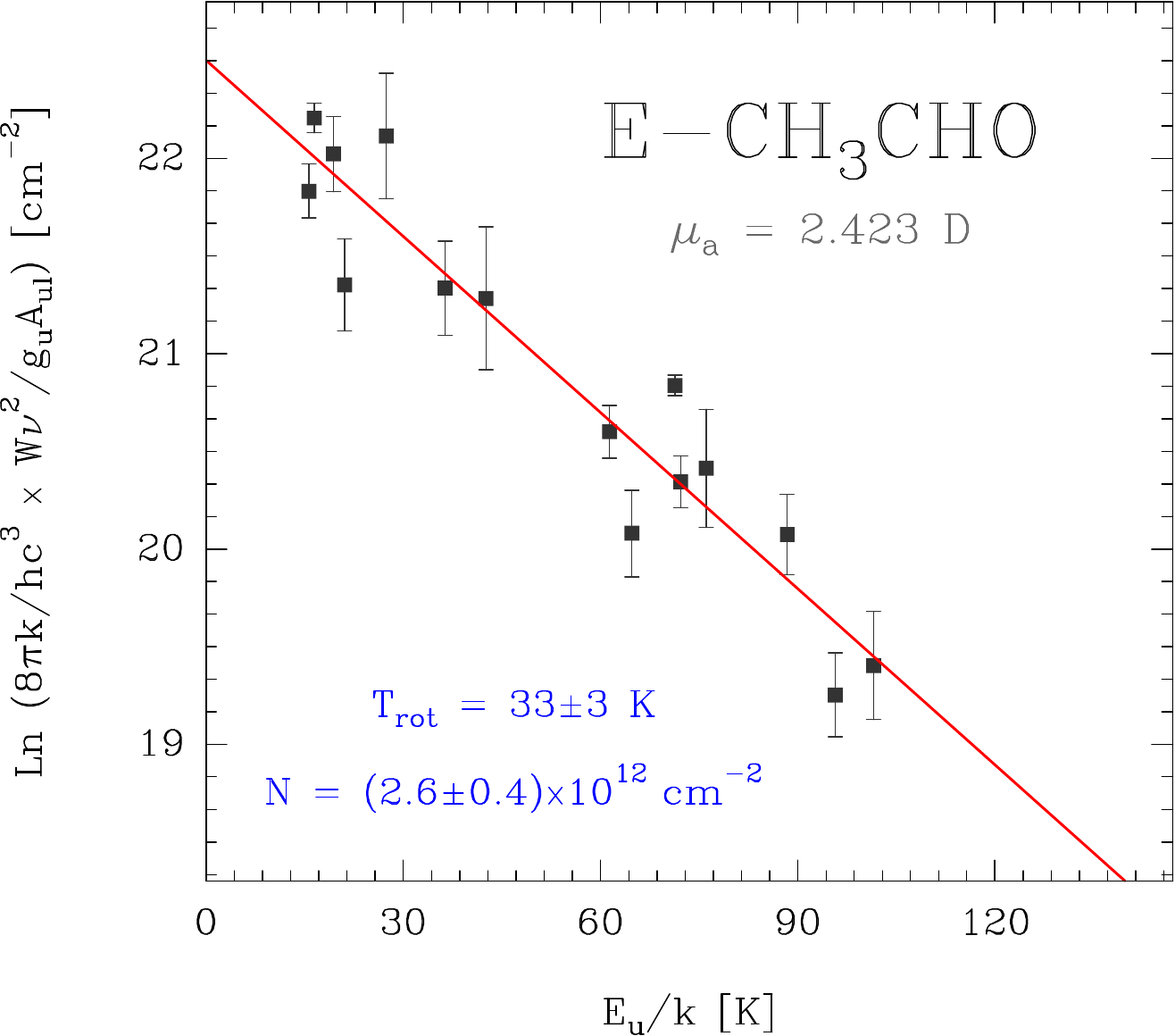} \\
\caption{Continued.}
\end{figure*}

\subsection{Rotational diagrams and column-density determination} \label{DR}

We determined the beam-averaged column density ($N$) and the rotational temperature ($T_{\rm rot}$) for each molecule found by building rotational diagrams \citep[e.g.][]{Goldsmith_1999,Cuadrado_2015a}. The large number of detected lines allows us to accurately determine rotational temperatures and column densities.
 Optically thin emission is assumed (consistent with our best LVG models of H$_{2}$CO, HC$_{3}$N, CH$_{3}$CN, and CH$_{3}$OH). We note that owing to their specific spectroscopy and prevailing excitation conditions at the edge of the Bar (subthermal excitation), a correct determination of the rotational temperatures and column densities for species such as H$_{2}$CO, H$_{2}$CS, H$_{2}$CCO, CH$_{3}$OH, and CH$_{3}$CN requires building specific diagrams for each rotational ladder
(see Sect.~\ref{effects}).
 For the molecules with known collisional rates, we determine \mbox{$N_{\rm LVG}$ $\approx$ $N_{\rm RD}$} (within a factor of~2). This comparison shows that the $N_{\rm RD}$ computed for the other molecules are accurate.

The rotational diagrams were built considering two limiting cases: (i) That the detected emission is extended (i.e. uniform beam filling), with \mbox{$\eta_{_{\rm bf}}$ = 1}; and (ii) that the emission is semi-extended, assuming that \mbox{$\theta_{_{\rm S}}$ = 9$''$} (the typical beam of the IRAM\,30\,m telescope at $\sim$1~mm)$^{6}$. We only considered lines not blended with other transitions and detected above a 3$\sigma$ level. The resulting rotational diagrams are shown in Fig.~\ref{fig:DR}. Rotational temperatures and column densities obtained by linear least squares fits are listed in Table \ref{Table_results}. The uncertainties shown in Table~\ref{Table_results} indicate the uncertainty obtained from the least squares fit to the rotational diagram. The uncertainty obtained in the determination of the line parameters with the Gaussian fitting programme are included in the uncertainty bars at each point of the rotational diagram.

For H$_{2}$CO, H$_{2}^{13}$CO, H$_{2}$CS, H$_{2}$CCO, A-CH$_{3}$OH, A-CH$_{3}$CN, and E-CH$_{3}$CN we have built specific rotational diagrams for different sets of transitions with the same $K_{a}$ quantum number value. The total column density of the molecule is obtained by adding the column density of each rotational ladder. For the other molecules, it is possible to fit the observed transitions by a single $T_{\rm rot}$ and $N$, independently of the $K_{a}$ value. The cis- and trans-HCOOH rotational diagrams are shown in \mbox{\citet{Cuadrado_2016}}.

For molecules with hyperfine structure, like that of HCO, each rotational transition was described without splits, only with a single rotational number $N$. For this purpose, the integrated intensity, $W$, level degeneracy, $g$, and line strength, $S$, of each transition was calculated as the sum of all allowed hyperfine components of each \mbox{$N$+1 $\rightarrow$ $N$} transition. The characteristic frequency, $\nu$, was determined using the weighted average with the relative strength of each line as weight, and the Einstein coefficient, $A$, was calculated using the weighted values in the usual relation.

Comparing the resulting rotational diagrams, we see that: (i) H$_{2}$CO and CH$_{3}$OH are, by far, the most abundant organic molecules (by a factor of $\gtrsim$5), followed by HCO, H$_{2}$CCO, CH$_{3}$CHO, H$_{2}$CS, HCOOH, CH$_{3}$CN, CH$_{2}$NH, HNCO, H$_{2}^{13}$CO, and HC$_{3}$N  (in decreasing order of abundance); (ii) the ortho-to-para (OTP) ratios obtained from the H$_{2}$CO, H$_{2}^{13}$CO, H$_{2}$CS, H$_{2}$CCO column densities are $\sim$3, as expected at high gas temperature; (iii) we obtain similar rotational temperatures for ortho-para and \mbox{A-E} species; (iv) and the inferred \mbox{[H$_{2}$CO]/[H$_{2}^{13}$CO]} ratio is \mbox{63 $\pm$ 3}, similar to the \mbox{$^{12}$C/$^{13}$C $\approx$ 67} isotopic ratio in Orion \citep{Langer_1990}. Hence, this confirms that even lines from the most abundant organic species studied in this work, are optically thin. 
                                        
In order to confirm the correct line identification and possible line blendings, we have modelled the spectrum of each molecule using the MADEX radiative transfer model assuming a Boltzmann distribution and a single excitation temperature for the rotational levels population (we recall that for the observed set of lines/ladders of a given molecule, we infer a single temperature component from the rotational diagrams).
Figures~\ref{fig:HCO_lines}, \ref{fig:H2CS_lines}, \ref{fig:HNCO_lines}, \ref{fig:CH2NH_lines}, \ref{fig:H2CCO_lines}, and \ref{fig:CH3CHO_lines} show the observational spectra (black histograms) and the modelled spectra (red lines) of each molecule. The obtained fits agree very well with the observations and rotational diagrams.

\subsection{Abundances} \label{Abundances}

In order to determine molecular abundances with respect to hydrogen nuclei\footnote{The molecular abundance with respect to hydrogen nuclei is given by ${\frac{N(X)}{N(H)+2N(H_{2})}}$.} towards the line survey position, we derived the beam-averaged H$_{2}$ column density from our observations of the optically thin C$^{18}$O lines \mbox{($J$ = 1 $\rightarrow$ 0}, \mbox{2 $\rightarrow$ 1}, and \mbox{3 $\rightarrow$ 2} transitions) assuming \mbox{$^{16}$O/$^{18}$O $\approx$ 500} \citep{Wilson_1994} and a
\mbox{[CO]/[H$_{2}$]} abundance of $\sim$10$^{-4}$ (lower than the canonical value due to photodissociation). The resulting H$_2$ column density, \mbox{$N$(H$_{2}$) $\simeq$ 3 $\times$ 10$^{22}$ cm$^{-2}$}, is in good agreement with previous estimations of $N$(H$_2$) close to the dissociation front \citep[see e.g.][]{Hogerheijde_1995}.
A hydrogen atom column density, $N$(H), of \mbox{$\sim$3 $\times$ 10$^{21}$ cm$^{-2}$} has been inferred from \HI\, observations towards the edge of the Orion Bar \citep{vanderWerf_2013}. Molecular abundances with respect to hydrogen nuclei are listed in Table \ref{Table_results}. The abundances of the detected species range \mbox{from 10$^{-9}$ to 10$^{-11}$.}

\begin{table}
\centering 
\caption{Upper limits for undetected COMs.}
\label{Table_undetected_lines}     
\begin{tabular}{l c c@{\vrule height 10pt depth 5pt width 0pt}}     
\hline\hline      
Molecule  &                   $N$(X) [cm$^{-2}$]       &  Abundance \tablefootmark{a} \\
\hline

HDCO\tablefootmark{\,b}                & (1.3 - 1.5) $\times$ 10$^{11}$    &  (2.1 - 2.4) $\times$ 10$^{-12}$  \\
CH$_{3}$O           & (2.0 - 4.0) $\times$ 10$^{12}$    &  (3.2 - 6.3) $\times$ 10$^{-11}$  \\
CH$_{3}$NC          & (5.0 - 8.0) $\times$ 10$^{10}$    &  (0.8 - 1.3) $\times$ 10$^{-12}$  \\ 
CH$_{3}$CCH         & (5.0 - 8.0) $\times$ 10$^{12}$    &  (0.8 - 1.3) $\times$ 10$^{-10}$  \\ 
CH$_{3}$OCH$_{3}$   & (7.0 - 9.0) $\times$ 10$^{12}$    &  (1.1 - 1.4) $\times$ 10$^{-10}$  \\
HCOOCH$_{3}$        & (9.0 - 9.5) $\times$ 10$^{12}$    &  (1.4 - 1.5) $\times$ 10$^{-10}$  \\
CH$_{3}$CH$_{2}$OH  & (1.0 - 1.2) $\times$ 10$^{12}$    &  (1.6 - 1.9) $\times$ 10$^{-11}$  \\
CH$_{2}$CHCN        & (9.0 - 9.5) $\times$ 10$^{12}$    &  (1.4 - 1.5) $\times$ 10$^{-10}$  \\
CH$_{3}$CH$_{2}$CN  & (7.0 - 5.0) $\times$ 10$^{11}$    &  (1.1 - 0.8) $\times$ 10$^{-11}$  \\
 \hline      
\end{tabular}
\tablefoot{
\tablefoottext{a}{The abundance of each species with respect to H nuclei (see Sect.~\ref{Abundances}).
}
\tablefoottext{b}{Detected by \citet{Parise_2009} towards clump $\#$3 \citep{Lis_2003} of the Orion Bar.}
}
\end{table}

\begin{table*}
\centering 
\caption{Abundances relative to H$_2$ in different environments (in units of 10$^{-10}$).}
\label{Table_sources}     
\begin{tabular}{l c c c c c c c c c@{\vrule height 9pt depth 5pt width 0pt}} 
\hline\hline      
Molecule         &  Orion Bar$^{1*}$ &  \multicolumn{2}{c}{Horsehead}                    & & \multicolumn{2}{c}{Orion KL}  &  L1157-B1$^7$                                & IRAS 16293-2422$^8$ &  B1-b$^9$           \\  \cline{3-4} \cline{6-7}
                 &      PDR          &    PDR$^{2*}$           &    Core$^{3*}$   & & HC$^4$        & CR$^5$        &  Outflow                                     & Hot corino          &  Dense core    \\

\hline
HCO              &    1.7            &     8.4$^{\rm b}$       &  $<$ 0.8$^{\rm b}$      & &    ---             &   $<$ 0.3$^{\rm 6,g}$  &    ---               &  $<$ 2.0           &   0.2$^{\rm n}$  \\
HNCO             &    0.2            &         ---             &   ---                   & &  780$^{\rm h}$     &  ---                   &   605$^{\rm k}$      &  1.7               &   0.7$^{\rm p}$     \\ 
H$_{2}$CO        &    9.0            &     2.9$^{\rm c}$       &  2.0$^{\rm c}$          & & 1200$^{\rm h}$     &  440$^{\rm h}$         &   4000$^{\rm l}$     &  7.0               &   4.6$^{\rm q}$           \\ 
H$_{2}$CS        &    0.8            &         ---             &   ---                   & &  150$^{\rm h}$     &  74$^{\rm h}$          &   1100$^{\rm l}$     &  ---               &   0.9$^{\rm q}$           \\ 
t-HCOOH$\dagger$      &    0.2$^{\rm a}$  &     0.5$^{\rm d}$       &  0.1$^{\rm d}$          & &  800$^{\rm i}$     &  ---                   &    ---               &  $<$ 3.0           &   0.1$^{\rm n}$     \\ 
CH$_{2}$NH       &    0.2            &       ---               &   ---                   & &  42$^{\rm h}$      &  ---                   &    ---               &  $<$ 5.0           &    ---          \\ 
H$_{2}$CCO       &    0.9            &     1.5$^{\rm d}$       &  0.5$^{\rm d}$          & &   ---              &  51$^{\rm h}$          &    ---               &  1.8               &   0.2$^{\rm n}$    \\ 
HC$_{3}$N        &    0.07           &     0.06$^{\rm e}$      &  0.08$^{\rm e}$         & &  81$^{\rm h}$      &  ---                   &  100$^{\rm l}$       &  0.3               &   3.1$^{\rm r}$    \\ 
CH$_{3}$OH       &    5.0            &     1.2$^{\rm f}$       &  2.3$^{\rm f}$          & &  22000$^{\rm h}$   &  12000$^{\rm h}$       &  115000$^{\rm l}$    &  44                &   31$^{\rm r,s}$     \\ 
CH$_{3}$CN       &    0.2            &     2.5$^{\rm e}$       &  0.08$^{\rm e}$         & &  300$^{\rm h}$     &  120$^{\rm h}$         &     ---              &  1.5               &   0.4$^{\rm r}$      \\ 
CH$_{3}$CHO      &    0.8            &     0.7$^{\rm d}$       &  0.2$^{\rm d}$          & &  9.5$^{\rm i}$     &   ---                  &  250$^{\rm m}$       &  1.0               &   0.1$^{\rm n}$      \\ 
CH$_{3}$CCH      &  $<$ 1.0          &     4.4$^{\rm d}$       &  3.0$^{\rm d}$          & &  24$^{\rm i}$      &  133$^{\rm 6,i}$       &    ---               &  6.5               &   5.0$^{\rm r}$     \\
\hline                                                                                                                                     
                                                                                     
\end{tabular}
\tablefoot{
\\
 $*$ Abundances with respect to total hydrogen nuclei. $\dagger$ To date, the cis conformer of HCOOH has only been detected towards the Orion Bar PDR \citep{Cuadrado_2016}, therefore, we only provide the abundances of the trans conformer.   \\
{\bf (1)} This work. Abundances calculated assuming uniform beam filling. Ref.: (a) \citet{Cuadrado_2016}.\\
{\bf (2)} Low-UV field Horsehead PDR ($\chi\approx$~60; \mbox{$N_{\rm H}$ = 3.8 $\times$ 10$^{22}$~cm$^{-2}$}) and {\bf (3)} condensation shielded from the UV field (dense core, \mbox{$N_{\rm H}$ = 6.4 $\times$ 10$^{22}$~cm$^{-2}$}) behind the Horsehead PDR edge. 
\mbox{Ref.: (b) \citet{Gerin_2009},} \mbox{(c) \citet{Guzman_2011},} \mbox{(d) \citet{Guzman_2014},} \mbox{(e) \citet{Gratier_2013},} \mbox{(f) \citet{Guzman_2013}.}  \\
{\bf (4)} Orion KL Hot Core (HC); and
{\bf (5)} Orion KL Compact Ridge (CR).  
{\bf (6)} Abundances calculated for the Extended Ridge.
 \mbox{Ref.: (g) \citet{Blake_1987},} (h) \citet{Crockett_2014}, (i) B. Tercero, private communication.\\
{\bf (7)} Peak B1 in the blue lobe of the L1157. \mbox{Ref.: (k) \citet{Mendoza_2014}}, (l) \citet{Bachiller_1997}, (m) \citet{Codella_2015}. \\
{\bf (8)} Low-mass protostar (hot corino) IRAS 16293-2422. The abundances are computed for a \mbox{$N$(H$_2$) = 2 $\times$ 10$^{23}$~cm$^{-2}$}. Abundances averaged over a $\sim$20$''$ beam \citep{vanDishoeck_1995}.   \\
{\bf (9)} Quiescent dark core \mbox{Barnard 1-b} (B1-b), \mbox{$N$(H$_2$) = 1.3 $\times$ 10$^{23}$~cm$^{-2}$} \citep[see e.g.][]{Hirano_1999,Lis_2002}. \mbox{Ref.: (n) \citet{Cernicharo_2012b}}, (p) \citet{Marcelino_2009}, (q) \citet{Marcelino_2005}, (r) N. Marcelino, private communication, (s) see also \citet{Oberg_2010}. \\
}
\end{table*}

\subsection{Undetected complex organic molecules and precursors} \label{Undetected_COMs}

The broadband frequency coverage of the survey allowed us to obtain upper limits for other chemically interesting
organic molecules that have not been detected towards the edge of the PDR, but could have been expected. In particular, we searched for 
HDCO, CH$_{3}$O, CH$_{3}$NC, CH$_{3}$CCH, CH$_{3}$OCH$_{3}$, 
	  	HCOOCH$_{3}$, CH$_{3}$CH$_{2}$OH, CH$_{2}$CHCN, and CH$_{3}$CH$_{2}$CN,
because they have been detected in other PDRs and star-forming regions (e.g. Horsehead, Orion KL, Barnard 1-b, or IRAS 16293-2422; see Table~\ref{Table_sources} for references). The CH$_{3}$O radical is thought to be an important intermediary for the H$_2$CO and CH$_3$OH chemistry but so far, has only been detected in cold and dense gas \citep{Cernicharo_2012b,Bacmann_2016}.
HDCO has been detected by \citet{Parise_2009} towards the lukewarm, dense and more FUV-shielded clump $\#$3 of \citet{Lis_2003}, but it is not detected towards the warmer and high FUV-illuminated line survey position, near the dissociation front.

 First, we estimated 3$\sigma$ line intensities using the relation 
\begin{equation}
\mathrm{\displaystyle{\int} T_{_ {MB}} dv=3\sigma \sqrt{2 \, \delta v \, \Delta v}  \, \, \, \, \, \, \, \, \, [K \, km \, s^{-1}],}
\end{equation}
\noindent where $\sigma$ is the rms of the observations per resolution channel [K], $\delta v$ is the velocity spectral resolution 
\mbox{[km s$^{-1}$]}, and $\Delta v$ is the assumed line widths \mbox{($\sim$2 km s$^{-1}$)}. Second, we used MADEX to create synthetic models that simulate the line emission to constrain their column densities. The column densities and 3$\sigma$   upper limit abundances assuming \mbox{$T_{\rm rot}$ = 20 $-$ 30 K} are listed in Table~\ref{Table_undetected_lines}. We also provide the following abundance ratio upper limits:
\mbox{[HDCO]/[H$_{2}$CO] $<$ 0.003} 
and \mbox{[CH$_{3}$NC]/[CH$_{3}$CN] $<$ 0.1}. 
  \citet{Parise_2009} estimated an 
 abundance ratio of \mbox{[HDCO]/[H$_{2}$CO] $>$ 0.006} towards clump $\#$3, which suggests that the deuteration diminishes from the lukewarm and shielded clumps to the warmer and more FUV-irradiated cloud edge.

\section{Discussion} \label{Discussion}

In order to shed more light on the possible chemical formation routes of COMs in strongly FUV-irradiated gas from an observational perspective, we compare the observed abundances of several COMs and precursors in different environments (see Table~\ref{Table_sources}). In addition to the Orion Bar, we consider
the Horsehead PDR (a low-FUV-flux PDR) and a nearby cold core also in the Horsehead nebula, the Orion~KL hot core and compact ridge (warm dense gas at roughly the same distance to the Bar), the L1157 outflow (shocked gas), the low-mass protostar  IRAS 16293-2422 (hot corino), and the quiescent dark cloud \mbox{Barnard 1-b}  (B1-b).

\subsection{COMs in different environments} \label{Environments}

\begin{figure}
\centering
\includegraphics[scale=0.38,angle=0]{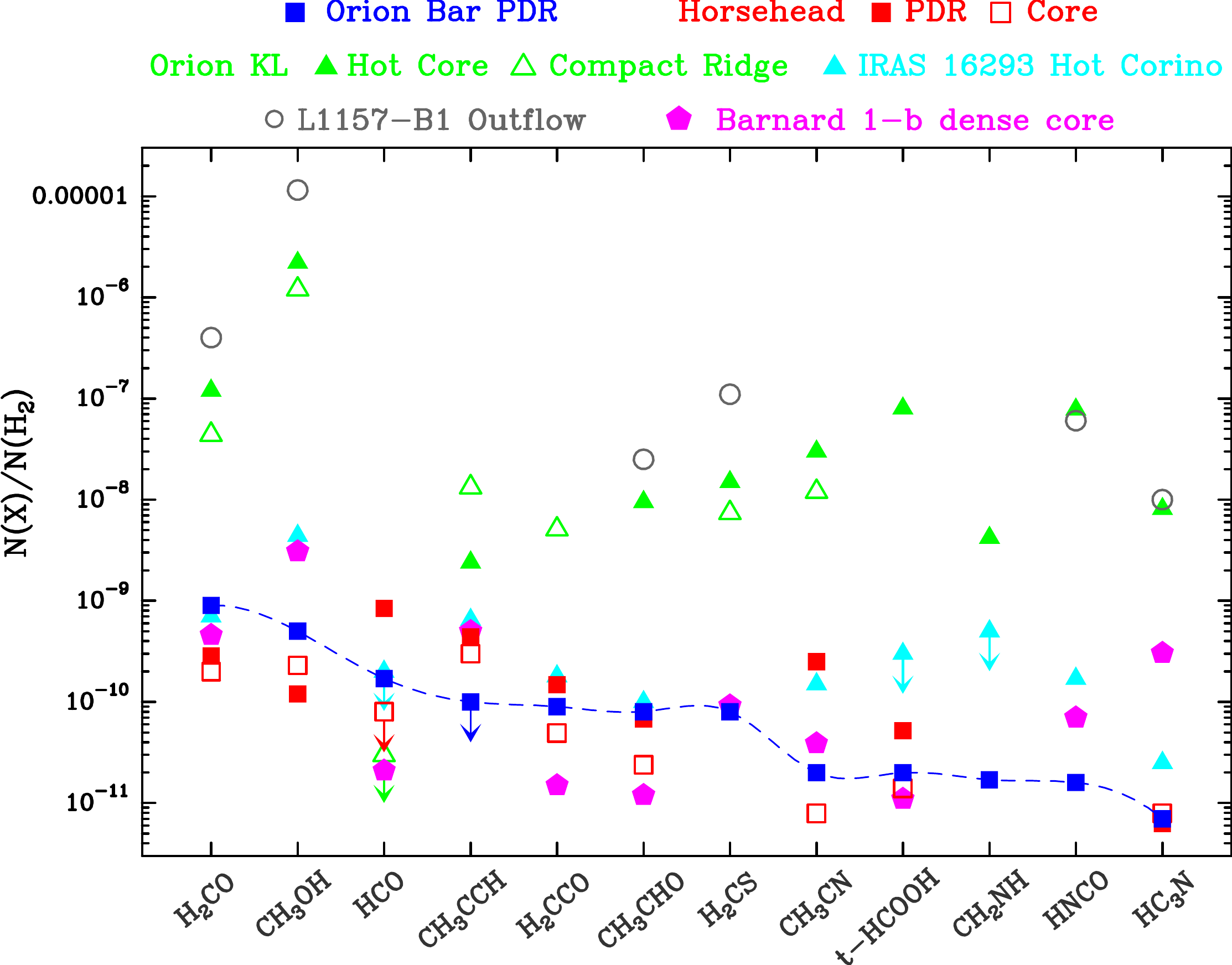} \vspace{0.2cm}
\caption{Molecular abundances with respect to H$_2$ in several sources. In the Orion Bar PDR, blue points represent the molecular abundance assuming uniform beam filling.}
\label{fig:abundances}
\end{figure}

Figure~\ref{fig:abundances} shows a comparison between the abundances of 12 organic molecules detected in the above sources.  As expected for widespread interstellar molecules, abundant \mbox{H$_{2}$CO} and \mbox{CH$_{3}$OH} are found in all sources \mbox{($N$(X)/$N$(H$_2$) $>$ 10$^{-10}$)}. 
In diffuse and translucent clouds (\mbox{$n$(H$_{2}$) $\simeq$ 10$^2$ $-$ 10$^3$~cm$^{-3}$}), only H$_{2}$CO has been detected so far \citep[][]{Liszt_2006}, suggesting that the formation of COMs is more efficient in dense gas.

In general, the abundances of COMs, and CH$_{3}$OH in particular, are much higher towards Orion KL (hot core and compact ridge) and towards the L1157-B1 outflow. This translates into low \mbox{[X$_{\rm COM}$]/[CH$_{3}$OH] $\ll$ 1} abundance ratios in those environments. 
The Horsehead PDR, however, shows \mbox{[X$_{\rm COM}$]/[CH$_{3}$OH] $>$ 1} ratios for \mbox{X$_{\rm COM}$ = HCO,} H$_{2}$CO, H$_{2}$CCO, CH$_{3}$CN, and CH$_{3}$CCH abundances.  
The Orion KL hot core and compact ridge show enhanced COM abundances. Owing to the very high dust temperatures ($T_{\rm d}$ $>$ 100\,K), ice sublimation and subsequent warm gas-phase chemistry dominates \citep[e.g.][]{Blake_1987}. On the other hand, the abundances inferred in the hot corino IRAS 16293-2422 (lower dust temperatures) are more similar to those  in the Orion Bar PDR and Horsehead. 
Interestingly, the L1157 outflow shows the highest abundances of CH$_3$OH, H$_{2}$CO, H$_{2}$CS, HC$_{3}$N, and CH$_3$CHO. This suggests that the combined effects of ice mantle sublimation, grain sputtering, and hot gas-phase chemistry in shocks results in a very efficient COM formation.

\begin{figure}
\centering
\includegraphics[scale=0.38,angle=0]{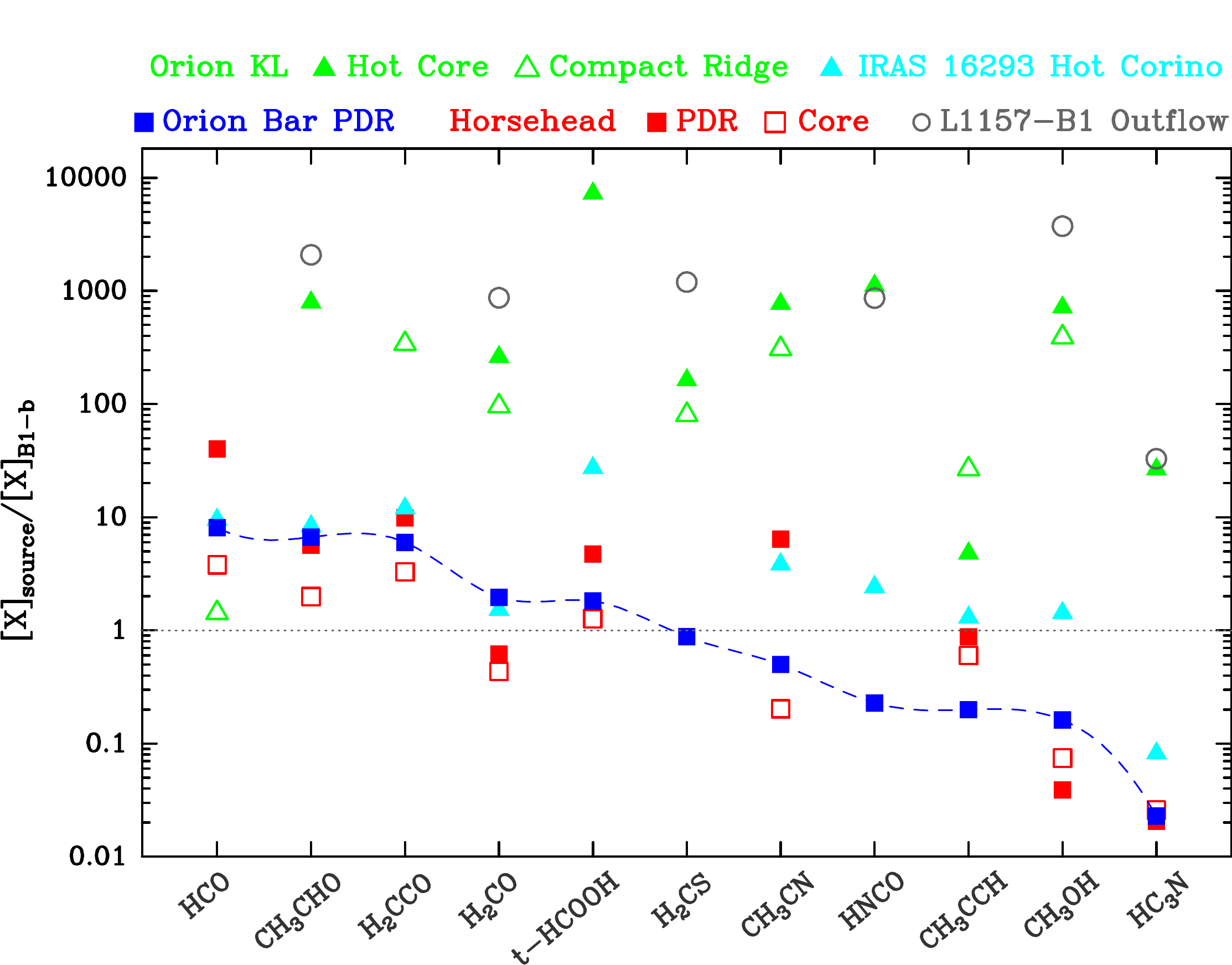}\vspace{0.2cm}
\caption{Comparison of the abundance ratios with respect to B1-b in several sources.}
\label{fig:abund_B1}
\end{figure}

Among the studied species, only the HCO radical and \mbox{cis-HCOOH} \citep{Cuadrado_2016} are more abundant in PDRs. This enhancement is thus a characteristic feature of FUV-illuminated gas.
Figure~\ref{fig:abund_B1} shows normalised abundances with respect to the abundances in B1-b (cold and gas shielded from strong FUV radiation). HCO, CH$_3$CHO, H$_2$CCO, H$_2$CO, and \mbox{t-HCOOH} are a factor of \mbox{$\sim$2 $-$ 10} more abundant in the PDRs than in the cold core.

\citet{Guzman_2014} observed two positions of the Horsehead, the moderately warm PDR ($T_{\rm k}$ $\simeq$ 60\,K) and a cold core behind the PDR and shielded from stellar FUV field. Interestingly, they found enhanced COM abundances in the PDR compared to the  core. Given the low-FUV field in the Horsehead, dust grains are relatively cold   even at the PDR edge \citep[\mbox{$T_{\rm d}$ $\lesssim$ 30~K},][]{Goicoechea_2009b}. These  dust temperatures are significantly below the sublimation temperatures of abundant interstellar ices \citep{Gibb_2000} such as H$_2$O ($\sim$100\,K), CH$_3$OH ($\sim$100\,K), and even H$_2$CO ($\sim$40\,K). Hence, grains must be coated with mantles, even in the Horsehead PDR, and photodesorption can be efficient, either desorbing specific COMs or their gas-phase 
precursors \citep{Guzman_2014}.  Indeed, PDR models adapted to the Horsehead and including
grain surface reactions and ice photodesorption were invoked to explain the observed  gas-phase CH$_3$OH and H$_2$CO abundances \citep{Guzman_2013}.
In addition, the low H$_2$CO ortho-to-para ratio of two in the PDR  \citep{Guzman_2011}
might support a cold surface origin for H$_2$CO.
 However, those Horsehead models required relatively high photodesorption yields (several 10$^{-4}$ to 10$^{-3}$ molecules per photon) that are apparently not supported by recent laboratory experiments of methanol photodesorption. In fact, little CH$_3$OH is actually seen to desorb in CH$_3$OH ice irradiation experiments \citep{Bertin_2016,Cruz-Diaz_2016}.

Also in the Horsehead, CH$_{3}$CN  is 30 times more abundant in the PDR  than in the shielded core \citep{Gratier_2013}. CH$_{3}$CN  is also an order of magnitude more abundant in the Horsehead than in the Orion Bar PDR. 
  Given the  high binding energy of CH$_{3}$CN ice (similar to CH$_3$OH, \citealt{Collings_2004}) this result is quite surprising (methanol is more abundant
than  CH$_{3}$CN in the Bar). 
In addition, \citet{Gratier_2013} detected CH$_{3}$NC towards the Horsehead PDR, with an isomeric ratio of \mbox{[CH$_{3}$NC]/[CH$_{3}$CN] = 0.15}, higher than our upper limit. 
These authors suggested that photodesorption triggers the CH$_{3}$CN abundance in the
Horsehead. 
 Therefore, photodesorption seems to enhance the gas-phase CH$_{3}$CN and  CH$_{3}$NC abundances, but not that of CH$_3$OH. This suggests that  photodesorption is a very selective process and thus, should be studied molecule by molecule in representative ISM ice analogs. In addition, gas phase reactions may allow the enhancement of CH$_{3}$CN with respect to CH$_3$OH.
  
 In the Orion Bar, a much more strongly FUV-irradiated PDR, 
 gas and dust  grains are warmer  \citep[from \mbox{$T_{\rm d}$ $\simeq$ 40 to 80 K};][]{Arab_2012}. Thus, ice mantle abundances must be small close to the cloud edge. Interestingly, we find higher H$_2$CO and
 CH$_3$OH  abundances in the Bar than in the Horsehead PDR (but not HCO). 
  In addition,  H$_2$CO is detected close to the irradiated edge of the Orion Bar
 and we determine a H$_2$CO ortho-to-para ratio of approximately
three (consistent with the high
 temperature limit).
Finally, we determine a \mbox{HCO/H$_2$CO/CH$_3$OH $\simeq$ 1/5/3} abundance ratio,
 whereas a \mbox{$\sim$7/2/1} ratio is found in the Horsehead PDR
 \citep{Gerin_2009, Guzman_2013}.
 The ratios typically inferred in cold dense gas are \mbox{$\sim$1/10/10} \citep{Bacmann_2016}. Although HCO is approximately
five times more abundant in the Horsehead, both PDRs show \mbox{[CH$_3$OH]/[H$_2$CO] $<$ 1} abundance ratios. We note that  H$_2$CO is also a product of photodissociation of CH$_{3}$OH. In warm but FUV-shielded environments (hot cores and shocks), CH$_3$OH is much more abundant than in PDRs and cold cores.

\subsection{Limits to steady-state PDR gas-phase chemistry} \label{Chemistry_models}

\begin{figure}
\centering
\includegraphics[scale=0.70,angle=0]{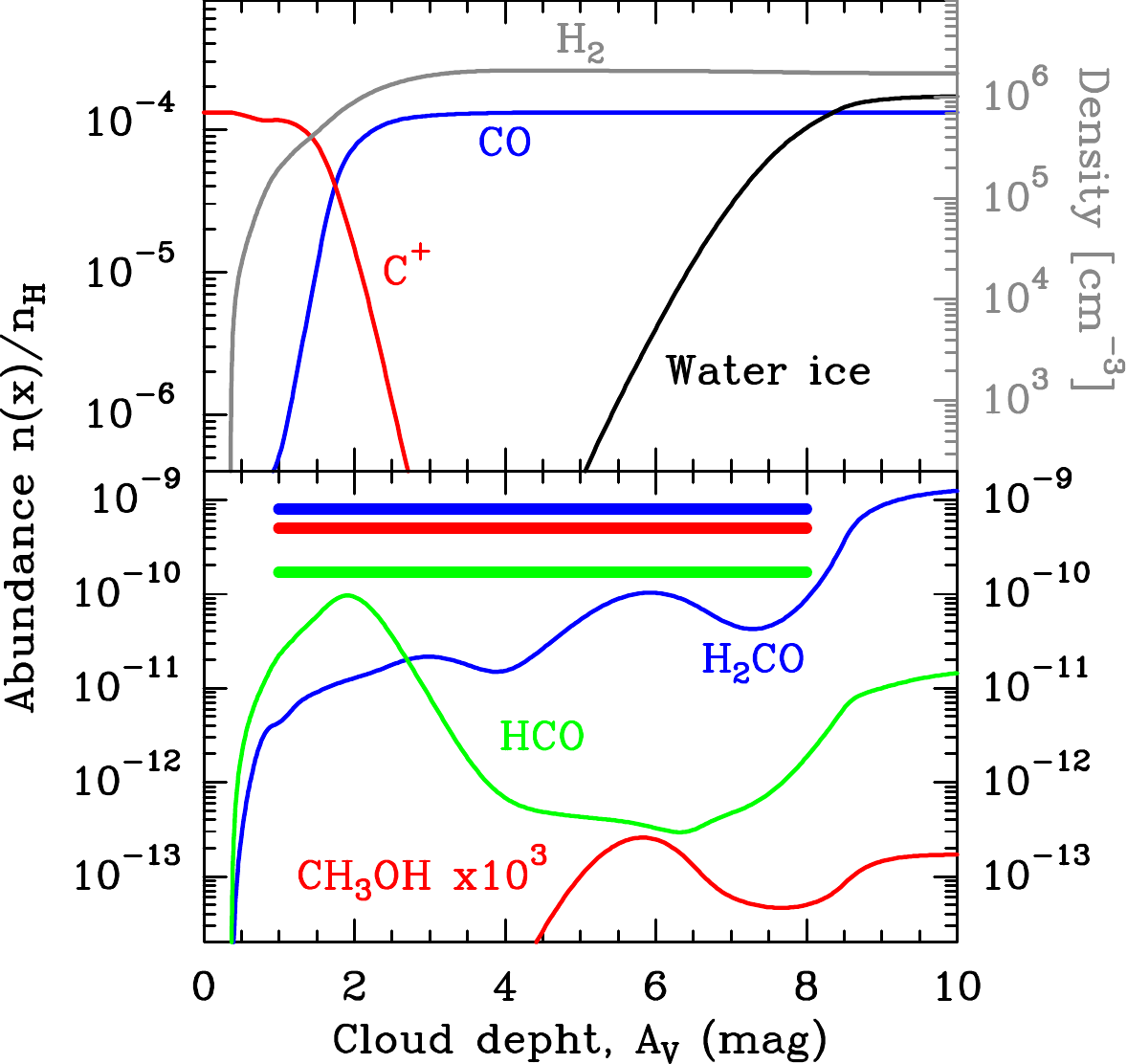} \vspace{0.3cm}
\caption{Isobaric PDR model for the Orion Bar with \mbox{$P_{\rm th}/k$ = 10$^8$\,K\,cm$^{-3}$} and \mbox{$\chi$ = 2 $\times$ 10$^4$}. The thick horizontal bars
show the inferred abundances of HCO, H$_2$CO, and CH$_3$OH towards the PDR edge.
In this model, these molecules are formed only through gas-phase reactions.}
\label{fig:pdr_mod}
\end{figure}

In order to explore the gas-phase production of HCO, H$_2$CO, and CH$_3$OH,
we show PDR model results adapted to the illumination conditions in the Bar using the Meudon PDR code \citep[e.g.][]{LePetit_2006,Goicoechea_2007}. This is a model of a high gas-pressure isobaric PDR with \mbox{$\chi$ = 2 $\times$ 10$^4$} and \mbox{$P_{\rm th}/k$ = 10$^{8}$~K\,cm$^{-3}$}. Only H$_2$O, CO, and atoms are allowed to deplete from the gas following their adsorption energies \citep[e.g.][]{Hollenbach_2009}.  However, we do not include surface chemistry (e.g. water ice only forms through water vapour freeze-out and desorbs thermally or by photodesorption; we refer to \citealt{Cuadrado_2015a} for details). Figure~\ref{fig:pdr_mod} shows the abundance profiles as
a function of cloud depth. Although the dust temperature is never low enough to
allow CO ice-mantle formation, water ice starts to be very abundant at \mbox{$A_{\rm V}$ $\gtrsim$ 5} (a similar behaviour can be expected for CH$_3$OH ice).  
We note that the typical gas temperature inferred from our
observations towards the line survey position \mbox{($T_{\rm k}$ $\approx$ 150\,K)} corresponds to \mbox{$A_{\rm V}$ $\approx$ 1.5 $-$ 2.0~mag} and \mbox{$T_{\rm d}$ $\approx$ 60\,K} in this model. Hence, we expect mostly bare grains at the edge of the PDR, only coated with a few monolayers of very polar ices such as water (with high adsorption energies).

The model shown in Fig.~\ref{fig:pdr_mod} roughly reproduces the HCO abundance enhancement at the PDR edge (driven by reaction \mbox{O + CH$_2$ $\rightarrow$ HCO + H}), but underpredicts the observed H$_2$CO and CH$_3$OH abundances. 
In these models, the H$_2$CO formation at \mbox{$A_{\rm V}$ $\approx$ 1.5 $-$ 2.0~mag} is largely dominated by reaction \mbox{O + CH$_3$ $\rightarrow$ H$_2$CO + H}.
This comparison suggests either the gas-phase model misses important formation routes (or rates) in hot molecular gas, or that (nearly bare) grain surface formation and subsequent desorption is important. \citet{Esplugues_2016} have shown PDR models including surface chemistry
on grains with different  ice content. In their models, chemical/reactive desorption \citep[the surface reaction exothermicity is used to break the adsorbate-surface bond, e.g.][]{Herbst_2015,Minissale_2016},
can dominate over photodesorption for some species (e.g. methanol).
Nevertheless, the H$_2$CO and CH$_3$OH abundances predicted by  \citet{Esplugues_2016} in a PDR with high FUV fluxes are still lower than our observed values.
Hence, at present, no model seems to reproduce the inferred abundances of molecules such as H$_2$CO and CH$_3$OH towards the Bar.

All these models, however, simulate a static PDR in which steady-state has been reached. 
A real PDR is likely to be more dynamic \citep{Goicoechea_2016}. Time-dependent flows of molecular gas and icy grains might advect from inside the cold molecular cloud to the warm PDR edge. There they can be reprocessed by the combination of high temperatures
and the presence of a strong FUV photon flux for some time before photodissociation.  Thus, time-dependent desorption 
and advection of COMs (or gas-phase precursors) from the molecular cloud interior to the PDR may contribute to enhance their abundances.

The non-detection of HDCO at the illuminated edge of the Bar \mbox{([HDCO]/[H$_2$CO] $<$ 3 $\times$ 10$^{-3}$)}, and also in the Horsehead PDR \citep{Guzman_2011}, shows that pure gas-phase H$_2$CO deuteration is not efficient at high gas temperatures. 
However, HDCO has been detected towards the lukewarm \mbox{($T_{\rm k}$ $\lesssim$ 70 K)}, dense and more FUV-shielded clump $\#$3 by \citet{Parise_2009}.
This agrees with specific chemical models in which
pure gas-phase deuteration is efficient in gas below $T_{\rm k}$ $\lesssim$ 70\,K \citep{Roueff_2007}. Time-dependent gas-phase models show that the [HDCO]/[H$_2$CO] abundance
ratio is particularly low at early cloud times \citep{Trevino-Morales_2014}.
In hot cores where high abundances of COMs have been detected, the gas deuteration
is very high as well. Both effects are related to the sublimation of ice
mantles formed and deuterated in a previous cold cloud stage.
In the context of a dynamic PDR, deuterated species could desorb from grain surfaces and advect to the PDR edge as well.
The very low [HDCO]/[H$_2$CO] upper limit abundance towards the edge of the Bar suggests that this mechanism is
not efficient for deuterated molecules, and that the gas temperature is too high to enhance the deuterium fractionation by pure gas-phase \mbox{reactions alone.}

\subsection{CH$_{3}$CCH non detection} \label{CH3CCH}

CH$_{3}$CCH is a widespread hydrocarbon present in the Horsehead PDR \mbox{($\sim$4.4 $\times$ 10$^{-10}$)} and cold core \mbox{($\sim$3 $\times$ 10$^{-10}$)}, 
the extended ridge of Orion~KL \mbox{($\sim$1 $\times$ 10$^{-8}$)},
the hot corino IRAS 16293-2422 \mbox{($\sim$7 $\times$ 10$^{-10}$)},
the Monoceros R2 ultra-compact \HII\, region \mbox{($\sim$2 $\times$ 10$^{-9}$;} \citealt{Ginard_2012}),
and 
even the nucleus of the starburst galaxy M82 \mbox{($\sim$1 $\times$ 10$^{-8}$;} \citealt{Fuente_2005,Aladro_2011}).
Unexpectedly,  CH$_{3}$CCH lines are not bright towards the Bar edge.
A few 2$\sigma$ line features seen in the 1\,mm range coincide with the expected frequencies of CH$_{3}$CCH. However, the detection cannot be confirmed with confidence
at the sensitivity level of our line survey.
Based on the tentative 2$\sigma$ features, we
estimate an upper limit abundance of  \mbox{[CH$_{3}$CCH] $<$ (0.8 $-$ 1.3) $\times$ 10$^{-10}$}.  
The unattenuated CH$_{3}$CCH  photodissociation rate is high (several 10$^{-9}$\,s$^{-1}$
for \mbox{$\chi$ = 1)} and CH$_{3}$CCH reacts with C$^+$ ions relatively fast (e.g. \citealt{Wakelam_2012}).  Both the FUV photon flux and the C$^+$ column density are particularly high towards the observed position at the edge of the Bar \citep{Ossenkopf_2013,Goicoechea_2015}. This combination likely explains the reduced CH$_{3}$CCH abundance. Indeed, in our PDR models,  CH$_{3}$CCH only reaches detectable abundances at $A_{\rm V}$ $>$ 8 \citep{Cuadrado_2015a}.

\section{Summary and conclusions} \label{Summary}

We have investigated the presence and abundance of complex organic molecules in the strongly FUV-irradiated edge of the Orion Bar PDR, near the H$_2$ dissociation front. We used the \mbox{IRAM 30 m} telescope to carry out a millimetre line survey, complemented with 8.5$''$ resolution maps of the molecular line emission at 0.9\,mm.
Despite being a very harsh environment, our observations show a relatively rich  spectrum, with more than 250 lines arising from COMs and related organic precursors with up to seven atoms: H$_{2}$CO, CH$_{3}$OH, HCO, H$_{2}$CCO, CH$_{3}$CHO, H$_{2}$CS, HCOOH, CH$_{3}$CN, CH$_{2}$NH, HNCO, H$_{2}^{13}$CO, and HC$_{3}$N  (in decreasing order of abundance). 
In particular, we obtained the following results:

\begin{itemize}

 \item The abundance of the detected species range from 10$^{-9}$ to 10$^{-11}$; H$_{2}$CO being the most abundant. The inferred rotational temperatures range from $\sim$10 to \mbox{$\sim$55 K}, significantly lower than the gas temperature and thus consistent with subthermal excitation.  H$_{2}^{13}$CO and H$_{2}$CCO have the lowest and the highest 
 rotational temperatures, respectively. We obtain similar rotational temperatures for the \mbox{ortho-para} and \mbox{A-E} species, and ortho-to-para ratios of approximately three.

  \item For some molecules, we constrain the beam-averaged physical conditions from LVG excitation models. As shown by previous works \citep[e.g.][]{Leurini_2006, Leurini_2010}, we conclude that not all COMs arise from the same PDR layer. In particular, while CH$_{3}$OH only arises from dense gas in the more shielded PDR interior (\mbox{$T_{\rm k}$ = 40 $-$ 50 K}), CH$_{3}$CN and H$_{2}$CO also trace warmer gas layers (\mbox{$T_{\rm k}$ = 150 $-$ 250 K}), more exposed to a strong FUV-field.

\item We determine a \mbox{HCO/H$_2$CO/CH$_3$OH $\simeq$ 1/5/3} abundance ratio.
Such relative abundances are not inferred in environments dominated by ice-mantle thermal
desorption or grain sputtering. Overall, pure gas-phase models have difficulty reproducing the observed H$_2$CO and especially CH$_{3}$OH abundances in PDRs.
Taking into account the  elevated gas and dust temperatures at the edge of the Bar (\mbox{$T_{\rm d}$ $\simeq$ 60 K}), we suggest the following scenarios for the formation of COMs: (i) hot gas reactions not included in current models; (ii) COMs are produced in the warm grain surfaces of nearly bare grains; or (iii) the PDR dynamics is such that COMs or specific precursors formed in cold icy grains deeper inside the molecular cloud desorb and advect into the PDR.

\end{itemize}

The presence of COMs in the interstellar medium is more widespread that initially expected. It includes very harsh environments such as shocked gas and now strongly FUV-irradiated gas.
COM formation reflects the complicated interplay between gas and grain surface chemistry in different environments. However, the specific formation pathways are not fully clear
and may even not be the same in different environments.
More laboratory experiments to study different grain surface processes and to investigate the products and rates of different desorption mechanisms (photodesorption, sublimation, chemical desorption, etc.) are needed to distinguish between the different possible scenarios. The census of increasingly complex organic molecule detections will obviously increase in the years to come.


\begin{acknowledgements} 
We thank our referee for pointing us towards several missing references in an earlier version of the manuscript.
We thank Nuria Marcelino for sharing the column densities of different COMs inferred towards \mbox{Barnard 1-b}.
We are very grateful to the IRAM staff for their help during the observations.
This work has been partially funded by MINECO grants (CSD2009-00038 and AYA2012-32032).
We thank the ERC for support under grant ERC-2013-Syg-610256-NANOCOSMOS.

\end{acknowledgements}


\bibliographystyle{aa}
\bibliography{references}

\begin{thebibliography}{159}
\expandafter\ifx\csname natexlab\endcsname\relax\def\natexlab#1{#1}\fi

\bibitem[{{Ag{\'u}ndez} {et~al.}(2008){Ag{\'u}ndez}, {Cernicharo}, \&
  {Goicoechea}}]{Agundez_2008a}
{Ag{\'u}ndez}, M., {Cernicharo}, J., \& {Goicoechea}, J.~R. 2008, \aap, 483,
  831

\bibitem[{{Aladro} {et~al.}(2011){Aladro}, {Mart{\'{\i}}n},
  {Mart{\'{\i}}n-Pintado}, {Mauersberger}, {Henkel}, {Oca{\~n}a Flaquer}, \&
  {Amo-Baladr{\'o}n}}]{Aladro_2011}
{Aladro}, R., {Mart{\'{\i}}n}, S., {Mart{\'{\i}}n-Pintado}, J., {et~al.} 2011,
  \aap, 535, A84

\bibitem[{Allegrini {et~al.}(1979)Allegrini, Johns, \&
  McKellar}]{Allegrini_1979}
Allegrini, M., Johns, J. W.~C., \& McKellar, A. R.~W. 1979, The Journal of
  Chemical Physics, 70, 2829

\bibitem[{{Andree-Labsch} {et~al.}(2017){Andree-Labsch}, {Ossenkopf-Okada}, \&
  {R{\"o}llig}}]{Andree-Labsch_2017}
{Andree-Labsch}, S., {Ossenkopf-Okada}, V., \& {R{\"o}llig}, M. 2017, \aap,
  598, A2

\bibitem[{{Anttila} {et~al.}(1993){Anttila}, {Horneman}, {Koivusaari}, \&
  {Paso}}]{Anttila_1993}
{Anttila}, R., {Horneman}, V.~M., {Koivusaari}, M., \& {Paso}, R. 1993, Journal
  of Molecular Spectroscopy, 157, 198

\bibitem[{{Arab} {et~al.}(2012){Arab}, {Abergel}, {Habart}, {Bernard-Salas},
  {Ayasso}, {Dassas}, {Martin}, \& {White}}]{Arab_2012}
{Arab}, H., {Abergel}, A., {Habart}, E., {et~al.} 2012, \aap, 541, A19

\bibitem[{{Austin} {et~al.}(1974){Austin}, {Levy}, {Gottlieb}, \&
  {Radford}}]{Austin_1974}
{Austin}, J.~A., {Levy}, D.~H., {Gottlieb}, C.~A., \& {Radford}, H.~E. 1974,
  \jcp, 60, 207

\bibitem[{{Avery} {et~al.}(1976){Avery}, {Broten}, {MacLeod}, {Oka}, \&
  {Kroto}}]{Avery_1976}
{Avery}, L.~W., {Broten}, N.~W., {MacLeod}, J.~M., {Oka}, T., \& {Kroto}, H.~W.
  1976, \apjl, 205, L173

\bibitem[{{Bachiller} \& {P{\'e}rez Guti{\'e}rrez}(1997)}]{Bachiller_1997}
{Bachiller}, R. \& {P{\'e}rez Guti{\'e}rrez}, M. 1997, \apjl, 487, L93

\bibitem[{{Bacmann} \& {Faure}(2016)}]{Bacmann_2016}
{Bacmann}, A. \& {Faure}, A. 2016, ArXiv e-prints

\bibitem[{{Bacmann} {et~al.}(2012){Bacmann}, {Taquet}, {Faure}, {Kahane}, \&
  {Ceccarelli}}]{Bacmann_2012}
{Bacmann}, A., {Taquet}, V., {Faure}, A., {Kahane}, C., \& {Ceccarelli}, C.
  2012, \aap, 541, L12

\bibitem[{{Batrla} \& {Wilson}(2003)}]{Batrla_2003}
{Batrla}, W. \& {Wilson}, T.~L. 2003, \aap, 408, 231

\bibitem[{{Beers} {et~al.}(1972){Beers}, {Klein}, {Kirchhoff}, \&
  {Johnson}}]{Beers_1972}
{Beers}, Y., {Klein}, G.~P., {Kirchhoff}, W.~H., \& {Johnson}, D.~R. 1972,
  Journal of Molecular Spectroscopy, 44, 553

\bibitem[{{Belloche} {et~al.}(2014){Belloche}, {Garrod}, {M{\"u}ller}, \&
  {Menten}}]{Belloche_2014}
{Belloche}, A., {Garrod}, R.~T., {M{\"u}ller}, H.~S.~P., \& {Menten}, K.~M.
  2014, Science, 345, 1584

\bibitem[{{Belloche} {et~al.}(2009){Belloche}, {Garrod}, {M{\"u}ller},
  {Menten}, {Comito}, \& {Schilke}}]{Belloche_2009}
{Belloche}, A., {Garrod}, R.~T., {M{\"u}ller}, H.~S.~P., {et~al.} 2009, \aap,
  499, 215

\bibitem[{{Bertin} {et~al.}(2016){Bertin}, {Romanzin}, {Doronin}, {Philippe},
  {Jeseck}, {Ligterink}, {Linnartz}, {Michaut}, \& {Fillion}}]{Bertin_2016}
{Bertin}, M., {Romanzin}, C., {Doronin}, M., {et~al.} 2016, \apjl, 817, L12

\bibitem[{{Bisschop} {et~al.}(2007){Bisschop}, {J{\o}rgensen}, {van Dishoeck},
  \& {de Wachter}}]{Bisschop_2007}
{Bisschop}, S.~E., {J{\o}rgensen}, J.~K., {van Dishoeck}, E.~F., \& {de
  Wachter}, E.~B.~M. 2007, \aap, 465, 913

\bibitem[{{Blake} {et~al.}(1984){Blake}, {Sastry}, \& {de Lucia}}]{Blake_1984}
{Blake}, G.~A., {Sastry}, K.~V.~L.~N., \& {de Lucia}, F.~C. 1984, \jcp, 80, 95

\bibitem[{{Blake} {et~al.}(1987){Blake}, {Sutton}, {Masson}, \&
  {Phillips}}]{Blake_1987}
{Blake}, G.~A., {Sutton}, E.~C., {Masson}, C.~R., \& {Phillips}, T.~G. 1987,
  \apj, 315, 621

\bibitem[{{Bockel{\'e}e-Morvan} {et~al.}(2004){Bockel{\'e}e-Morvan},
  {Crovisier}, {Mumma}, \& {Weaver}}]{Bockelee_2004}
{Bockel{\'e}e-Morvan}, D., {Crovisier}, J., {Mumma}, M.~J., \& {Weaver}, H.~A.
  2004, {The composition of cometary volatiles. Comets II} (University of
  Arizona Press), 391--423

\bibitem[{{Bocquet} {et~al.}(1996){Bocquet}, {Demaison}, {Poteau}, {Liedtke},
  {Belov}, {Yamada}, {Winnewisser}, {Gerke}, {Gripp}, \&
  {K{\"o}hler}}]{Bocquet_1996}
{Bocquet}, R., {Demaison}, J., {Poteau}, L., {et~al.} 1996, Journal of
  Molecular Spectroscopy, 177, 154

\bibitem[{{Bottinelli} {et~al.}(2004){Bottinelli}, {Ceccarelli}, {Lefloch},
  {Williams}, {Castets}, {Caux}, {Cazaux}, {Maret}, {Parise}, \&
  {Tielens}}]{Bottinelli_2004}
{Bottinelli}, S., {Ceccarelli}, C., {Lefloch}, B., {et~al.} 2004, \apj, 615,
  354

\bibitem[{{Boucher} {et~al.}(1977){Boucher}, {Burie}, {Demaison}, {Dubrulle},
  {Legrand}, \& {Segard}}]{Boucher_1977}
{Boucher}, D., {Burie}, J., {Demaison}, J., {et~al.} 1977, Journal of Molecular
  Spectroscopy, 64, 290

\bibitem[{{Bowater} {et~al.}(1971){Bowater}, {Brown}, \&
  {Carrington}}]{Bowater_1971}
{Bowater}, I.~C., {Brown}, J.~M., \& {Carrington}, A. 1971, \jcp, 54, 4957

\bibitem[{{Brown} {et~al.}(1975){Brown}, {Crofts}, {Godfrey}, {Gardner},
  {Robinson}, \& {Whiteoak}}]{Brown_1975}
{Brown}, R.~D., {Crofts}, J.~G., {Godfrey}, P.~D., {et~al.} 1975, \apjl, 197,
  L29

\bibitem[{{Brown} {et~al.}(1990){Brown}, {Godfrey}, {McNaughton}, {Pierlot}, \&
  {Taylor}}]{Brown_1990}
{Brown}, R.~D., {Godfrey}, P.~D., {McNaughton}, D., {Pierlot}, A.~P., \&
  {Taylor}, W.~H. 1990, Journal of Molecular Spectroscopy, 140, 340

\bibitem[{{Br{\"u}nken} {et~al.}(2003){Br{\"u}nken}, {M{\"u}ller}, {Lewen}, \&
  {Winnewisser}}]{Brunken_2003}
{Br{\"u}nken}, S., {M{\"u}ller}, H.~S.~P., {Lewen}, F., \& {Winnewisser}, G.
  2003, Physical Chemistry Chemical Physics (Incorporating Faraday
  Transactions), 5

\bibitem[{{Burton} {et~al.}(1990){Burton}, {Hollenbach}, \&
  {Tielens}}]{Burton_1990}
{Burton}, M.~G., {Hollenbach}, D.~J., \& {Tielens}, A.~G.~G.~M. 1990, \apj,
  365, 620

\bibitem[{{Caselli} \& {Ceccarelli}(2012)}]{Caselli_2012}
{Caselli}, P. \& {Ceccarelli}, C. 2012, \aapr, 20, 56

\bibitem[{{Cazaux} {et~al.}(2003){Cazaux}, {Tielens}, {Ceccarelli}, {Castets},
  {Wakelam}, {Caux}, {Parise}, \& {Teyssier}}]{Cazaux_2003}
{Cazaux}, S., {Tielens}, A.~G.~G.~M., {Ceccarelli}, C., {et~al.} 2003, \apjl,
  593, L51

\bibitem[{{Cazzoli} \& {Puzzarini}(2006)}]{Cazzoli_2006}
{Cazzoli}, G. \& {Puzzarini}, C. 2006, Journal of Molecular Spectroscopy, 240,
  153

\bibitem[{{Cernicharo}(2012)}]{Cernicharo_2012}
{Cernicharo}, J. 2012, in EAS Publications Series, Vol.~58, EAS Publications
  Series, 251--261

\bibitem[{{Cernicharo} {et~al.}(1986){Cernicharo}, {Bachiller}, \&
  {Duvert}}]{Cernicharo_1986}
{Cernicharo}, J., {Bachiller}, R., \& {Duvert}, G. 1986, \aap, 160, 181

\bibitem[{{Cernicharo} {et~al.}(2000){Cernicharo}, {Gu{\'e}lin}, \&
  {Kahane}}]{Cernicharo_2000}
{Cernicharo}, J., {Gu{\'e}lin}, M., \& {Kahane}, C. 2000, \aaps, 142, 181

\bibitem[{{Cernicharo} {et~al.}(2001){Cernicharo}, {Heras}, {Tielens}, {Pardo},
  {Herpin}, {Gu{\'e}lin}, \& {Waters}}]{Cernicharo_2001}
{Cernicharo}, J., {Heras}, A.~M., {Tielens}, A.~G.~G.~M., {et~al.} 2001, \apjl,
  546, L123

\bibitem[{{Cernicharo} {et~al.}(2016){Cernicharo}, {Kisiel}, {Tercero},
  {Kolesnikov{\'a}}, {Medvedev}, {L{\'o}pez}, {Fortman}, {Winnewisser}, {de
  Lucia}, {Alonso}, \& {Guillemin}}]{Cernicharo_2016}
{Cernicharo}, J., {Kisiel}, Z., {Tercero}, B., {et~al.} 2016, \aap, 587, L4

\bibitem[{{Cernicharo} {et~al.}(2012){Cernicharo}, {Marcelino}, {Roueff},
  {Gerin}, {Jim{\'e}nez-Escobar}, \& {Mu{\~n}oz Caro}}]{Cernicharo_2012b}
{Cernicharo}, J., {Marcelino}, N., {Roueff}, E., {et~al.} 2012, \apjl, 759, L43

\bibitem[{{Chen} {et~al.}(1991){Chen}, {Bocquet}, {Wlodarczak}, \&
  {Boucher}}]{Chen_1991}
{Chen}, W., {Bocquet}, R., {Wlodarczak}, G., \& {Boucher}, D. 1991,
  International Journal of Infrared and Millimeter Waves, 12, 987

\bibitem[{{Codella} {et~al.}(2015){Codella}, {Fontani}, {Ceccarelli}, {Podio},
  {Viti}, {Bachiller}, {Benedettini}, \& {Lefloch}}]{Codella_2015}
{Codella}, C., {Fontani}, F., {Ceccarelli}, C., {et~al.} 2015, \mnras, 449, L11

\bibitem[{{Collings} {et~al.}(2004){Collings}, {Anderson}, {Chen}, {Dever},
  {Viti}, {Williams}, \& {McCoustra}}]{Collings_2004}
{Collings}, M.~P., {Anderson}, M.~A., {Chen}, R., {et~al.} 2004, \mnras, 354,
  1133

\bibitem[{{Creswell} {et~al.}(1977){Creswell}, {Winnewisser}, \&
  {Gerry}}]{Creswell_1977}
{Creswell}, R.~A., {Winnewisser}, G., \& {Gerry}, M.~C.~L. 1977, Journal of
  Molecular Spectroscopy, 65, 420

\bibitem[{{Crockett} {et~al.}(2014){Crockett}, {Bergin}, {Neill}, {Favre},
  {Schilke}, {Lis}, {Bell}, {Blake}, {Cernicharo}, {Emprechtinger},
  {Esplugues}, {Gupta}, {Kleshcheva}, {Lord}, {Marcelino}, {McGuire},
  {Pearson}, {Phillips}, {Plume}, {van der Tak}, {Tercero}, \&
  {Yu}}]{Crockett_2014}
{Crockett}, N.~R., {Bergin}, E.~A., {Neill}, J.~L., {et~al.} 2014, \apj, 787,
  112

\bibitem[{{Cronin} \& {Chang}(1993)}]{Cronin_1993}
{Cronin}, J.~R. \& {Chang}, S. 1993, in NATO Advanced Science Institutes (ASI)
  Series C, Vol. 416, NATO Advanced Science Institutes (ASI) Series C, ed.
  J.~M. {Greenberg}, C.~X. {Mendoza-G{\'o}mez}, \& V.~{Pirronello}, 209--258

\bibitem[{{Cruz-Diaz} {et~al.}(2016){Cruz-Diaz}, {Mart{\'{\i}}n-Dom{\'e}nech},
  {Mu{\~n}oz Caro}, \& {Chen}}]{Cruz-Diaz_2016}
{Cruz-Diaz}, G.~A., {Mart{\'{\i}}n-Dom{\'e}nech}, R., {Mu{\~n}oz Caro}, G.~M.,
  \& {Chen}, Y.-J. 2016, ArXiv e-prints

\bibitem[{{Cuadrado} {et~al.}(2015){Cuadrado}, {Goicoechea}, {Pilleri},
  {Cernicharo}, {Fuente}, \& {Joblin}}]{Cuadrado_2015a}
{Cuadrado}, S., {Goicoechea}, J.~R., {Pilleri}, P., {et~al.} 2015, \aap, 575,
  A82

\bibitem[{{Cuadrado} {et~al.}(2016){Cuadrado}, {Goicoechea}, {Roncero},
  {Aguado}, {Tercero}, \& {Cernicharo}}]{Cuadrado_2016}
{Cuadrado}, S., {Goicoechea}, J.~R., {Roncero}, O., {et~al.} 2016, \aap, 596,
  L1

\bibitem[{{de Zafra}(1971)}]{Zafra_1971}
{de Zafra}, R.~L. 1971, \apj, 170, 165

\bibitem[{{DeLeon} \& {Muenter}(1985)}]{DeLeon_1985}
{DeLeon}, R.~L. \& {Muenter}, J.~S. 1985, \jcp, 82, 1702

\bibitem[{{Dickinson}(1972)}]{Dickinson_1972}
{Dickinson}, D.~F. 1972, \aplett, 12, 235

\bibitem[{{Dore} {et~al.}(2012){Dore}, {Bizzocchi}, \& {Degli
  Esposti}}]{Dore_2012}
{Dore}, L., {Bizzocchi}, L., \& {Degli Esposti}, C. 2012, \aap, 544, A19

\bibitem[{{Dore} {et~al.}(2010){Dore}, {Bizzocchi}, {Degli Esposti}, \&
  {Gauss}}]{Dore_2010}
{Dore}, L., {Bizzocchi}, L., {Degli Esposti}, C., \& {Gauss}, J. 2010, Journal
  of Molecular Spectroscopy, 263, 44

\bibitem[{{Draine} {et~al.}(1983){Draine}, {Roberge}, \&
  {Dalgarno}}]{Draine_1983}
{Draine}, B.~T., {Roberge}, W.~G., \& {Dalgarno}, A. 1983, \apj, 264, 485

\bibitem[{Eliet {et~al.}(2012)Eliet, Cuisset, Guinet, Hindle, Mouret, Bocquet,
  \& Demaison}]{Eliet_2012}
Eliet, S., Cuisset, A., Guinet, M., {et~al.} 2012, Journal of Molecular
  Spectroscopy, 279, 12

\bibitem[{{Esplugues} {et~al.}(2016){Esplugues}, {Cazaux}, {Meijerink},
  {Spaans}, \& {Caselli}}]{Esplugues_2016}
{Esplugues}, G.~B., {Cazaux}, S., {Meijerink}, R., {Spaans}, M., \& {Caselli},
  P. 2016, \aap, 591, A52

\bibitem[{{Fabricant} {et~al.}(1977){Fabricant}, {Krieger}, \&
  {Muenter}}]{Fabricant_1977}
{Fabricant}, B., {Krieger}, D., \& {Muenter}, J.~S. 1977, \jcp, 67, 1576

\bibitem[{{Fuente} {et~al.}(2005){Fuente}, {Garc{\'{\i}}a-Burillo}, {Gerin},
  {Teyssier}, {Usero}, {Rizzo}, \& {de Vicente}}]{Fuente_2005}
{Fuente}, A., {Garc{\'{\i}}a-Burillo}, S., {Gerin}, M., {et~al.} 2005, \apjl,
  619, L155

\bibitem[{{Fuente} {et~al.}(2008){Fuente}, {Garc{\'{\i}}a-Burillo}, {Usero},
  {Gerin}, {Neri}, {Faure}, {Le Bourlot}, {Gonz{\'a}lez-Garc{\'{\i}}a},
  {Rizzo}, {Alonso-Albi}, \& {Tennyson}}]{Fuente_2008}
{Fuente}, A., {Garc{\'{\i}}a-Burillo}, S., {Usero}, A., {et~al.} 2008, \aap,
  492, 675

\bibitem[{Gadhi {et~al.}(1995)Gadhi, Lahrouni, Legrand, \&
  Demaison}]{Gadhi_1995}
Gadhi, J., Lahrouni, A., Legrand, J., \& Demaison, J. 1995, Journal de chimie
  physique, 92, 1984

\bibitem[{{Gerin} {et~al.}(2009){Gerin}, {Goicoechea}, {Pety}, \&
  {Hily-Blant}}]{Gerin_2009}
{Gerin}, M., {Goicoechea}, J.~R., {Pety}, J., \& {Hily-Blant}, P. 2009, \aap,
  494, 977

\bibitem[{{Gibb} {et~al.}(2000){Gibb}, {Whittet}, {Schutte}, {Boogert},
  {Chiar}, {Ehrenfreund}, {Gerakines}, {Keane}, {Tielens}, {van Dishoeck}, \&
  {Kerkhof}}]{Gibb_2000}
{Gibb}, E.~L., {Whittet}, D.~C.~B., {Schutte}, W.~A., {et~al.} 2000, \apj, 536,
  347

\bibitem[{{Ginard} {et~al.}(2012){Ginard}, {Gonz{\'a}lez-Garc{\'{\i}}a},
  {Fuente}, {Cernicharo}, {Alonso-Albi}, {Pilleri}, {Gerin},
  {Garc{\'{\i}}a-Burillo}, {Ossenkopf}, {Rizzo}, {Kramer}, {Goicoechea},
  {Pety}, {Bern{\'e}}, \& {Joblin}}]{Ginard_2012}
{Ginard}, D., {Gonz{\'a}lez-Garc{\'{\i}}a}, M., {Fuente}, A., {et~al.} 2012,
  \aap, 543, A27

\bibitem[{{Goicoechea} {et~al.}(2009){Goicoechea}, {Compi{\`e}gne}, \&
  {Habart}}]{Goicoechea_2009b}
{Goicoechea}, J.~R., {Compi{\`e}gne}, M., \& {Habart}, E. 2009, \apjl, 699,
  L165

\bibitem[{{Goicoechea} {et~al.}(2017){Goicoechea}, {Cuadrado}, {Pety}, \&
  et~al.}]{Goicoechea_2017}
{Goicoechea}, J.~R., {Cuadrado}, S., {Pety}, J., \& et~al. 2017, \aap, 601, L9

\bibitem[{{Goicoechea} {et~al.}(2011){Goicoechea}, {Joblin}, {Contursi},
  {Bern{\'e}}, {Cernicharo}, {Gerin}, {Le Bourlot}, {Bergin}, {Bell}, \&
  {R{\"o}llig}}]{Goicoechea_2011}
{Goicoechea}, J.~R., {Joblin}, C., {Contursi}, A., {et~al.} 2011, \aap, 530,
  L16

\bibitem[{{Goicoechea} \& {Le Bourlot}(2007)}]{Goicoechea_2007}
{Goicoechea}, J.~R. \& {Le Bourlot}, J. 2007, \aap, 467, 1

\bibitem[{{Goicoechea} {et~al.}(2016){Goicoechea}, {Pety}, {Cuadrado},
  {Cernicharo}, {Chapillon}, \& {Fuente}}]{Goicoechea_2016}
{Goicoechea}, J.~R., {Pety}, J., {Cuadrado}, S., {et~al.} 2016, Nature, 537,
  207

\bibitem[{{Goicoechea} {et~al.}(2015){Goicoechea}, {Teyssier}, {Etxaluze},
  {Goldsmith}, {Ossenkopf}, {Gerin}, {Bergin}, {Black}, {Cernicharo},
  {Cuadrado}, {Encrenaz}, {Falgarone}, {Fuente}, {Hacar}, {Lis}, {Marcelino},
  {Melnick}, {M{\"u}ller}, {Persson}, {Pety}, {R{\"o}llig}, {Schilke}, {Simon},
  {Snell}, \& {Stutzki}}]{Goicoechea_2015}
{Goicoechea}, J.~R., {Teyssier}, D., {Etxaluze}, M., {et~al.} 2015, \apj, 812,
  75

\bibitem[{{Goldsmith} \& {Langer}(1999)}]{Goldsmith_1999}
{Goldsmith}, P.~F. \& {Langer}, W.~D. 1999, \apj, 517, 209

\bibitem[{{Gottlieb}(1973)}]{Gottlieb_1973}
{Gottlieb}, C.~A. 1973, in Molecules in the Galactic Environment, ed. M.~A.
  {Gordon} \& L.~E. {Snyder}, 181

\bibitem[{{Gratier} {et~al.}(2013){Gratier}, {Pety}, {Guzm{\'a}n}, {Gerin},
  {Goicoechea}, {Roueff}, \& {Faure}}]{Gratier_2013}
{Gratier}, P., {Pety}, J., {Guzm{\'a}n}, V., {et~al.} 2013, \aap, 557, A101

\bibitem[{{Green}(1986)}]{Green_1986}
{Green}, S. 1986, \apj, 309, 331

\bibitem[{{Green}(1991)}]{Green_1991}
{Green}, S. 1991, \apjs, 76, 979

\bibitem[{{Greve} {et~al.}(1998){Greve}, {Kramer}, \& {Wild}}]{Greve_1998}
{Greve}, A., {Kramer}, C., \& {Wild}, W. 1998, \aaps, 133, 271

\bibitem[{{Guzm{\'a}n} {et~al.}(2011){Guzm{\'a}n}, {Pety}, {Goicoechea},
  {Gerin}, \& {Roueff}}]{Guzman_2011}
{Guzm{\'a}n}, V., {Pety}, J., {Goicoechea}, J.~R., {Gerin}, M., \& {Roueff}, E.
  2011, \aap, 534, A49

\bibitem[{{Guzm{\'a}n} {et~al.}(2013){Guzm{\'a}n}, {Goicoechea}, {Pety},
  {Gratier}, {Gerin}, {Roueff}, {Le Petit}, {Le Bourlot}, \&
  {Faure}}]{Guzman_2013}
{Guzm{\'a}n}, V.~V., {Goicoechea}, J.~R., {Pety}, J., {et~al.} 2013, \aap, 560,
  A73

\bibitem[{{Guzm{\'a}n} {et~al.}(2015){Guzm{\'a}n}, {Pety}, {Goicoechea},
  {Gerin}, {Roueff}, {Gratier}, \& {{\"O}berg}}]{Guzman_2015}
{Guzm{\'a}n}, V.~V., {Pety}, J., {Goicoechea}, J.~R., {et~al.} 2015, \apjl,
  800, L33

\bibitem[{{Guzm{\'a}n} {et~al.}(2014){Guzm{\'a}n}, {Pety}, {Gratier},
  {Goicoechea}, {Gerin}, {Roueff}, {Le Petit}, \& {Le Bourlot}}]{Guzman_2014}
{Guzm{\'a}n}, V.~V., {Pety}, J., {Gratier}, P., {et~al.} 2014, Faraday
  Discussions, 168, 103

\bibitem[{{Habart} {et~al.}(2010){Habart}, {Dartois}, {Abergel}, {Baluteau},
  {Naylor}, {Polehampton}, {Joblin}, {Ade}, {Anderson}, {Andr{\'e}}, {Arab},
  {Bernard}, {Blagrave}, {Bontemps}, {Boulanger}, {Cohen}, {Compiegne}, {Cox},
  {Davis}, {Emery}, {Fulton}, {Gry}, {Huang}, {Jones}, {Kirk}, {Lagache},
  {Lim}, {Madden}, {Makiwa}, {Martin}, {Miville-Desch{\^e}nes}, {Molinari},
  {Moseley}, {Motte}, {Okumura}, {Pinheiro Gon{\c c}alves}, {Rodon}, {Russeil},
  {Saraceno}, {Sidher}, {Spencer}, {Swinyard}, {Ward-Thompson}, {White}, \&
  {Zavagno}}]{Habart_2010}
{Habart}, E., {Dartois}, E., {Abergel}, A., {et~al.} 2010, \aap, 518, L116

\bibitem[{{Herbst}(2015)}]{Herbst_2015}
{Herbst}, E. 2015, in European Physical Journal Web of Conferences, Vol.~84,
  European Physical Journal Web of Conferences, 06002

\bibitem[{{Herbst} \& {van Dishoeck}(2009)}]{Herbst_2009}
{Herbst}, E. \& {van Dishoeck}, E.~F. 2009, \araa, 47, 427

\bibitem[{{Hirano} {et~al.}(1999){Hirano}, {Kamazaki}, {Mikami}, {Ohashi}, \&
  {Umemoto}}]{Hirano_1999}
{Hirano}, N., {Kamazaki}, T., {Mikami}, H., {Ohashi}, N., \& {Umemoto}, T.
  1999, in Star Formation 1999, ed. T.~{Nakamoto}, 181--182

\bibitem[{{Hocking} {et~al.}(1975){Hocking}, {Gerry}, \&
  {Winnewisser}}]{Hocking_1975}
{Hocking}, W.~H., {Gerry}, M.~C.~L., \& {Winnewisser}, G. 1975, Canadian
  Journal of Physics, 53, 1869

\bibitem[{{Hogerheijde} {et~al.}(1995){Hogerheijde}, {Jansen}, \& {van
  Dishoeck}}]{Hogerheijde_1995}
{Hogerheijde}, M.~R., {Jansen}, D.~J., \& {van Dishoeck}, E.~F. 1995, \aap,
  294, 792

\bibitem[{{Hollenbach} {et~al.}(2009){Hollenbach}, {Kaufman}, {Bergin}, \&
  {Melnick}}]{Hollenbach_2009}
{Hollenbach}, D., {Kaufman}, M.~J., {Bergin}, E.~A., \& {Melnick}, G.~J. 2009,
  \apj, 690, 1497

\bibitem[{{Ikeda} {et~al.}(2001){Ikeda}, {Ohishi}, {Nummelin}, {Dickens},
  {Bergman}, {Hjalmarson}, \& {Irvine}}]{Ikeda_2001}
{Ikeda}, M., {Ohishi}, M., {Nummelin}, A., {et~al.} 2001, \apj, 560, 792

\bibitem[{{Jansen} {et~al.}(1995){Jansen}, {Spaans}, {Hogerheijde}, \& {van
  Dishoeck}}]{Jansen_1995}
{Jansen}, D.~J., {Spaans}, M., {Hogerheijde}, M.~R., \& {van Dishoeck}, E.~F.
  1995, \aap, 303, 541

\bibitem[{{Johns} {et~al.}(1992){Johns}, {Nemes}, {Yamada}, {Wang},
  {Dom{\'e}nech}, {Santos}, {Cancio}, {Bermejo}, {Ortigoso}, \&
  {Escribano}}]{Johns_1992}
{Johns}, J.~W.~C., {Nemes}, L., {Yamada}, K.~M.~T., {et~al.} 1992, Journal of
  Molecular Spectroscopy, 156, 501

\bibitem[{{Johnson} {et~al.}(1971){Johnson}, {Powell}, \&
  {Kirchhoff}}]{Johnson_1971}
{Johnson}, D.~R., {Powell}, F.~X., \& {Kirchhoff}, W.~H. 1971, Journal of
  Molecular Spectroscopy, 39, 136

\bibitem[{{Johnson} \& {Strandberg}(1952)}]{Johnson_1952}
{Johnson}, H.~R. \& {Strandberg}, M.~W.~P. 1952, \jcp, 20, 687

\bibitem[{{Kirchhoff} {et~al.}(1973){Kirchhoff}, {Johnson}, \&
  {Lovas}}]{Kirchhoff_1973}
{Kirchhoff}, W.~H., {Johnson}, D.~R., \& {Lovas}, F.~J. 1973, Journal of
  Physical and Chemical Reference Data, 2, 1

\bibitem[{{Kleiner} {et~al.}(1996){Kleiner}, {Lovas}, \&
  {Godefroid}}]{Kleiner_1996}
{Kleiner}, I., {Lovas}, F.~J., \& {Godefroid}, M. 1996, Journal of Physical and
  Chemical Reference Data, 25, 1113

\bibitem[{Kukolich {et~al.}(1971)Kukolich, Nelson, \&
  Yamanashi}]{Kukolich_1971}
Kukolich, S., Nelson, A., \& Yamanashi, B. 1971, Journal of the American
  Chemical Society, 93, 6769

\bibitem[{{Kukolich}(1982)}]{Kukolich_1982}
{Kukolich}, S.~G. 1982, \jcp, 76, 97

\bibitem[{{Kukolich} {et~al.}(1973){Kukolich}, {Ruben}, {Wang}, \&
  {Williams}}]{Kukolich_1973}
{Kukolich}, S.~G., {Ruben}, D.~J., {Wang}, J.~H.~S., \& {Williams}, J.~R. 1973,
  \jcp, 58, 3155

\bibitem[{{Langer} \& {Penzias}(1990)}]{Langer_1990}
{Langer}, W.~D. \& {Penzias}, A.~A. 1990, \apj, 357, 477

\bibitem[{{Lapinov} {et~al.}(2007){Lapinov}, {Golubiatnikov}, {Markov}, \&
  {Guarnieri}}]{Lapinov_2007}
{Lapinov}, A.~V., {Golubiatnikov}, G.~Y., {Markov}, V.~N., \& {Guarnieri}, A.
  2007, Astronomy Letters, 33, 121

\bibitem[{{Le Petit} {et~al.}(2006){Le Petit}, {Nehm{\'e}}, {Le Bourlot}, \&
  {Roueff}}]{LePetit_2006}
{Le Petit}, F., {Nehm{\'e}}, C., {Le Bourlot}, J., \& {Roueff}, E. 2006, \apjs,
  164, 506

\bibitem[{{Lees} \& {Baker}(1968)}]{Lees_1968}
{Lees}, R.~M. \& {Baker}, J.~G. 1968, \jcp, 48, 5299

\bibitem[{{Lees} {et~al.}(1973){Lees}, {Lovas}, {Kirchhoff}, \&
  {Johnson}}]{Lees_1973}
{Lees}, R.~M., {Lovas}, F.~J., {Kirchhoff}, W.~H., \& {Johnson}, D.~R. 1973,
  Journal of Physical and Chemical Reference Data, 2, 205

\bibitem[{{Leurini} {et~al.}(2010){Leurini}, {Parise}, {Schilke}, {Pety}, \&
  {Rolffs}}]{Leurini_2010}
{Leurini}, S., {Parise}, B., {Schilke}, P., {Pety}, J., \& {Rolffs}, R. 2010,
  \aap, 511, A82

\bibitem[{{Leurini} {et~al.}(2006){Leurini}, {Rolffs}, {Thorwirth}, {Parise},
  {Schilke}, {Comito}, {Wyrowski}, {G{\"u}sten}, {Bergman}, {Menten}, \&
  {Nyman}}]{Leurini_2006}
{Leurini}, S., {Rolffs}, R., {Thorwirth}, S., {et~al.} 2006, \aap, 454, L47

\bibitem[{{Lis} {et~al.}(2002){Lis}, {Roueff}, {Gerin}, {Phillips}, {Coudert},
  {van der Tak}, \& {Schilke}}]{Lis_2002}
{Lis}, D.~C., {Roueff}, E., {Gerin}, M., {et~al.} 2002, \apjl, 571, L55

\bibitem[{{Lis} \& {Schilke}(2003)}]{Lis_2003}
{Lis}, D.~C. \& {Schilke}, P. 2003, \apjl, 597, L145

\bibitem[{{Liszt} {et~al.}(2006){Liszt}, {Lucas}, \& {Pety}}]{Liszt_2006}
{Liszt}, H.~S., {Lucas}, R., \& {Pety}, J. 2006, \aap, 448, 253

\bibitem[{{Maeda} {et~al.}(2008){Maeda}, {Medvedev}, {Winnewisser}, {De Lucia},
  {Herbst}, {M{\"u}ller}, {Koerber}, {Endres}, \& {Schlemmer}}]{Maeda_2008}
{Maeda}, A., {Medvedev}, I.~R., {Winnewisser}, M., {et~al.} 2008, \apjs, 176,
  543

\bibitem[{{Marcelino} {et~al.}(2005){Marcelino}, {Cernicharo}, {Roueff},
  {Gerin}, \& {Mauersberger}}]{Marcelino_2005}
{Marcelino}, N., {Cernicharo}, J., {Roueff}, E., {Gerin}, M., \&
  {Mauersberger}, R. 2005, \apj, 620, 308

\bibitem[{{Marcelino} {et~al.}(2009){Marcelino}, {Cernicharo}, {Tercero}, \&
  {Roueff}}]{Marcelino_2009}
{Marcelino}, N., {Cernicharo}, J., {Tercero}, B., \& {Roueff}, E. 2009, \apjl,
  690, L27

\bibitem[{{Marconi} {et~al.}(1998){Marconi}, {Testi}, {Natta}, \&
  {Walmsley}}]{Marconi_1998}
{Marconi}, A., {Testi}, L., {Natta}, A., \& {Walmsley}, C.~M. 1998, \aap, 330,
  696

\bibitem[{{Mbosei} {et~al.}(2000){Mbosei}, {Fayt}, {Dr{\'e}an}, \&
  {Cosl{\'e}ou}}]{Mbosei_2000}
{Mbosei}, L., {Fayt}, A., {Dr{\'e}an}, P., \& {Cosl{\'e}ou}, J. 2000, Journal
  of Molecular Structure, 517, 271

\bibitem[{{Meier} \& {Turner}(2005)}]{Meier_2005}
{Meier}, D.~S. \& {Turner}, J.~L. 2005, \apj, 618, 259

\bibitem[{{Meier} \& {Turner}(2012)}]{Meier_2012}
{Meier}, D.~S. \& {Turner}, J.~L. 2012, \apj, 755, 104

\bibitem[{{Mendoza} {et~al.}(2014){Mendoza}, {Lefloch}, {L{\'o}pez-Sepulcre},
  {Ceccarelli}, {Codella}, {Boechat-Roberty}, \& {Bachiller}}]{Mendoza_2014}
{Mendoza}, E., {Lefloch}, B., {L{\'o}pez-Sepulcre}, A., {et~al.} 2014, \mnras,
  445, 151

\bibitem[{{Menten} {et~al.}(2007){Menten}, {Reid}, {Forbrich}, \&
  {Brunthaler}}]{Menten_2007}
{Menten}, K.~M., {Reid}, M.~J., {Forbrich}, J., \& {Brunthaler}, A. 2007, \aap,
  474, 515

\bibitem[{{Minissale} {et~al.}(2016){Minissale}, {Moudens}, {Baouche},
  {Chaabouni}, \& {Dulieu}}]{Minissale_2016}
{Minissale}, M., {Moudens}, A., {Baouche}, S., {Chaabouni}, H., \& {Dulieu}, F.
  2016, \mnras, 458, 2953

\bibitem[{{M{\"u}ller} {et~al.}(2009){M{\"u}ller}, {Drouin}, \&
  {Pearson}}]{Muller_2009}
{M{\"u}ller}, H.~S.~P., {Drouin}, B.~J., \& {Pearson}, J.~C. 2009, \aap, 506,
  1487

\bibitem[{{M{\"u}ller} {et~al.}(2000){M{\"u}ller}, {Gendriesch},
  {Margul{\`e}s}, {Lewen}, {Winnewisser}, {Bocquet}, {Demaison}, {W{\"o}tzel},
  \& {M{\"a}der}}]{Muller_2000b}
{M{\"u}ller}, H.~S.~P., {Gendriesch}, R., {Margul{\`e}s}, L., {et~al.} 2000,
  Physical Chemistry Chemical Physics (Incorporating Faraday Transactions), 2

\bibitem[{{M{\"u}ller} {et~al.}(2005){M{\"u}ller}, {Schl{\"o}der}, {Stutzki},
  \& {Winnewisser}}]{Muller_2005}
{M{\"u}ller}, H.~S.~P., {Schl{\"o}der}, F., {Stutzki}, J., \& {Winnewisser}, G.
  2005, Journal of Molecular Structure, 742, 215

\bibitem[{{M{\"u}ller} {et~al.}(2001){M{\"u}ller}, {Thorwirth}, {Roth}, \&
  {Winnewisser}}]{Muller_2001}
{M{\"u}ller}, H.~S.~P., {Thorwirth}, S., {Roth}, D.~A., \& {Winnewisser}, G.
  2001, \aap, 370, L49

\bibitem[{{Nagy} {et~al.}(2017){Nagy}, {Choi}, {Ossenkopf-Okada}, {van der
  Tak}, {Bergin}, {Gerin}, {Joblin}, {R{\"o}llig}, {Simon}, \&
  {Stutzki}}]{Nagy_2017}
{Nagy}, Z., {Choi}, Y., {Ossenkopf-Okada}, V., {et~al.} 2017, \aap, 599, A22

\bibitem[{{Nagy} {et~al.}(2013){Nagy}, {Van der Tak}, {Ossenkopf}, {Gerin}, {Le
  Petit}, {Le Bourlot}, {Black}, {Goicoechea}, {Joblin}, {R{\"o}llig}, \&
  {Bergin}}]{Nagy_2013}
{Nagy}, Z., {Van der Tak}, F.~F.~S., {Ossenkopf}, V., {et~al.} 2013, \aap, 550,
  A96

\bibitem[{{{\"O}berg} {et~al.}(2010){{\"O}berg}, {Bottinelli}, {J{\o}rgensen},
  \& {van Dishoeck}}]{Oberg_2010}
{{\"O}berg}, K.~I., {Bottinelli}, S., {J{\o}rgensen}, J.~K., \& {van Dishoeck},
  E.~F. 2010, \apj, 716, 825

\bibitem[{{Ossenkopf} {et~al.}(2013){Ossenkopf}, {R{\"o}llig}, {Neufeld},
  {Pilleri}, {Lis}, {Fuente}, {van der Tak}, \& {Bergin}}]{Ossenkopf_2013}
{Ossenkopf}, V., {R{\"o}llig}, M., {Neufeld}, D.~A., {et~al.} 2013, \aap, 550,
  A57

\bibitem[{{Parise} {et~al.}(2009){Parise}, {Leurini}, {Schilke}, {Roueff},
  {Thorwirth}, \& {Lis}}]{Parise_2009}
{Parise}, B., {Leurini}, S., {Schilke}, P., {et~al.} 2009, \aap, 508, 737

\bibitem[{{Parmar} {et~al.}(1991){Parmar}, {Lacy}, \&
  {Achtermann}}]{Parmar_1991}
{Parmar}, P.~S., {Lacy}, J.~H., \& {Achtermann}, J.~M. 1991, \apjl, 372, L25

\bibitem[{{Pavone} {et~al.}(1990){Pavone}, {Zink}, {Prevedelli}, {Inguscio}, \&
  {Fusina}}]{Pavone_1990}
{Pavone}, F.~S., {Zink}, L.~R., {Prevedelli}, M., {Inguscio}, M., \& {Fusina},
  L. 1990, Journal of Molecular Spectroscopy, 144, 45

\bibitem[{{Pickett} {et~al.}(1998){Pickett}, {Poynter}, {Cohen}, {Delitsky},
  {Pearson}, \& {M{\"u}ller}}]{Pickett_1998}
{Pickett}, H.~M., {Poynter}, R.~L., {Cohen}, E.~A., {et~al.} 1998, \jqsrt, 60,
  883

\bibitem[{{Rabli} \& {Flower}(2010)}]{Rabli_2010}
{Rabli}, D. \& {Flower}, D.~R. 2010, \mnras, 406, 95

\bibitem[{{Remijan} {et~al.}(2006){Remijan}, {Hollis}, {Snyder}, {Jewell}, \&
  {Lovas}}]{Remijan_2006}
{Remijan}, A.~J., {Hollis}, J.~M., {Snyder}, L.~E., {Jewell}, P.~R., \&
  {Lovas}, F.~J. 2006, \apjl, 643, L37

\bibitem[{{Requena-Torres} {et~al.}(2006){Requena-Torres},
  {Mart{\'{\i}}n-Pintado}, {Rodr{\'{\i}}guez-Franco}, {Mart{\'{\i}}n},
  {Rodr{\'{\i}}guez-Fern{\'a}ndez}, \& {de Vicente}}]{Requena-Torres_2006}
{Requena-Torres}, M.~A., {Mart{\'{\i}}n-Pintado}, J.,
  {Rodr{\'{\i}}guez-Franco}, A., {et~al.} 2006, \aap, 455, 971

\bibitem[{{Roueff} {et~al.}(2007){Roueff}, {Parise}, \& {Herbst}}]{Roueff_2007}
{Roueff}, E., {Parise}, B., \& {Herbst}, E. 2007, \aap, 464, 245

\bibitem[{{Saito}(1972)}]{Saito_1972}
{Saito}, S. 1972, \apjl, 178, L95

\bibitem[{{Sastry} {et~al.}(1981){Sastry}, {Lees}, \& {Van der
  Linde}}]{Sastry_1981}
{Sastry}, K.~V.~L.~N., {Lees}, R.~M., \& {Van der Linde}, J. 1981, Journal of
  Molecular Spectroscopy, 88, 228

\bibitem[{{Schilke} {et~al.}(2001){Schilke}, {Pineau des For{\^e}ts},
  {Walmsley}, \& {Mart{\'{\i}}n-Pintado}}]{Schilke_2001}
{Schilke}, P., {Pineau des For{\^e}ts}, G., {Walmsley}, C.~M., \&
  {Mart{\'{\i}}n-Pintado}, J. 2001, \aap, 372, 291

\bibitem[{{Simon} {et~al.}(1997){Simon}, {Stutzki}, {Sternberg}, \&
  {Winnewisser}}]{Simon_1997}
{Simon}, R., {Stutzki}, J., {Sternberg}, A., \& {Winnewisser}, G. 1997, \aap,
  327, L9

\bibitem[{{Snyder} {et~al.}(1969){Snyder}, {Buhl}, {Zuckerman}, \&
  {Palmer}}]{Snyder_1969}
{Snyder}, L.~E., {Buhl}, D., {Zuckerman}, B., \& {Palmer}, P. 1969, Physical
  Review Letters, 22, 679

\bibitem[{{Stoerzer} {et~al.}(1995){Stoerzer}, {Stutzki}, \&
  {Sternberg}}]{Stoerzer_1995}
{Stoerzer}, H., {Stutzki}, J., \& {Sternberg}, A. 1995, \aap, 296, L9

\bibitem[{{Tercero} {et~al.}(2015){Tercero}, {Cernicharo}, {L{\'o}pez},
  {Brouillet}, {Kolesnikov{\'a}}, {Motiyenko}, {Margul{\`e}s}, {Alonso}, \&
  {Guillemin}}]{Tercero_2015}
{Tercero}, B., {Cernicharo}, J., {L{\'o}pez}, A., {et~al.} 2015, \aap, 582, L1

\bibitem[{{Tercero} {et~al.}(2010){Tercero}, {Cernicharo}, {Pardo}, \&
  {Goicoechea}}]{Tercero_2010}
{Tercero}, B., {Cernicharo}, J., {Pardo}, J.~R., \& {Goicoechea}, J.~R. 2010,
  \aap, 517, A96

\bibitem[{{Tercero} {et~al.}(2013){Tercero}, {Kleiner}, {Cernicharo}, {Nguyen},
  {L{\'o}pez}, \& {Mu{\~n}oz Caro}}]{Tercero_2013}
{Tercero}, B., {Kleiner}, I., {Cernicharo}, J., {et~al.} 2013, \apjl, 770, L13

\bibitem[{{Thorwirth} {et~al.}(2000){Thorwirth}, {M{\"u}ller}, \&
  {Winnewisser}}]{Thorwirth_2000}
{Thorwirth}, S., {M{\"u}ller}, H.~S.~P., \& {Winnewisser}, G. 2000, Journal of
  Molecular Spectroscopy, 204, 133

\bibitem[{{Tielens} \& {Hollenbach}(1985{\natexlab{a}})}]{Tielens_1985b}
{Tielens}, A.~G.~G.~M. \& {Hollenbach}, D. 1985{\natexlab{a}}, \apj, 291, 747

\bibitem[{{Tielens} \& {Hollenbach}(1985{\natexlab{b}})}]{Tielens_1985a}
{Tielens}, A.~G.~G.~M. \& {Hollenbach}, D. 1985{\natexlab{b}}, \apj, 291, 722

\bibitem[{{Tielens} {et~al.}(1993){Tielens}, {Meixner}, {van der Werf},
  {Bregman}, {Tauber}, {Stutzki}, \& {Rank}}]{Tielens_1993}
{Tielens}, A.~G.~G.~M., {Meixner}, M.~M., {van der Werf}, P.~P., {et~al.} 1993,
  Science, 262, 86

\bibitem[{{Trevi{\~n}o-Morales} {et~al.}(2014){Trevi{\~n}o-Morales}, {Pilleri},
  {Fuente}, {Kramer}, {Roueff}, {Gonz{\'a}lez-Garc{\'{\i}}a}, {Cernicharo},
  {Gerin}, {Goicoechea}, {Pety}, {Bern{\'e}}, {Ossenkopf}, {Ginard},
  {Garc{\'{\i}}a-Burillo}, {Rizzo}, \& {Viti}}]{Trevino-Morales_2014}
{Trevi{\~n}o-Morales}, S.~P., {Pilleri}, P., {Fuente}, A., {et~al.} 2014, \aap,
  569, A19

\bibitem[{{Turner}(1971)}]{Turner_1971}
{Turner}, B.~E. 1971, \apjl, 163, L35

\bibitem[{{{\v S}ime{\v c}kov{\'a}} {et~al.}(2004){{\v S}ime{\v c}kov{\'a}},
  {Urban}, {Fuchs}, {Lewen}, {Winnewisser}, {Morino}, \&
  {Yamada}}]{Simeckova_2004}
{{\v S}ime{\v c}kov{\'a}}, M., {Urban}, {\v S}., {Fuchs}, U., {et~al.} 2004,
  Journal of Molecular Spectroscopy, 226, 123

\bibitem[{{van der Tak} {et~al.}(2013){van der Tak}, {Nagy}, {Ossenkopf},
  {Makai}, {Black}, {Faure}, {Gerin}, \& {Bergin}}]{van_der_Tak_2013}
{van der Tak}, F.~F.~S., {Nagy}, Z., {Ossenkopf}, V., {et~al.} 2013, \aap, 560,
  A95

\bibitem[{{van der Werf} {et~al.}(2013){van der Werf}, {Goss}, \&
  {O'Dell}}]{vanderWerf_2013}
{van der Werf}, P.~P., {Goss}, W.~M., \& {O'Dell}, C.~R. 2013, \apj, 762, 101

\bibitem[{{van der Werf} {et~al.}(1996){van der Werf}, {Stutzki}, {Sternberg},
  \& {Krabbe}}]{vanderWerf_1996}
{van der Werf}, P.~P., {Stutzki}, J., {Sternberg}, A., \& {Krabbe}, A. 1996,
  \aap, 313, 633

\bibitem[{{van der Wiel} {et~al.}(2009){van der Wiel}, {van der Tak},
  {Ossenkopf}, {Spaans}, {Roberts}, {Fuller}, \& {Plume}}]{Wiel_2009}
{van der Wiel}, M.~H.~D., {van der Tak}, F.~F.~S., {Ossenkopf}, V., {et~al.}
  2009, \aap, 498, 161

\bibitem[{{van Dishoeck} {et~al.}(1995){van Dishoeck}, {Blake}, {Jansen}, \&
  {Groesbeck}}]{vanDishoeck_1995}
{van Dishoeck}, E.~F., {Blake}, G.~A., {Jansen}, D.~J., \& {Groesbeck}, T.~D.
  1995, \apj, 447, 760

\bibitem[{{Wakelam} {et~al.}(2012){Wakelam}, {Herbst}, {Loison}, {Smith},
  {Chandrasekaran}, {Pavone}, {Adams}, {Bacchus-Montabonel}, {Bergeat},
  {B{\'e}roff}, {Bierbaum}, {Chabot}, {Dalgarno}, {van Dishoeck}, {Faure},
  {Geppert}, {Gerlich}, {Galli}, {H{\'e}brard}, {Hersant}, {Hickson},
  {Honvault}, {Klippenstein}, {Le Picard}, {Nyman}, {Pernot}, {Schlemmer},
  {Selsis}, {Sims}, {Talbi}, {Tennyson}, {Troe}, {Wester}, \&
  {Wiesenfeld}}]{Wakelam_2012}
{Wakelam}, V., {Herbst}, E., {Loison}, J.-C., {et~al.} 2012, \apjs, 199, 21

\bibitem[{{Walmsley} {et~al.}(2000){Walmsley}, {Natta}, {Oliva}, \&
  {Testi}}]{Walmsley_2000}
{Walmsley}, C.~M., {Natta}, A., {Oliva}, E., \& {Testi}, L. 2000, \aap, 364,
  301

\bibitem[{{Wernli} {et~al.}(2007){Wernli}, {Wiesenfeld}, {Faure}, \&
  {Valiron}}]{Wernli_2007}
{Wernli}, M., {Wiesenfeld}, L., {Faure}, A., \& {Valiron}, P. 2007, \aap, 464,
  1147

\bibitem[{{Wilson} \& {Rood}(1994)}]{Wilson_1994}
{Wilson}, T.~L. \& {Rood}, R. 1994, \araa, 32, 191

\bibitem[{{Xu} {et~al.}(2008){Xu}, {Fisher}, {Lees}, {Shi}, {Hougen},
  {Pearson}, {Drouin}, {Blake}, \& {Braakman}}]{Xu_2008}
{Xu}, L.-H., {Fisher}, J., {Lees}, R.~M., {et~al.} 2008, Journal of Molecular
  Spectroscopy, 251, 305

\bibitem[{{Yamada} {et~al.}(1995){Yamada}, {Moravec}, \&
  {Winnewisser}}]{Yamada_1995}
{Yamada}, K.~M.~T., {Moravec}, A., \& {Winnewisser}, G. 1995, Zeitschrift
  Naturforschung Teil A, 50, 1179

\bibitem[{{Young Owl} {et~al.}(2000){Young Owl}, {Meixner}, {Wolfire},
  {Tielens}, \& {Tauber}}]{YoungOwl_2000}
{Young Owl}, R.~C., {Meixner}, M.~M., {Wolfire}, M., {Tielens}, A.~G.~G.~M., \&
  {Tauber}, J. 2000, \apj, 540, 886

\bibitem[{{Ziurys} \& {McGonagle}(1993)}]{Ziurys_1993}
{Ziurys}, L.~M. \& {McGonagle}, D. 1993, \apjs, 89, 155

\end{thebibliography}


\newpage
\appendix

\section{Possible line emission contamination from the telescope side lobes} \label{errorbeam}

\begin{figure}
\centering
\includegraphics[scale=0.55,angle=0]{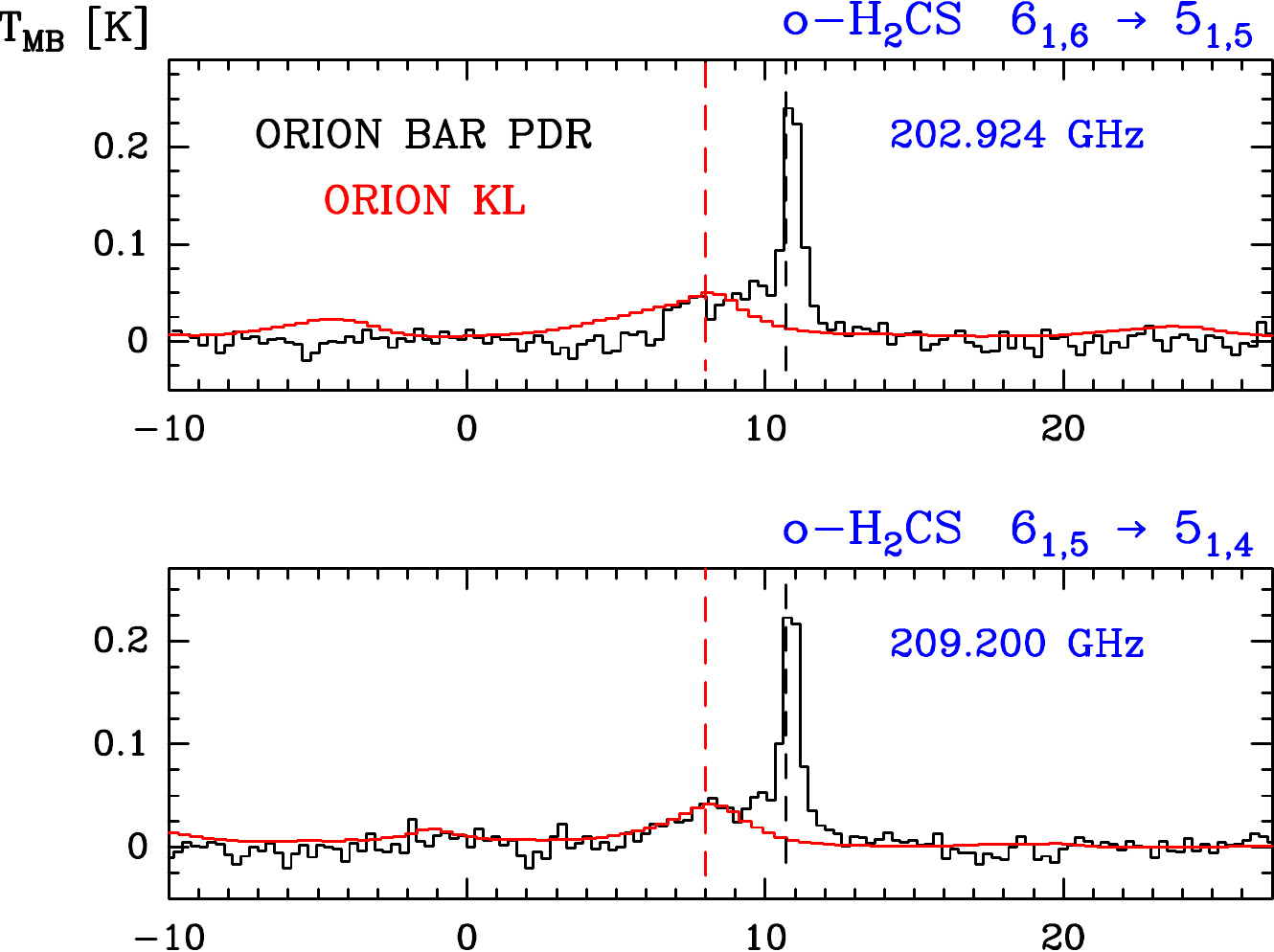}\vspace{0.3cm}
\includegraphics[scale=0.55,angle=0]{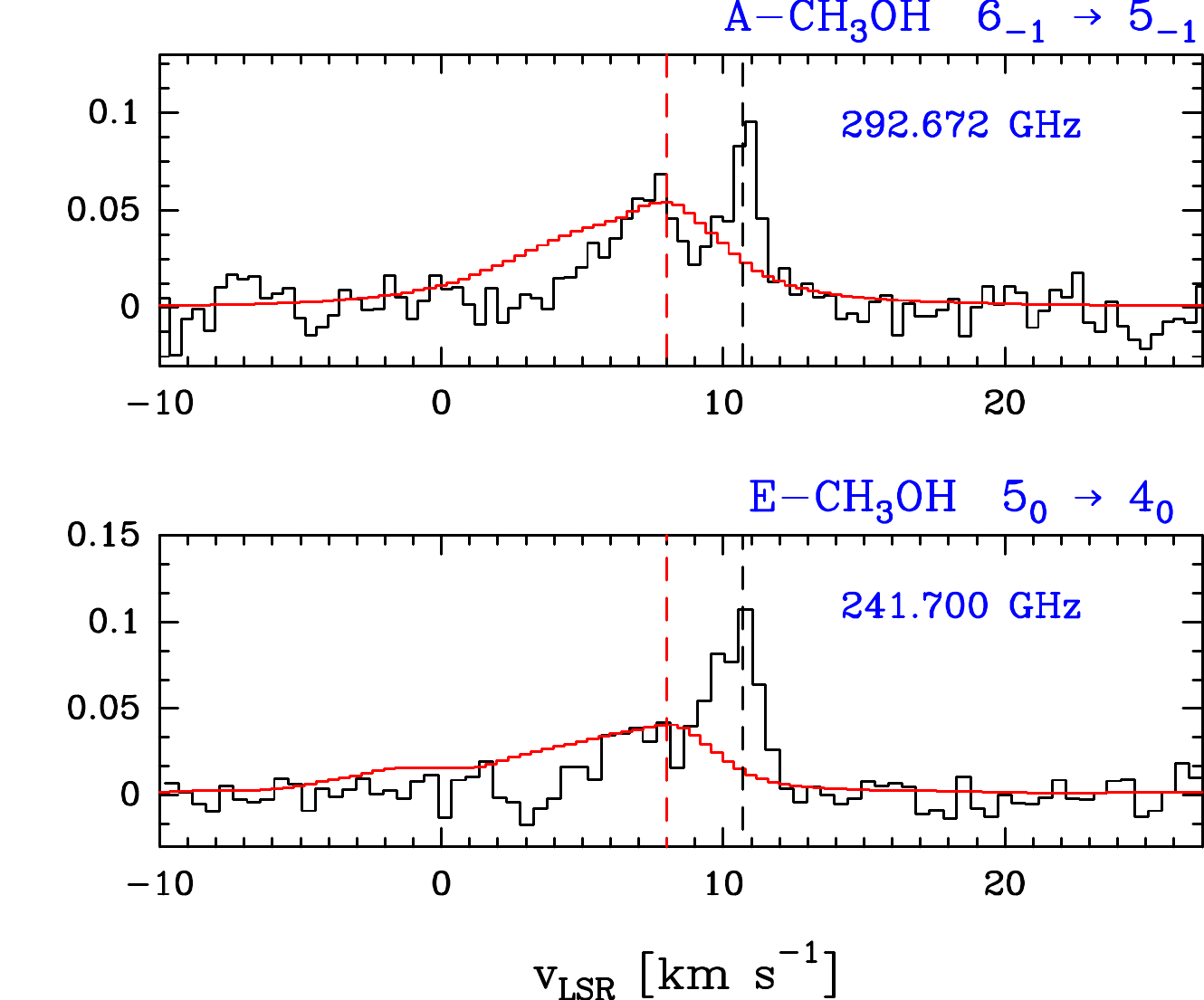}
\caption{Orion Bar spectra (black) at different frequencies in the 1\,mm range. 
For each line, 
the spectrum towards Orion~BN/KL is also shown in red \citep{Tercero_2010,Tercero_2015}.
The CH$_{3}$OH spectra towards Orion~BN/KL have been divided by $\sim$400, and the H$_{2}$CS spectra by $\sim$200. The black and red dashed lines indicate the LSR
velocity of the Orion Bar PDR \mbox{(10.7 km s$^{-1}$)} and Orion extended ridge and/or south hot core \mbox{($\sim$8 km s$^{-1}$}, see e.g. \citealt{Cernicharo_2016}), respectively.
}
\label{fig:contam_OrionKL}
\end{figure}

In order to determine whether or not the emission from the bright Orion BN/KL region  contributes to our detected signal, we compared the 1~mm spectrum of the Orion Bar with that of Orion~BN/KL \citep[located at $\sim$2$'$ north of the Bar, e.g.][]{Tercero_2010}. Figure~\ref{fig:contam_OrionKL} shows several H$_{2}$CS and CH$_{3}$OH lines observed towards the Bar (black histogram) and Orion BN/KL (red histogram; see \citealt{Tercero_2010,Tercero_2015}). We note that the amplitude of Orion BN/KL spectra are divided by a given value to match the line intensities from the Bar. This comparison shows that while the line emission from the Orion Bar peaks at \mbox{$v_{\rm LSR}$ $\simeq$ 10.7 km s$^{-1}$}, and can be fully attributed to gas in the PDR \citep{Goicoechea_2016}, the blue-shifted shoulder emission seen in some lines at  \mbox{$\sim$8 km s$^{-1}$} is likely
produced by the intense emission from Orion BN/KL, and perhaps by the extended Orion cloud component (for the low-excitation lines). This emission is detected through the telescope extended side lobes and results in a blue-shifted emission shoulder in some spectra towards the Bar.
For the IRAM~30~m telescope, the contribution of the secondary lobes increases with frequency \citep{Greve_1998}. Given the specific emission velocity and narrow line widths of the emission from the Bar, and the fact that this effect is stronger at $\sim$1~mm, we conclude that our line assignments and line intensity extraction in the PDR is correct.

\section{Identified lines of complex organic molecules} \label{Tables}

A summary of the main line spectroscopic parameters is presented in Tables \ref{Table_HCO}-\ref{Table_CH3CHO}. 
Line frequency (in MHz), 
energy of the upper level of each transition (\mbox{$E_{\rm u}/k$} in K), 
Einstein coefficient for spontaneous emission (\mbox{$A_{\rm ul}$} in \mbox{s$^{-1}$}), 
intrinsic line strength \mbox{($S_{\rm ul}$)}, 
and the level degeneracy \mbox{($g_{\rm u}$)} 
from MADEX spectral catalogue, and JPL and CDMS molecular databases are shown.
The velocity-integrated line intensity \mbox{($\int T_{_{\rm MB}}$d$v$} in \mbox{mK km s$^{-1}$)}, 
LSR velocity ($v_{_{\rm LSR}}$ in km s$^{-1}$), 
FWHM line width \mbox{($\Delta$$v$}  
in \mbox{km s$^{-1}$)}, and the line peak temperature ($T_{_{\rm MB}}$ in \mbox{mK)} 
were obtained from Gaussian fits. Parentheses indicate the uncertainty. 
When two or more transitions were found to overlap, the total profile was fitted. 
Fully overlapping transitions are marked with connecting symbols in the tables.

     
      \begin{table*}[!h] 
      \begin{center}
      \caption{Line parameters of HCO.}  \label{Table_HCO}  
      \begin{tabular}{c c r c c c c c c c c@{\vrule height 10pt depth 5pt width 0pt}}     
      \hline\hline       
      
  Transition &  Frequency &  $E_{\rm u}/k$ &  $A_{\rm ul}$ &  $S_{\rm ul}$ &  $g_{\rm u}$ &  $\displaystyle{\int} T_{_{\rm MB}}$d$v$ &  $v_{_{\rm LSR}}$ &  $\Delta v$ &  $T_{_{\rm MB}}$ & $S/N$ \rule[-0.3cm]{0cm}{0.8cm}\ \\ \cline{1-1}

 $(N_{K_{\rm a},K_{\rm c}},J,F)_{\rm u} \rightarrow (N_{K_{\rm a},K_{\rm c}},J,F)_{\rm l}$   & [MHz] & [K] & [$\mathrm{s^{-1}}$] & & & [$\mathrm{mK\, km\, s^{-1}}$] & [$\mathrm{km\, s^{-1}}$] & [$\mathrm{km\, s^{-1}}$]  & [$\mathrm{mK}$]  & \\
         
          \hline

    1$_{0,1}$, 3/2, 2 $\rightarrow$ 0$_{0,0}$, 1/2, 1 &  86670.760$^F$  &   4.2 & 4.68 $\times$ 10$^{-6}$  & 1.66 &  5 &  497(12)  &  10.5($<$0.1)  &   3.1(0.1)   & 149  &  52  \\                                                                                                                                                                                                                                                                                        
    1$_{0,1}$, 3/2, 1 $\rightarrow$ 0$_{0,0}$, 1/2, 0 &  86708.360$^F$  &   4.2 & 4.59 $\times$ 10$^{-6}$  & 0.98 &  3 &  286(7)   &  10.5($<$0.1)  &   2.8(0.1)   & 95   &  33  \\                                                                                                                                                                                                                                                                                        
    1$_{0,1}$, 1/2, 1 $\rightarrow$ 0$_{0,0}$, 1/2, 1 &  86777.460$^F$  &   4.2 & 4.60 $\times$ 10$^{-6}$  & 0.98 &  3 &  294(7)   &  10.5($<$0.1)  &   3.0(0.1)   & 92   &  32 \\                                                                                                                                                                                                                                                                                       
    1$_{0,1}$, 1/2, 0 $\rightarrow$ 0$_{0,0}$, 1/2, 1 &  86805.780$^F$  &   4.2 & 4.71 $\times$ 10$^{-6}$  & 0.33 &  1 &   75(6)   &  10.5(0.1)     &   2.3(0.2)   & 31   &  10 \\                                                                                                                                                                                                                                                                                        
    2$_{0,2}$, 5/2, 3 $\rightarrow$ 1$_{0,1}$, 3/2, 2 & 173377.377$^W$  &  12.5 & 4.50 $\times$ 10$^{-5}$  & 2.80 &  7 &  683(86)  &    ---         &     ---      & ---    &  8  \\                                                                                                                                                                                                                                                                                        
    2$_{0,2}$, 5/2, 2 $\rightarrow$ 1$_{0,1}$, 3/2, 1 & 173406.082$^W$  &  12.5 & 4.42 $\times$ 10$^{-5}$  & 1.96 &  5 &  346(70)  &    ---         &     ---      & ---    &  6  \\                                                                                                                                                                                                                                                                                       
    2$_{0,2}$, 3/2, 2 $\rightarrow$ 1$_{0,1}$, 1/2, 1 & 173443.065$^W$  &  12.5 & 3.38 $\times$ 10$^{-5}$  & 1.50 &  5 &  409(94)  &    ---         &     ---      & ---    &  5  \\                                                                                                                                                                                                                                                                                        
    2$_{0,2}$, 3/2, 1 $\rightarrow$ 1$_{0,1}$, 1/2, 0 & 173474.400$^W$  &  12.5 & 2.50 $\times$ 10$^{-5}$  & 0.66 &  3 &  185(79)  &    ---         &     ---      & ---    &  4  \\                                                                                                                                                                                                                                                                                        
    3$_{0,3}$, 7/2, 4 $\rightarrow$ 2$_{0,2}$, 5/2, 3 & 260060.329$^F$  &  25.0 & 1.63 $\times$ 10$^{-4}$  & 3.85 &  9 &  576(15)  &  10.5($<$0.1)  &   2.0(0.1)   & 275  &  23   \\                                                                                                                                                                                                                                                                                       
    3$_{0,3}$, 7/2, 3 $\rightarrow$ 2$_{0,2}$, 5/2, 2 & 260082.192$^F$  &  25.0 & 1.60 $\times$ 10$^{-4}$  & 2.95 &  7 &  446(15)  &  10.5($<$0.1)  &   2.0(0.1)   & 209  &  16  \\                                                                                                                                                                                                                                                                                       
    3$_{0,3}$, 5/2, 3 $\rightarrow$ 2$_{0,2}$, 3/2, 2 & 260133.586$^F$  &  25.0 & 1.45 $\times$ 10$^{-4}$  & 2.67 &  7 &  347(17)  &  10.6($<$0.1)  &   1.7(0.4)   & 195  &  15  \\                                                                                                                                                                                                                                                                                       
    3$_{0,3}$, 5/2, 2 $\rightarrow$ 2$_{0,2}$, 3/2, 1 & 260155.769$^W$  &  25.0 & 1.37 $\times$ 10$^{-4}$  & 1.80 &  5 &  382(25)  &    ---         &     ---      & ---    &  19  \\                                                                                                                                                                                                                                                                                                                                                                                                                                                                                                                                                                                                                                                   
    4$_{0,4}$, 9/2, 5 $\rightarrow$ 3$_{0,3}$, 7/2, 4 & 346708.493$^F$  &  41.6 & 3.99 $\times$ 10$^{-4}$  & 4.88 & 11 &  437(38)  &  10.6(0.1)     &   2.6(0.3)   & 156  &  7  \\                                                                                                                                                                                                                                                                                       
    4$_{0,4}$, 9/2, 4 $\rightarrow$ 3$_{0,3}$, 7/2, 3 & 346725.172$^F$  &  41.6 & 3.95 $\times$ 10$^{-4}$  & 3.95 &  9 &  308(36)  &  10.6(0.1)     &   2.0(0.3)   & 142  &  7  \\                                                                                                                                                                                                                                                                                       
    4$_{0,4}$, 7/2, 4 $\rightarrow$ 3$_{0,3}$, 5/2, 3 & 346787.898$^F$  &  41.6 & 3.76 $\times$ 10$^{-4}$  & 3.76 &  9 &  362(41)  &  10.5(0.2)     &   3.0(0.5)   & 112  &  6   \\                                                                                                                                                                                                                                                                                       
    4$_{0,4}$, 7/2, 3 $\rightarrow$ 3$_{0,3}$, 5/2, 2 & 346804.597$^F$  &  41.6 & 3.67 $\times$ 10$^{-4}$  & 2.85 &  7 &  223(29)  &  10.6(0.1)     &   1.6(0.3)   & 127  &  7  \\                                                                                                                                                                                                                                                                                   
                                                                                                                                                                                                                                                                                                                                                                       
      \hline                                                                                                                                                                                                                                                                                                                                                                                    
                                                                                                                                                                                                                                                                                                                                                                                                                                                                           
     \end{tabular}                                                                                                                                                                                                                                                                                                                                                                                                                                                                                                         
      \end{center}   
      \tablefoot{Frequencies, \mbox{$E_{\rm u}/k$}, \mbox{$A_{\rm ul}$}, \mbox{$S_{\rm ul}$}, and \mbox{$g_{\rm u}$} from JPL catalogue.
      {\bf Labels}: $^F$ Detected with FTS backend. $^W$ The lines detected with  WILMA
      backend just give information about the integrated line intensity (see \citealt{Cuadrado_2015a}). $^{*}$ Symmetry (ortho-para or E-A).
      }
      \end{table*}


        \begin{table*}[!h] 
        \begin{center}
         \vspace*{0.5cm}
        \caption{Line parameters of H$_{2}$CO.}  \label{Table_H2CO}  
\begin{tabular}{ c c c r c c c c c c c c@{\vrule height 10pt depth 5pt width 0pt}}      
        \hline\hline       
        
   Transition & Sym.$^{*}$ & Frequency &  $E_{\rm u}/k$ &  $A_{\rm ul}$ &  $S_{\rm ul}$ &  $g_{\rm u}$ &  $\displaystyle{\int} T_{_{\rm MB}}$d$v$ &  $v_{_{\rm LSR}}$ &  $\Delta v$ &  $T_{_{\rm MB}}$ & $S/N$ \rule[-0.3cm]{0cm}{0.8cm}\ \\ \cline{1-1}

      $(J_{K_{\rm a},K_{\rm c}})_{\rm u} \rightarrow (J_{K_{\rm a},K_{\rm c}})_{\rm l}$  &  & [MHz] & [K] & [$\mathrm{s^{-1}}$] & & & [$\mathrm{mK\, km\, s^{-1}}$] & [$\mathrm{km\, s^{-1}}$] & [$\mathrm{km\, s^{-1}}$]  & [$\mathrm{mK}$]  & \\
       
        \hline

  2$_{1,2}\rightarrow$ 1$_{1,1}$  &   $ortho$ & 140839.516$^W$  &   6.8 &  5.30 $\times$ 10$^{-5}$ &  1.5 &  5  &  13488(28)  &  ---        &       ---  &    --- & 423    \\ 
  2$_{0,2}\rightarrow$ 1$_{0,1}$  &   $para $ & 145602.951$^W$  &  10.5 &  7.80 $\times$ 10$^{-5}$ &  2.0 &  5  &   5611(67)  &  ---        &       ---  &    --- &  95   \\ 
  2$_{1,1}\rightarrow$ 1$_{1,0}$  &   $ortho$ & 150498.335$^W$  &   7.5 &  6.46 $\times$ 10$^{-5}$ &  1.5 &  5  &  10850(97)  &  ---        &       ---  &    --- & 208   \\
  3$_{1,3}\rightarrow$ 2$_{1,2}$  &   $ortho$ & 211211.449$^F$  &  16.9 &  2.27 $\times$ 10$^{-4}$ &  2.7 &  7  &  13732(26)  &  10.6(0.1)  &   2.3(0.1) &  5519  & 338    \\ 
  3$_{0,3}\rightarrow$ 2$_{0,2}$  &   $para $ & 218222.187$^F$  &  21.0 &  2.81 $\times$ 10$^{-4}$ &  3.0 &  7  &   6567(22)  &  10.6(0.1)  &   2.2(0.1) &  2771  & 311    \\ 
  3$_{2,2}\rightarrow$ 2$_{2,1}$  &   $para $ & 218475.634$^F$  &  68.1 &  1.57 $\times$ 10$^{-4}$ &  1.7 &  7  &   2600(9)   &  10.7(0.1)  &   2.1(0.1) &  1148  & 175    \\
\hline

       \end{tabular}     
       \end{center}    
                 \tablefoot{Frequencies, \mbox{$E_{\rm u}/k$}, \mbox{$A_{\rm ul}$}, \mbox{$S_{\rm ul}$}, and \mbox{$g_{\rm u}$} from MADEX code, 
                 that fit to all rotational lines reported by \citet{Bocquet_1996}, \citet{Brunken_2003}, and \citet{Eliet_2012}. }
       \end{table*}


       \begin{table*}[!h] 
        \begin{center}
          \vspace{0.5cm}
        \caption{Line parameters of H$_{2}^{13}$CO.}  \label{Table_H2-13CO}  
        \begin{tabular}{ c c c r c c c c c c c c@{\vrule height 10pt depth 5pt width 0pt}}    
        \hline\hline       
        
   Transition & Sym.$^{*}$ & Frequency &  $E_{\rm u}/k$ &  $A_{\rm ul}$ &  $S_{\rm ul}$ &  $g_{\rm u}$ &  $\displaystyle{\int} T_{_{\rm MB}}$d$v$ &  $v_{_{\rm LSR}}$ &  $\Delta v$ &  $T_{_{\rm MB}}$ & $S/N$ \rule[-0.3cm]{0cm}{0.8cm}\ \\ \cline{1-1}

      $(J_{K_{\rm a},K_{\rm c}})_{\rm u} \rightarrow (J_{K_{\rm a},K_{\rm c}})_{\rm l}$   &  & [MHz] & [K] & [$\mathrm{s^{-1}}$] & & & [$\mathrm{mK\, km\, s^{-1}}$] & [$\mathrm{km\, s^{-1}}$] & [$\mathrm{km\, s^{-1}}$]  & [$\mathrm{mK}$]  & \\
       
        \hline    
     2$_{1,2}\rightarrow$ 1$_{1,1}$  &   $ortho$ & 137449.954$^W$  &   6.6 &  4.92 $\times$ 10$^{-5}$ &  1.5 &  5  &   298(53)  &     ---     &     ---    &  --- & 7            \\ 
     2$_{1,1}\rightarrow$ 1$_{1,0}$  &   $ortho$ & 146635.669$^W$  &   7.3 &  5.98 $\times$ 10$^{-5}$ &  1.5 &  5  &    73(29)  &     ---     &     ---    &  --- & 5            \\
     3$_{1,3}\rightarrow$ 2$_{1,2}$  &   $ortho$ & 206131.619$^F$  &  16.5 &  2.11 $\times$ 10$^{-4}$ &  2.6 &  7  &   251(12)  &  10.5(0.1)  &   2.4(0.1) &  97  & 11     \\ 
     3$_{0,3}\rightarrow$ 2$_{0,2}$  &   $para $ & 212811.190$^F$  &  20.4 &  2.61 $\times$ 10$^{-4}$ &  3.0 &  7  &    84(13)  &  10.3(0.1)  &   2.1(0.4) &  37  & 4     \\ 
     3$_{1,2}\rightarrow$ 2$_{1,1}$  &   $ortho$ & 219908.481$^F$  &  17.8 &  2.56 $\times$ 10$^{-4}$ &  2.7 &  7  &   151(12)  &  10.7(0.1)  &   2.0(0.2) &  70  & 7      \\
     4$_{1,4}\rightarrow$ 3$_{1,3}$  &   $ortho$ & 274762.103$^F$  &  29.7 &  5.46 $\times$ 10$^{-4}$ &  3.8 &  9  &   180(16)  &  10.8(0.1)  &   2.3(0.3) &  74  & 6      \\
\hline      
       \end{tabular}  
       \end{center} 
       \tablefoot{Frequencies, \mbox{$E_{\rm u}/k$}, \mbox{$A_{\rm ul}$}, \mbox{$S_{\rm ul}$}, and \mbox{$g_{\rm u}$} from MADEX code, 
                 that fit to all rotational lines reported by \citet{Muller_2000b}.}

       \end{table*}


         \begin{table*}[!h] 
         \begin{center}
         \caption{Line parameters of H$_{2}$CS.}  \label{Table_H2CS}  
         \begin{tabular}{ c c c r c c c c c c c c@{\vrule height 10pt depth 5pt width 0pt}}      
         \hline\hline       
         
   Transition & Sym.$^{*}$ & Frequency &  $E_{\rm u}/k$ &  $A_{\rm ul}$ &  $S_{\rm ul}$ &  $g_{\rm u}$ &  $\displaystyle{\int} T_{_{\rm MB}}$d$v$ &  $v_{_{\rm LSR}}$ &  $\Delta v$ &  $T_{_{\rm MB}}$ & S/N \rule[-0.3cm]{0cm}{0.8cm}\ \\ \cline{1-1}

       $(J_{K_{\rm a},K_{\rm c}})_{\rm u} \rightarrow (J_{K_{\rm a},K_{\rm c}})_{\rm l}$  &  & [MHz] & [K] & [$\mathrm{s^{-1}}$] & & & [$\mathrm{mK\, km\, s^{-1}}$] & [$\mathrm{km\, s^{-1}}$] & [$\mathrm{km\, s^{-1}}$]  & [$\mathrm{mK}$] &  \\
        
         \hline

      3$_{1,3}\rightarrow$ 2$_{1,2}$    &  $ortho$   &   101477.810$^F$   &      8.1  &    1.26 $\times$ 10$^{-5}$   &   2.67  &    7  &     79(8)   &    10.8(0.1)     &    1.1(0.1)  &  68    & 10    \\
      3$_{2,2}\rightarrow$ 2$_{2,1}$    &  $para$    &   103039.907$^F$   &     62.6  &    8.25 $\times$ 10$^{-6}$   &   1.67  &    7  &   \multirow{2}{*}[0cm]{$\Big \rangle$ 53(5)}   &  \multirow{2}{*}[-0.01cm]{10.9(0.1)}   &   \multirow{2}{*}[-0.01cm]{1.8(0.2)}    &   \multirow{2}{*}[-0.01cm]{28}        &  \multirow{2}{*}[-0.01cm]{9}    \\
      3$_{0,3}\rightarrow$ 2$_{0,2}$    &  $para$    &   103040.452$^F$   &      9.9  &    1.48 $\times$ 10$^{-5}$   &   3.00  &    7  &                 &         &      &      &    \\                              
      3$_{2,1}\rightarrow$ 2$_{2,0}$    &  $para$    &   103051.847$^F$   &     62.6  &    8.25 $\times$ 10$^{-6}$   &   1.67  &    7  &     19(7)   &    10.5(0.3)     &    1.5(1.3)  &  10     &  3    \\
      3$_{1,2}\rightarrow$ 2$_{1,1}$    &  $ortho$   &   104617.040$^F$   &      8.4  &    1.38 $\times$ 10$^{-5}$   &   2.67  &    7  &     85(5)   &    10.7($<$0.1)  &    1.3(0.1)  &  59    & 14     \\
\hline                                                                                                   
        \end{tabular}                                     
        \end{center}  
        
        \end{table*}

         \begin{table*}[!h] 
         \begin{center}
           \setcounter{table}{3}
         \caption{continued.}  
         \begin{tabular}{ c c c r c c c c c c c c@{\vrule height 10pt depth 5pt width 0pt}}    
         \hline\hline       
         
   Transition & Sym.$^{*}$ & Frequency &  $E_{\rm u}/k$ &  $A_{\rm ul}$ &  $S_{\rm ul}$ &  $g_{\rm u}$ &  $\displaystyle{\int} T_{_{\rm MB}}$d$v$ &  $v_{_{\rm LSR}}$ &  $\Delta v$ &  $T_{_{\rm MB}}$ & S/N \rule[-0.3cm]{0cm}{0.8cm}\ \\ \cline{1-1}

       $(J_{K_{\rm a},K_{\rm c}})_{\rm u} \rightarrow (J_{K_{\rm a},K_{\rm c}})_{\rm l}$  &  & [MHz] & [K] & [$\mathrm{s^{-1}}$] & & & [$\mathrm{mK\, km\, s^{-1}}$] & [$\mathrm{km\, s^{-1}}$] & [$\mathrm{km\, s^{-1}}$]  & [$\mathrm{mK}$] &  \\
        
         \hline

      7$_{1,6}\rightarrow$ 6$_{1,5}$    &  $ortho$   &   244048.504$^F$   &     45.2  &    2.10 $\times$ 10$^{-4}$   &   6.86  &   15  &    185(11)  &    10.9($<$0.1)  &    0.8(0.1)  &  260   & 14   \\
      8$_{1,8}\rightarrow$ 7$_{1,7}$    &  $ortho$   &   270521.931$^F$   &     56.8  &    2.90 $\times$ 10$^{-4}$   &   7.88  &   17  &    128(16)  &    10.8(0.1)     &    0.9(0.2)  &  139   &  6   \\
      8$_{0,8}\rightarrow$ 7$_{0,7}$    &  $para$    &   274521.931$^F$   &     59.3  &    3.08 $\times$ 10$^{-4}$   &   8.00  &   17  &     60(9)   &    11.1(0.1)     &    0.7(0.1)  &  77    &  5   \\
      8$_{2,6}\rightarrow$ 7$_{2,5}$    &  $para$    &   274953.744$^F$   &    112.0  &    2.90 $\times$ 10$^{-4}$   &   7.50  &   17  &     36(9)   &    11.0(0.1)     &    0.9(0.2)  &  39    &  3   \\
      8$_{1,7}\rightarrow$ 7$_{1,6}$    &  $ortho$   &   278887.661$^F$   &     58.6  &    3.18 $\times$ 10$^{-4}$   &   7.87  &   17  &    195(13)  &    10.9($<$0.1)  &    0.7(0.1)  &  245   & 14    \\
      9$_{1,9}\rightarrow$ 8$_{1,8}$    &  $ortho$   &   304307.709$^F$   &     71.4  &    4.17 $\times$ 10$^{-4}$   &   8.89  &   19  &     75(8)   &    10.7(0.1)     &    0.7(0.1)  &  101   &  5   \\
\hline                                                                                                   
        \end{tabular}                                     
        \end{center}  
         \tablefoot{Frequencies, \mbox{$E_{\rm u}/k$}, \mbox{$A_{\rm ul}$}, \mbox{$S_{\rm ul}$}, and \mbox{$g_{\rm u}$} from CDMS catalogue.}
        \end{table*}

      
      \begin{table*}[!h] 
      \vspace*{1cm}
      \begin{center}
      \caption{Line parameters of HNCO.}  \label{Table_HNCO}  
      \begin{tabular}{c c r c r c c c c c c@{\vrule height 10pt depth 5pt width 0pt}}      
      \hline\hline       
      
   Transition &  Frequency &  $E_{\rm u}/k$ &  $A_{\rm ul}$ &  $S_{\rm ul}$ &  $g_{\rm u}$ &  $\displaystyle{\int} T_{_{\rm MB}}$d$v$ &  $v_{_{\rm LSR}}$ &  $\Delta v$ &  $T_{_{\rm MB}}$ & $S/N$ \rule[-0.3cm]{0cm}{0.8cm}\ \\ \cline{1-1}

 $(J_{K_{\rm a},K_{\rm c}})_{\rm u} \rightarrow (J_{K_{\rm a},K_{\rm c}})_{\rm l}$   & [MHz] & [K] & [$\mathrm{s^{-1}}$] & & & [$\mathrm{mK\, km\, s^{-1}}$] & [$\mathrm{km\, s^{-1}}$] & [$\mathrm{km\, s^{-1}}$]  & [$\mathrm{mK}$] & \\
         
          \hline
     
        4$_{0,4}\rightarrow$ 3$_{0,3}$   &   87925.237$^F$ &  10.5  & 9.03 $\times$ 10$^{-6}$ &  4.0  &  9  &    49(6)   &  10.8(0.1)  &   1.7(0.2) & 27  & 8 \\  
        5$_{0,5}\rightarrow$ 4$_{0,4}$   &  109905.749$^F$ &  15.8  & 1.80 $\times$ 10$^{-5}$ &  5.0  & 11  &   110(9)   &  10.6(0.1)  &   1.9(0.2) & 54  & 10 \\
        7$_{0,7}\rightarrow$ 6$_{0,6}$   &  153865.086$^W$ &  29.5  & 5.08 $\times$ 10$^{-5}$ &  7.0  & 15  &   125(39)  &   ---       &    ---     & ---   & 3  \\
  10$_{0,10}\rightarrow$ 9$_{0,9}$ \ \ \ &  219798.274$^F$ &  58.0  & 1.51 $\times$ 10$^{-4}$ & 10.0  & 21  &   154(16)  &  10.6(0.2)  &   3.3(0.4) & 44  & 6   \\
      11$_{0,11}\rightarrow$ 10$_{0,10}$ &  241774.032$^F$ &  69.6  & 2.02 $\times$ 10$^{-4}$ & 11.0  & 23  &   118(21)  &  10.7(0.1)  &   1.6(0.3) & 69  & 5  \\
      13$_{0,13}\rightarrow$ 12$_{0,12}$ &  285721.951$^F$ &  96.0  & 3.36 $\times$ 10$^{-4}$ & 13.0  & 27  &   129(26)  &  10.4(0.3)  &   2.8(0.7) & 43  & 5   \\
    
         \hline                                
         \end{tabular}                         
         \end{center} 
          \tablefoot{Frequencies, \mbox{$E_{\rm u}/k$}, \mbox{$A_{\rm ul}$}, \mbox{$S_{\rm ul}$}, and \mbox{$g_{\rm u}$} from CDMS catalogue.}
         \end{table*}    

        
              \begin{table*}[!h] 
               \vspace*{1cm}
             \begin{center}
             \caption{Line parameters of CH$_{2}$NH.}  \label{Table_CH2NH}  
             \begin{tabular}{c c r c c c c c c c c@{\vrule height 10pt depth 5pt width 0pt}}     
             \hline\hline       
             
  Transition &  Frequency &  $E_{\rm u}/k$ &  $A_{\rm ul}$ &  $S_{\rm ul}$ &  $g_{\rm u}$ &  $\displaystyle{\int} T_{_{\rm MB}}$d$v$ &  $v_{_{\rm LSR}}$ &  $\Delta v$ &  $T_{_{\rm MB}}$ & $S/N$ \rule[-0.3cm]{0cm}{0.8cm}\ \\ \cline{1-1}

       $(J_{K_{\rm a},K_{\rm c}})_{\rm u} \rightarrow (J_{K_{\rm a},K_{\rm c}})_{\rm l}$    & [MHz] & [K] & [$\mathrm{s^{-1}}$] & & & [$\mathrm{mK\, km\, s^{-1}}$] & [$\mathrm{km\, s^{-1}}$] & [$\mathrm{km\, s^{-1}}$]  & [$\mathrm{mK}$] &  \\
                
         \hline                                                
  4$_{0,4}\rightarrow$ 3$_{1,3}$ &   105794.062$^F$  &    30.6 &  1.62 $\times$ 10$^{-5}$ &   5.0  &  9  &       55(13) &   11.3(0.5)  &   4.2(1.3)  & 12  & 3      \\   
  1$_{1,1}\rightarrow$ 0$_{0,0}$ &   225554.609$^F$  &    10.8 &  2.79 $\times$ 10$^{-4}$ &   3.0  &  3  &      120(26) &   10.5(0.2)  &   2.9(0.9)  & 39  & 4     \\    
  4$_{1,4}\rightarrow$ 3$_{1,3}$ &   245125.866$^F$  &    37.3 &  3.85 $\times$ 10$^{-4}$ &  11.2  &  9  &      106(14) &   11.1(0.1)  &   2.3(0.4)  & 43  & 4    \\     
  6$_{0,6}\rightarrow$ 5$_{1,5}$ &   251421.265$^F$  &    64.1 &  2.73 $\times$ 10$^{-4}$ &   9.2  & 13  &       72(13) &   11.0(0.3)  &   3.1(0.6)  & 22  & 4        \\ 
  4$_{0,4}\rightarrow$ 3$_{0,3}$ &   254685.137$^F$  &    30.6 &  4.60 $\times$ 10$^{-4}$ &  12.0  &  9  &      314(12) &   10.8(0.1)  &   2.0(0.1)  & 147 & 14      \\    
  4$_{1,3}\rightarrow$ 3$_{1,2}$ &   266270.024$^F$  &    39.8 &  4.93 $\times$ 10$^{-4}$ &  11.2  &  9  &      117(12) &   10.9(0.1)  &   2.0(0.2)  & 54  & 5     \\

             \hline       
             \end{tabular}  
             \end{center} 
             \tablefoot{Frequencies, \mbox{$E_{\rm u}/k$}, \mbox{$A_{\rm ul}$}, \mbox{$S_{\rm ul}$}, and \mbox{$g_{\rm u}$} from CDMS catalogue.
            }
             \end{table*}


          \begin{table*}[!h] 
          \begin{center}
          \caption{Line parameters of H$_{2}$CCO.}  \label{Table_H2CCO}  
          \begin{tabular}{ c c c r c c c c c c c c@{\vrule height 10pt depth 5pt width 0pt}}     
          \hline\hline       
          
   Transition & Sym.$^{*}$ & Frequency &  $E_{\rm u}/k$ &  $A_{\rm ul}$ &  $S_{\rm ul}$ &  $g_{\rm u}$ &  $\displaystyle{\int} T_{_{\rm MB}}$d$v$ &  $v_{_{\rm LSR}}$ &  $\Delta v$ &  $T_{_{\rm MB}}$ & $S/N$ \rule[-0.3cm]{0cm}{0.8cm}\ \\ \cline{1-1}

       $(J_{K_{\rm a},K_{\rm c}})_{\rm u} \rightarrow (J_{K_{\rm a},K_{\rm c}})_{\rm l}$  &  & [MHz] & [K] & [$\mathrm{s^{-1}}$] & & & [$\mathrm{mK\, km\, s^{-1}}$] & [$\mathrm{km\, s^{-1}}$] & [$\mathrm{km\, s^{-1}}$]  & [$\mathrm{mK}$] &  \\
          
          \hline

         4$_{2,3}\rightarrow$ 3$_{2,2}$      & $para $    &   80820.400$^F$   &   61.9   &   4.14 $\times$ 10$^{-6}$  &    3.0   &   9  &    13(9)   &    10.9(0.7)   &    1.6(0.7)  &    8  &  2    \\
         4$_{1,3}\rightarrow$ 3$_{1,2}$      & $ortho$    &   81586.239$^F$   &    8.8   &   5.33 $\times$ 10$^{-6}$  &    3.8   &   9  &    44(8)   &    11.0(0.3)   &    2.5(0.6)  &   13  &  4      \\
         5$_{1,5}\rightarrow$ 4$_{1,4}$      & $ortho$    &  100094.511$^F$   &   13.5   &   1.03 $\times$ 10$^{-5}$  &    4.8   &  11  &    61(6)   &    10.4(0.2)   &    1.9(0.5)  &   30  &  3       \\    
         5$_{3,3}\rightarrow$ 4$_{3,2}$      & $ortho$    &  101002.349$^F$   &  117.9   &   7.06 $\times$ 10$^{-6}$  &    3.2   &  11  &   \multirow{2}{*}[0cm]{$\Big \rangle$ 25(8)}   &  \multirow{2}{*}[-0.01cm]{10.3(0.2)}   &   \multirow{2}{*}[-0.01cm]{1.0(0.3)}    &   \multirow{2}{*}[-0.01cm]{25}        &  \multirow{2}{*}[-0.01cm]{3}    \\   
         5$_{3,2}\rightarrow$ 4$_{3,1}$      & $ortho$    &  101002.354$^F$   &  117.9   &   7.06 $\times$ 10$^{-6}$  &    3.2   &  11  &       &       &      &    &        \\
         5$_{2,4}\rightarrow$ 4$_{2,3}$      & $para $    &  101024.430$^F$   &   66.7   &   9.27 $\times$ 10$^{-6}$  &    4.2   &  11  &    24(6)   &    10.2(0.2)   &    1.7(0.8)  &   15  &  2     \\
\hline                                 
         \end{tabular}  
         \end{center}  
      \tablefoot{Frequencies, \mbox{$E_{\rm u}/k$}, \mbox{$A_{\rm ul}$}, \mbox{$S_{\rm ul}$}, and \mbox{$g_{\rm u}$} from 
      MADEX code, 
                 that fit to all rotational lines reported in the CDMS database. }
     \end{table*}


            \begin{table*}[!h] 
             \begin{center}
             \caption{Line parameters of HC$_{3}$N.}  \label{Table_HC3N}  
             \begin{tabular}{c c r c c c c c c c c@{\vrule height 10pt depth 5pt width 0pt}}      
             \hline\hline       
  Transition &  Frequency &  $E_{\rm u}/k$ &  $A_{\rm ul}$ &  $S_{\rm ul}$ &  $g_{\rm u}$ &  $\displaystyle{\int} T_{_{\rm MB}}$d$v$ &  $v_{_{\rm LSR}}$ &  $\Delta v$ &  $T_{_{\rm MB}}$ & $S/N$ \rule[-0.3cm]{0cm}{0.8cm}\ \\ \cline{1-1}

        $J_{\rm u}$ $\rightarrow$ $J_{\rm l}$      & [MHz] & [K] & [$\mathrm{s^{-1}}$] & & & [$\mathrm{mK\, km\, s^{-1}}$] & [$\mathrm{km\, s^{-1}}$] & [$\mathrm{km\, s^{-1}}$]  & [$\mathrm{mK}$] &  \\

                 \hline
     
     9 $\rightarrow$  8   &   81881.462$^F$   &     19.6   &   4.22 $\times$ 10$^{-5}$  &     9.0  &    19  &     50(19)  &   10.8(0.2)  &     2.1(0.6)  & 23(s)  & 6 \\
10 $\rightarrow$  9  \ \  &   90978.989$^F$   &     24.0   &   5.81 $\times$ 10$^{-5}$  &    10.0  &    21  &     86(11)  &   10.5(0.2)  &     2.9(0.4)  & 28     & 6 \\
    11 $\rightarrow$  10  &  100076.385$^F$   &     28.8   &   7.77 $\times$ 10$^{-5}$  &    11.0  &    23  &     80(14)  &   10.5(0.2)  &     1.8(0.3)  & 42     & 6 \\
    12 $\rightarrow$  11  &  109173.637$^F$   &     34.1   &   1.01 $\times$ 10$^{-4}$  &    12.0  &    25  &     76(8)   &   10.7(0.1)  &     2.0(0.2)  & 37     & 10  \\
    15 $\rightarrow$  14  &  136464.402$^W$   &     52.4   &   1.99 $\times$ 10$^{-4}$  &    15.0  &    31  &    164(59)  &    ---       &      ---      & ---    & 3  \\
    16 $\rightarrow$  15  &  145560.949$^W$   &     59.4   &   2.42 $\times$ 10$^{-4}$  &    16.0  &    33  &     95(59)  &    ---       &      ---      & ---    & 4   \\
\hline                                                                
             \end{tabular}                                                         
             \end{center}  
             \tablefoot{Frequencies, \mbox{$E_{\rm u}/k$}, \mbox{$A_{\rm ul}$}, \mbox{$S_{\rm ul}$}, and \mbox{$g_{\rm u}$} from MADEX code, 
                              that fit to all rotational lines reported by 
             \citet{Zafra_1971,Mbosei_2000,Creswell_1977,Chen_1991,Yamada_1995,Thorwirth_2000}.}
            \end{table*}

 
          \begin{table*}[!h] 
          \begin{center}
          \vspace{1cm}
          \caption{Line parameters of CH$_{3}$CN.}  \label{Table_CH3CN}
          \begin{tabular}{ c c c r c c c c c c c c@{\vrule height 10pt depth 5pt width 0pt}}     
          \hline\hline

    Transition & Sym.$^{*}$ & Frequency &  $E_{\rm u}/k$ &  $A_{\rm ul}$ &  $S_{\rm ul}$ &  $g_{\rm u}$ &  $\displaystyle{\int} T_{_{\rm MB}}$d$v$ &  $v_{_{\rm LSR}}$ &  $\Delta v$ &  $T_{_{\rm MB}}$ & $S/N$ \rule[-0.3cm]{0cm}{0.8cm}\ \\ \cline{1-1}

      $(J_{\rm K})_{\rm u}$ $\rightarrow$ $(J_{\rm K})_{\rm l}$  &  & [MHz] & [K] & [$\mathrm{s^{-1}}$] & & & [$\mathrm{mK\, km\, s^{-1}}$] & [$\mathrm{km\, s^{-1}}$] & [$\mathrm{km\, s^{-1}}$]  & [$\mathrm{mK}$]  & \\

          \hline

   5$_{4}\rightarrow$ 4$_{4}$   &  E  &   91958.726$^F$    &   119.5  &  2.28 $\times$ 10$^{-5}$  &    1.8   &  11  &    26(13)   &  12.5(0.5)  &   2.0(1.0) &   13   &  3   \\
   5$_{3}\rightarrow$ 4$_{3}$   &  A  &   91971.130$^F$    &    77.5  &  4.06 $\times$ 10$^{-5}$  &    6.4   &  22  &   103(9)    &  10.2(0.2)  &   4.0(0.4) &   24   &  6     \\
   5$_{2}\rightarrow$ 4$_{2}$   &  E  &   91979.993$^F$    &    33.8  &  5.33 $\times$ 10$^{-5}$  &    4.2   &  11  &    84(16)   &  10.6(0.3)  &   3.0(0.7) &   26   &  6      \\
   5$_{1}\rightarrow$ 4$_{1}$   &  E  &   91985.313$^F$    &    12.4  &  6.09 $\times$ 10$^{-5}$  &    4.8   &  11  &   114(12)   &  10.5(0.1)  &   2.6(0.4) &   40   &  9     \\
   5$_{0}\rightarrow$ 4$_{0}$   &  A  &   91987.087$^F$    &    13.2  &  6.35 $\times$ 10$^{-5}$  &    5.0   &  11  &   138(13)   &  10.9(0.1)  &   2.8(0.4) &   46   &  11    \\                                       
   6$_{3}\rightarrow$ 5$_{3}$   &  A  &  110364.353$^F$    &    82.8  &  8.35 $\times$ 10$^{-5}$  &    9.0   &  26  &   151(11)   &  10.8(0.1)  &   2.7(0.2) &   53   &  8  \\
\hline                                                                                                                                                                                           
         \end{tabular}                                                                                                                                                                                    
         \end{center}                                                                                                                                                                                      
         \end{table*}

              \begin{table*}[!h] 
          \begin{center}
           \setcounter{table}{8}
          \caption{continued}
          \begin{tabular}{ c c c r c c c c c c c c@{\vrule height 10pt depth 5pt width 0pt}}      
          \hline\hline

    Transition & Sym.$^{*}$ & Frequency &  $E_{\rm u}/k$ &  $A_{\rm ul}$ &  $S_{\rm ul}$ &  $g_{\rm u}$ &  $\displaystyle{\int} T_{_{\rm MB}}$d$v$ &  $v_{_{\rm LSR}}$ &  $\Delta v$ &  $T_{_{\rm MB}}$ & $S/N$ \rule[-0.3cm]{0cm}{0.8cm}\ \\ \cline{1-1}

     $(J_{\rm K})_{\rm u}$ $\rightarrow$ $(J_{\rm K})_{\rm l}$  &  & [MHz] & [K] & [$\mathrm{s^{-1}}$] & & & [$\mathrm{mK\, km\, s^{-1}}$] & [$\mathrm{km\, s^{-1}}$] & [$\mathrm{km\, s^{-1}}$]  & [$\mathrm{mK}$]  & \\

          \hline

   9$_{2}\rightarrow$ 8$_{2}$   &  E  &  165556.321$^W$    &    60.3  &  3.66 $\times$ 10$^{-4}$  &    8.6   &  19  &  108(19)  & ---   & ---    & ---   &   7 \\               
   9$_{1}\rightarrow$ 8$_{1}$   &  E  &  165565.890$^W$    &    38.8  &  3.81 $\times$ 10$^{-4}$  &    8.9   &  19  &     \multirow{2}{*}[0cm]{$\Big \rangle$ 419(24)}   &  \multirow{2}{*}[-0.01cm]{---}   &   \multirow{2}{*}[-0.01cm]{---}    &   \multirow{2}{*}[-0.01cm]{---}        &  \multirow{2}{*}[-0.01cm]{14}    \\                     
   9$_{0}\rightarrow$ 8$_{0}$   &  A  &  165569.080$^W$    &    39.7  &  3.86 $\times$ 10$^{-4}$  &    9.0   &  19  &            &            &             &       &           \\   
  11$_{3}\rightarrow$ 10$_{3}$  &  A  &  202320.442$^F$    &   122.6  &  6.58 $\times$ 10$^{-4}$  &   20.3   &  46  &   140(11)  &  10.7(0.1)  &   2.0(0.2) &   64  &  8  \\  
  11$_{2}\rightarrow$ 10$_{2}$  &  E  &  202339.920$^F$    &    78.8  &  6.87 $\times$ 10$^{-4}$  &   10.6   &  23  &    78(9)   &  10.7(0.1)  &   1.5(0.2) &   49  &  6 \\
  11$_{1}\rightarrow$ 10$_{1}$  &  E  &  202351.610$^F$    &    57.4  &  7.05 $\times$ 10$^{-4}$  &   10.9   &  23  &   103(13)  &  10.7(0.2)  &   2.9(0.4) &   33  &  5 \\
\hline                                                                                                                                                                                             
         \end{tabular}                                                                                                                                                                                      
         \end{center}    
          \tablefoot{Frequencies, \mbox{$E_{\rm u}/k$}, \mbox{$A_{\rm ul}$}, \mbox{$S_{\rm ul}$}, and \mbox{$g_{\rm u}$} from MADEX code, 
           that fit to all rotational lines reported by 
      \citet{Kukolich_1973,Boucher_1977,Kukolich_1982,Pavone_1990,Cazzoli_2006,Simeckova_2004,Muller_2009}.}
         \end{table*}


            \begin{table*}[!h] 
          \begin{center}
          \caption{Line parameters of CH$_{3}$OH.}  \label{Table_CH3OH}
          \begin{tabular}{ c c c r c c c c c c c c@{\vrule height 10pt depth 5pt width 0pt}}    
          \hline\hline       
  
     Transition & Sym.$^{*}$ & Frequency &  $E_{\rm u}/k$ &  $A_{\rm ul}$ &  $S_{\rm ul}$ &  $g_{\rm u}$ &  $\displaystyle{\int} T_{_{\rm MB}}$d$v$ &  $v_{_{\rm LSR}}$ &  $\Delta v$ &  $T_{_{\rm MB}}$ & $S/N$ \rule[-0.3cm]{0cm}{0.8cm}\ \\ \cline{1-1}

        $(J_{\rm K})_{\rm u}$ $\rightarrow$ $(J_{\rm K})_{\rm l}$  &  & [MHz] & [K] & [$\mathrm{s^{-1}}$] & & & [$\mathrm{mK\, km\, s^{-1}}$] & [$\mathrm{km\, s^{-1}}$] & [$\mathrm{km\, s^{-1}}$]  & [$\mathrm{mK}$]  & \\
         
            \hline

     5$_{-1}\rightarrow$ 4$_{ 0}$  \ \  &  E  &   84521.172$^F$ &   32.5 &   1.97 $\times$ 10$^{-6}$ &   1.5 &   11   &    149(8)  &   10.6(0.1) &    1.6(0.1) &  86     & 18   \\                                                                                                                          
     8$_{ 0}\rightarrow$ 7$_{ 1}$       &  A  &   95169.391$^F$ &   83.5 &   2.13 $\times$ 10$^{-6}$ &   1.8 &   17   &     59(6)  &   10.4(0.1) &    1.5(0.1) &  54     & 14    \\                                                                                                                         
     2$_{ 1}\rightarrow$ 1$_{ 1}$       &  A  &   95914.310$^F$ &   21.4 &   2.49 $\times$ 10$^{-6}$ &   1.5 &    5   &     28(12) &   10.8(0.3) &    1.7(0.7) &  15     &  2    \\                                                                                                                        
     2$_{-1}\rightarrow$ 1$_{-1}$       &  E  &   96739.358$^F$ &    4.6 &   2.56 $\times$ 10$^{-6}$ &   1.5 &    5   &    107(9)  &   10.7(0.1) &    1.5(0.1) &  67     &  9    \\                                                                                                                        
     2$_{ 0}\rightarrow$ 1$_{ 0}$       &  A  &   96741.371$^F$ &    7.0 &   3.41 $\times$ 10$^{-6}$ &   2.0 &    5   &    206(9)  &   10.8(0.1) &    1.7(0.2) & 114     & 16   \\
\hline     
        \end{tabular}
        \end{center} 
        \end{table*}

             \begin{table*}[!h] 
          \begin{center}
          \setcounter{table}{9}
          \caption{continued.}
          \begin{tabular}{ c c c r c c c c c c c c@{\vrule height 10pt depth 5pt width 0pt}}      
          \hline\hline       

       Transition & Sym.$^{*}$ & Frequency &  $E_{\rm u}/k$ &  $A_{\rm ul}$ &  $S_{\rm ul}$ &  $g_{\rm u}$ &  $\displaystyle{\int} T_{_{\rm MB}}$d$v$ &  $v_{_{\rm LSR}}$ &  $\Delta v$ &  $T_{_{\rm MB}}$ & $S/N$ \rule[-0.3cm]{0cm}{0.8cm}\ \\ \cline{1-1}

          $(J_{\rm K})_{\rm u}$ $\rightarrow$ $(J_{\rm K})_{\rm l}$  &  & [MHz] & [K] & [$\mathrm{s^{-1}}$] & & & [$\mathrm{mK\, km\, s^{-1}}$] & [$\mathrm{km\, s^{-1}}$] & [$\mathrm{km\, s^{-1}}$]  & [$\mathrm{mK}$]  & \\
         
         \hline
 
     5$_{ 0}\rightarrow$ 4$_{ 0}$       &  A  &  241791.352$^F$ &   34.8 &   6.05 $\times$ 10$^{-5}$ &   5.0 &   11   &    646(43) &   10.5(0.2) &    2.2(0.2) & 273   & 27        \\                                                                                                                     
     5$_{-4}\rightarrow$ 4$_{-4}$       &  E  &  241813.255$^F$ &  114.8 &   2.18 $\times$ 10$^{-5}$ &   1.8 &   11   &     38(13) &   10.7(0.2) &    1.2(0.6) &  28   & 2        \\                                                                                                                                                                        
     5$_{ 1}\rightarrow$ 4$_{ 1}$       &  E  &  241879.025$^F$ &   48.0 &   5.96 $\times$ 10$^{-5}$ &   5.0 &   11   &    205(25) &   10.5(0.3) &    3.3(0.6) &  56   &  7       \\                                                                                                                                                                            
     5$_{-2}\rightarrow$ 4$_{-2}$       &  E  &  241904.147$^F$ &   52.8 &   5.09 $\times$ 10$^{-5}$ &   4.2 &   11   &   \multirow{2}{*}[0cm]{$\Big \rangle$ 179(21)}   &  \multirow{2}{*}[-0.01cm]{10.5(0.1)}   &   \multirow{2}{*}[-0.01cm]{2.4(0.3)}    &   \multirow{2}{*}[-0.01cm]{67}        &  \multirow{2}{*}[-0.01cm]{7}    \\
\hline     
       \end{tabular}
       \end{center} 
       \tablefoot{Frequencies, \mbox{$E_{\rm u}/k$}, \mbox{$A_{\rm ul}$}, \mbox{$S_{\rm ul}$}, and \mbox{$g_{\rm u}$} from JPL catalogue.}
       \end{table*}

                    
          \begin{table*}[!h] 
         \begin{center}
         \vspace{0.5cm}
         \caption{Line parameters of CH$_{3}$CHO.}  \label{Table_CH3CHO}  
         \begin{tabular}{ c c c r c c c c c c c c@{\vrule height 10pt depth 5pt width 0pt}}    
         \hline\hline       
         
   Transition & Sym.$^{*}$ & Frequency &  $E_{\rm u}/k$ &  $A_{\rm ul}$ &  $S_{\rm ul}$ &  $g_{\rm u}$ &  $\displaystyle{\int} T_{_{\rm MB}}$d$v$ &  $v_{_{\rm LSR}}$ &  $\Delta v$ &  $T_{_{\rm MB}}$ & $S/N$ \rule[-0.3cm]{0cm}{0.8cm}\ \\ \cline{1-1}

       $(J_{K_{\rm a},K_{\rm c}})_{\rm u} \rightarrow (J_{K_{\rm a},K_{\rm c}})_{\rm l}$  &  & [MHz] & [K] & [$\mathrm{s^{-1}}$] & & & [$\mathrm{mK\, km\, s^{-1}}$] & [$\mathrm{km\, s^{-1}}$] & [$\mathrm{km\, s^{-1}}$]  & [$\mathrm{mK}$] &  \\
        
         \hline

   5$_{1,5}\rightarrow$ 4$_{1,4}$      &   A  &   93580.909$^F$    &   15.7  &  2.63 $\times$ 10$^{-5}$  &   5.2   &   11   &  60(6)  &  10.4(0.1)  &   2.7(0.3) &  21  & 6   \\      
   5$_{1,5}\rightarrow$ 4$_{1,4}$      &   E  &   93595.235$^F$    &   15.7  &  2.63 $\times$ 10$^{-5}$  &   5.2   &   11   &  51(7)  &  10.8(0.2)  &   2.3(0.4) &  21  & 6  \\       
   5$_{0,5}\rightarrow$ 4$_{0,4}$      &   E  &   95947.437$^F$    &   13.8  &  2.96 $\times$ 10$^{-5}$  &   5.4   &   11   &  29(6)  &  10.7(0.3)  &   2.0(0.4) &  13  & 3 \\ 
   5$_{0,5}\rightarrow$ 4$_{0,4}$      &   A  &   95963.459$^F$    &   13.8  &  2.95 $\times$ 10$^{-5}$  &   5.4   &   11   &  54(13) &  10.7(0.2)  &   2.0(0.5) &  25  & 3    \\     
   5$_{1,4}\rightarrow$ 4$_{1,3}$      &   E  &   98863.314$^F$    &   16.5  &  3.10 $\times$ 10$^{-5}$  &   5.2   &   11   &  79(6)  &  11.1(0.3)  &   2.9(0.5) &  26  & 3   \\               
   5$_{1,4}\rightarrow$ 4$_{1,3}$      &   A  &   98900.944$^F$    &   16.5  &  3.11 $\times$ 10$^{-5}$  &   5.2   &   11   &  63(10) &  11.1(0.1)  &   1.6(0.3) &  38  & 7  \\
\hline                                                                                                    
                                                                                               
        \end{tabular}                                     
        \end{center} 
          \end{table*}

     \begin{table*}[!h] 
         \begin{center}
         \setcounter{table}{10}
         \caption{continued.}  
         \begin{tabular}{ c c c r c c c c c c c c@{\vrule height 10pt depth 5pt width 0pt}}    
         \hline\hline       
         
   Transition & Sym.$^{*}$ & Frequency &  $E_{\rm u}/k$ &  $A_{\rm ul}$ &  $S_{\rm ul}$ &  $g_{\rm u}$ &  $\displaystyle{\int} T_{_{\rm MB}}$d$v$ &  $v_{_{\rm LSR}}$ &  $\Delta v$ &  $T_{_{\rm MB}}$ & S/N \rule[-0.3cm]{0cm}{0.8cm}\ \\ \cline{1-1}

       $(J_{K_{\rm a},K_{\rm c}})_{\rm u} \rightarrow (J_{K_{\rm a},K_{\rm c}})_{\rm l}$  &  & [MHz] & [K] & [$\mathrm{s^{-1}}$] & & & [$\mathrm{mK\, km\, s^{-1}}$] & [$\mathrm{km\, s^{-1}}$] & [$\mathrm{km\, s^{-1}}$]  & [$\mathrm{mK}$] &  \\
        
         \hline
   7$_{1,6}\rightarrow$ 6$_{1,5}$      &   A  &  138319.628$^W$    &   28.8  &  8.90 $\times$ 10$^{-5}$  &   7.4   &   15   & 166(62) &  ---  & ---  & ---         &  3   \\     
   8$_{1,8}\rightarrow$ 7$_{1,7}$      &   E  &  149505.128$^W$    &   34.6  &  1.14 $\times$ 10$^{-4}$  &   8.5   &   17   &     \multirow{2}{*}[0cm]{$\Big \rangle$ 193(96)}   &  \multirow{2}{*}[-0.01cm]{---}   &   \multirow{2}{*}[-0.01cm]{---}    &   \multirow{2}{*}[-0.01cm]{---}        &  \multirow{2}{*}[-0.01cm]{2}    \\     
   8$_{1,8}\rightarrow$ 7$_{1,7}$      &   A  &  149507.462$^W$    &   34.6  &  1.14 $\times$ 10$^{-4}$  &   8.5   &   17   &  &   &   &         &   \\       
   8$_{0,8}\rightarrow$ 7$_{0,7}$      &   E  &  152607.614$^W$    &   33.1  &  1.23 $\times$ 10$^{-4}$  &   8.6   &   17   & 184(74) &  ---  & ---  & ---         & 4    \\     
   8$_{1,7}\rightarrow$ 7$_{1,6}$      &   E  &  157937.697$^W$    &   36.4  &  1.34 $\times$ 10$^{-4}$  &   8.5   &   17   &  86(21) &  ---  & ---  & ---         & 4        \\         
   8$_{1,7}\rightarrow$ 7$_{1,6}$      &   A  &  157974.590$^W$    &   36.4  &  1.34 $\times$ 10$^{-4}$  &   8.5   &   17   & 105(24) &  ---  & ---  & ---         & 6     \\
\hline                                                                                                    
                                                                                               
        \end{tabular}                                     
        \end{center} 
         \tablefoot{$^B$ Blended with CF$^+$ 2 $\rightarrow$ 1. Frequencies, \mbox{$E_{\rm u}/k$}, \mbox{$A_{\rm ul}$}, \mbox{$S_{\rm ul}$}, and \mbox{$g_{\rm u}$} from JPL catalogue.} 
          \end{table*}


\end{document}